\documentclass[aps,prb,twocolumn,groupedaddress,showpacs]{revtex4}

\usepackage{graphicx} \usepackage{amsmath,amssymb,amsfonts}

\newcommand{\EqLabel}[1]{\label{#1}} \newcommand{\mb}[1]{\mathbf{#1}}

\begin{document}

\title{The Green's function of the Holstein polaron}

\author{Glen L. Goodvin} \author{Mona Berciu} \author{George
A. Sawatzky} \affiliation{Department of Physics and Astronomy,
University of British Columbia, Vancouver, BC, V6T 1Z1}
\date{\today}

\begin{abstract}
We present a novel, highly efficient yet accurate analytical
approximation for the Green's function of a Holstein polaron. It is
obtained by summing all the self-energy diagrams, but with  each
self-energy diagram  averaged over the momenta of its free
propagators. The result becomes exact for both zero bandwidth and
for zero electron-phonon coupling, and is accurate everywhere in the
parameter space. The resulting Green's function satisfies exactly
the first six spectral weight sum rules. All higher sum rules are
satisfied with great accuracy, becoming asymptotically exact for
coupling both much larger and much smaller than the free particle
bandwidth. Comparison with existing numerical data also confirms
this accuracy. We use this approximation to analyze in detail the
redistribution of the spectral weight as the coupling strength
varies.
\end{abstract}
\pacs{71.38.-k, 72.10.Di, 63.20.Kr}

\maketitle

\section{Introduction}

There is considerable interest to understand the effects on the
properties of a particle coming from interactions with an
environment. Examples of such problems abound in condensed matter;
the problem discussed here is that of an electron coupled to lattice
vibrations, i.e. of electron-phonon coupling. For example, such
coupling is believed to be relevant for understanding certain
aspects of the high-temperature superconductors' behavior in the
underdoped limit (where the ``particle'' coupling to phonons is the
doping hole already dressed by interactions with the electrons in
the lower Hubbard
band)\cite{lanzara:2001,shen:2004,mishchenko:2004,zhou:06} but there
are many examples of other materials characterized by
strong-electron phonon coupling, including polymers like
polyacetylene, nanotubes, C$_{60}$ molecules and other
fullerenes.\cite{hengs:99,gunn:94,su:79} Other problems of the same
general type regard electrons coupled to spin-waves of a
magnetically ordered system,\cite{mishchenko:01} or to orbitrons,
for example in manganites,\cite{isihara:05,hotta:06,uhrig:06} or to
some combination thereof.

In this work, we focus on the simplest Hamiltonian describing an
electron on a lattice interacting with an optical phonon mode,
namely the Holstein model:\cite{holstein:59}
%%%%%%%%%%%%%%%%%%%%%%%%%%%%%% EQUATION %%%%%%%%%%%%%%%%%%%%%%%%%%%%%%
\begin{equation}
\EqLabel{eq:holstein} {\cal H}=\sum_{\mb{k}} \left(
\varepsilon_{\mb{k}} c_{\mb{k} }^{\dagger} c_{\mb{k}}+\Omega
b_{\mb{k}}^{\dagger}b_{\mb{k}} \right)
\\ + \frac{g}{\sqrt{N}} \sum_{\mb{k}, \mb{q}}
c_{\mb{k}-\mb{q}}^{\dagger} c_{\mb{k}} \left(b_{\mb{q}}^{\dagger} +
b_{\mb{-q}} \right).
\end{equation}
%%%%%%%%%%%%%%%%%%%%%%%%%%%%%%%%%%%%%%%%%%%%%%%%%%%%%%%%%%%%%%%%%%%%%%
The first term is the kinetic energy of the electron, with
$c_{\mb{k}}^{\dagger}$ and $c_{\mb{k}}$ being the electron creation
and annihilation operators. For the single dressed particle (known
as polaron, in this case) problem of interest to us, the spin of the
particle is irrelevant and we suppress its index.
$\varepsilon_{\mb{k}}$ is the free-particle dispersion. In all
results shown here, we assume nearest-neighbor hopping on a
$d$-dimensional simple cubic lattice of constant $a$ and a total of $N$
sites with periodic boundary conditions, so that
\begin{equation}
\varepsilon_{\mb{k}}=-2t \sum_{i=1}^d \cos(k_i a),
\end{equation}
but our results are valid for any other dispersion.  The second term
describes a branch of optical phonons of energy $\Omega$ (we set
$\hbar=1$ throughout this paper). $b_{\mb{q}}^{\dagger}$ and
$b_{\mb{q}}$ are the phonon creation and annihilation operators. The
last term is the on-site linear electron-phonon coupling
$g\sum_{i}^{} c^\dag_i c_i (b^\dag_i + b_i)$, written in
$\mb{k}$-space. All sums over momenta are over the first Brillouin
zone, $-{\pi\over a} < k_i \leq {\pi\over a}$.

The quantity of interest to us is the Green's function of the single
dressed particle, or polaron, defined as:\cite{mahan:1981}
\begin{equation}
\label{eq:green_time_ordered} G(\mb{k},\tau)=-i \langle 0| T
[c_{\mb{k}}^{\dagger}(\tau) c_{\mb{k}}(0)] | 0 \rangle,
\end{equation}
where $|0\rangle$ is the ground state of the zero-particle system,
which is the vacuum. $T$ is the time ordering operator, and $
c_{\mb{k}}(\tau)=e^{i {\cal H} \tau} c_{\mb{k}} e^{-i{\cal H} \tau}
$. Since ${\cal H}|0\rangle =0$ and $c_{\mb{k}}|0\rangle = 0$, Eq.
(\ref{eq:green_time_ordered}) simplifies to:
\begin{equation} \label{eq:green_simplified}
G(\mb{k},\tau) = -i \Theta(\tau) \langle 0| c_{\mb{k}} e^{-i{\cal H}
  \tau} c_{\mb{k}}^{\dagger} |0 \rangle,
\end{equation}
where $\Theta(\tau)$ is the Heaviside function. In other words, only
the retarded part contributes. This Green's function gives the
amplitude of probability that an electron introduced in the system
at $\tau=0$ and removed at a later time $\tau$, leaves the system in
its ground state. The usefulness of this quantity can be appreciated
using its Lehmann representation:\cite{mahan:1981}
%%%%%%%%%%%%%%%%%%%%%%%%%%%%%% EQUATION %%%%%%%%%%%%%%%%%%%%%%%%%%%%%%
\begin{equation}
\EqLabel{leh} G(\mb{k},\omega) = \sum_{\alpha}^{} \frac{|\langle
\alpha | c_{\mb{k}}^\dag|0\rangle|^2}{\omega - E_\alpha + i\eta},
\end{equation}
%%%%%%%%%%%%%%%%%%%%%%%%%%%%%%%%%%%%%%%%%%%%%%%%%%%%%%%%%%%%%%%%%%%%%%
where $\{ |\alpha\rangle\}$ and $\{E_\alpha\}$ are the complete set
of one-particle eigenstates and eigenenergies, ${\cal
H}|\alpha\rangle = E_\alpha |\alpha\rangle$, $\sum_{\mb k}^{}
c^\dag_{\mb k} c_{\mb k} |\alpha\rangle = |\alpha\rangle$. Thus, the
poles of the Green's function give the {\em whole one-particle
spectrum}, while the associated residues, called quasi-particle
($qp$) weights, give partial information on the nature of the
eigenstates. Moreover, the imaginary part of this Green's
function, called the spectral weight, is directly measured
experimentally through angle-resolved photoemission
spectroscopy.\cite{damascelli:2003}

There has already been a large amount of work devoted to
understanding the properties of the Holstein polaron, and we briefly
review some of it here. Most of it is numerically-intensive work.
Some examples are (i) exact diagonalization (ED)
methods.\cite{mello:97,marsiglio:1995} These are usually hampered by
the fact that even for a finite lattice, the Hilbert space is
infinite due to the infinite number of possible phonon
configurations. Thus, some truncation scheme is needed, but for
small phonon frequencies and/or large couplings, the CPU resources
needed and run times become prohibitive. This has led to a number of
(ii) proposals based on variational approaches to decide which
phonon configurations should be
included,\cite{bonca:1999,romero:1999,romero:1999-2,romero:1999-3,cataudella:1999,cataudella:2000,cataud:04,filippis:2005,barisic:2004}
as well as (iii) various Quantum Monte Carlo (QMC)
methods.\cite{QMC} A brief review of these is provided in Ref.
\onlinecite{hohenadler:2004}. Of special interest are the so-called
diagrammatic Monte-Carlo
simulations\cite{prokofev:1998,mishchenko:2000,macridin:2003} which
calculate directly the Green's function in imaginary time, by {\em
numerically} summing all diagrams in the perturbational expansion.
We make extensive use of comparisons with low-energy results of this
method from Ref. \onlinecite{macridin:2003}. While this method
allows in principle the exact calculation of the Green's function,
the requirement of convergence for the propagator series usually
means that only low-energy properties are shown. Finally, there are
methods suitable for some particular cases, such as density-matrix
renormalization group for one-dimensional
systems,\cite{jeckelmann:1998} and dynamic mean-field theory (DMFT)
for infinite-dimensional systems.\cite{ciuchi:1997}

Of course, given the long history of this problem, many analytical
techniques have been applied to it with varying degrees of success
(for a review, see Ref. \onlinecite{Alex:95}). First of all, the
Green's function is known exactly in two asymptotic limits. If there
is no coupling, $g=0$, then the Green's function is that of the free
electron. The ground-state is at $\epsilon_{\mb 0} = -2dt$ and the
spectrum consists of a continuous band extending from $[-2dt, 2dt]$
(for the tight-binding model). The so-called impurity limit, with
$t=0$, also has an exact solution, given by the Lang-Firsov
formula\cite{lang:1963}
\begin{equation} \label{eq:green_LF_zerohopping}
G(\omega)=e^{-{g^2\over
    \Omega^2}}\sum_{n=0}^{\infty}\frac{1}{n!}\left(\frac{g}{\Omega}\right)^{2n}
    \frac{1}{\omega+{g^2\over \Omega}-n\Omega+i\eta}.
\end{equation}
This can be viewed as the strong-coupling limit, since for $t=0$,
$g$ becomes the important energy scale in the system.  In this limit
the electron is localized at one site in real space, therefore it is
fully delocalized in ${\mb k}$-space, and the Green's function is
independent of ${\mb k}$. The spectrum has the ground-state (GS) at
$E_0= - g^2/\Omega$ and an infinite sequence of equidistant levels
spaced by $\Omega$ above it. This is extremely different from the
free-particle spectrum, and it is of considerable interest to
understand not only how the ground-state evolves from $-2dt$ to
$-g^2/\Omega$ as the coupling $g$ is increased, but also the
evolution of all the higher-energy spectral weight from a
continuous, finite-width band to an infinite set of discrete levels.
Note that it is customary to describe the effective coupling as the
ratio of the two asymptotic ground-state energies, using as a new
parameter
$$ \lambda = {g^2\over 2dt \Omega}.
$$

Most of the numerical methods reviewed above calculate only GS or
low-energy properties, given the significant CPU time and numerical
resources needed to calculate the whole spectrum. Very recently,
several sets of whole-spectrum results have become
available,\cite{filippis:2005,hohenadler:2004,hohenadler:2003,hohenadler:2005,
bayo} however only for a few points in the parameter space, and
generally for low dimensions.

It is of obvious interest to find an analytical approximation for
the Green's function that is simple to estimate, so that the whole
parameter space can be studied easily, but also with high accuracy.
This is precisely what we propose here (a short version of this work
has been published in Ref. \onlinecite{berciu:2006}). We call our
approximation the momentum average (MA) approximation; its essence
consists in analytically summing {\em all the diagrams} in the
diagrammatic expansion, but with each diagram simplified in a
certain way. Before introducing this method, we briefly review here
the other two simple (in terms of computational effort) analytical
approximations for the Green's function of the Holstein polaron,
available in the literature.

The first is the self-consistent Born approximation (SCBA), which
consists of summing exactly only the non-crossed diagrams.  The
percentage of diagrams kept decreases fast with increasing order
(see Table \ref{tab1}). If the coupling is small, the sum is
dominated by the low order diagrams and SCBA works reasonably well.
At strong coupling, the contribution of higher order diagrams
becomes essential, and SCBA is expected to fail (see below). In this
approximation, the Green's function is written in terms of a
self-energy:
$$ G_{\textrm{SCBA}}(\mb{k}, \omega) = {1\over \omega - \epsilon_{\mb k} -
\Sigma_{\textrm{SCBA}}(\omega) + i \eta},
$$ with the self-consistency condition
$$ \Sigma_{\textrm{SCBA}}(\omega) = \frac{g^2}{N} \sum_{\mb{q}}
G_{\textrm{SCBA}}(\mb{k}-\mb{q},\omega-\Omega).
$$ Note that $\Sigma_{\textrm{SCBA}}(\omega)$ is independent of ${\mb
k}$. This is a consequence of the simplicity of the Holstein model:
if either the coupling $g$ or the dispersion $\Omega$ were functions
of the phonon momentum ${\mb q}$, the SCBA self-energy would depend
explicitly on ${\mb k}$.\cite{slezak:2006}
$\Sigma_{\textrm{SCBA}}(\omega)$ can be expressed\cite{berciu:2006}
as a function of the average of the free propagator over the
Brillouin zone (BZ) and can be evaluated very efficiently.

\begin{table}[b]
\begin{ruledtabular}
 \begin{tabular}{||c|c|c|c|c|c|c|c|c||}
 order & 1 & 2& 3& 4& 5 & 6 & 7& 8\\ \hline total & 1 & 2& 10& 74&
706& 8162& 110410 & 1708394\\ \hline SCBA & 1 & 1& 2& 5& 14 & 42 &
132
& 429\\
\end{tabular}
\caption{Comparison between the total number of diagrams of a given
  order in the proper self-energy $\Sigma(\mb{k},\omega)$ vs. the number of
  diagrams of a given order kept within SCBA.}
\label{tab1}
\end{ruledtabular}
\end{table}

The other simple analytical approximation for the Green's function
of the Holstein model is the generalized Lang-Firsov (LF)
expression.\cite{alexandrov:1992,kornilovitch:2002} It is
reminiscent of the Lang-Firsov expression of Eq.
(\ref{eq:green_LF_zerohopping}):
\begin{equation} \label{eq:green_LF}
G_{\textrm{LF}}(\mb{k},\omega) = e^{-{g^2\over
    \Omega^2}}\sum_{n=0}^{\infty}
    \frac{\frac{1}{n!}\left(\frac{g}{\Omega}\right)^{2n}
    }{\omega-e^{-{g^2\over
    \Omega^2}}\varepsilon_{\mb{k}}+{g^2\over
    \Omega}-n\Omega+i\eta}.
\end{equation}
This expression is exact in both asymptotic limits $\lambda=0$
($g=0$) and $\lambda \rightarrow \infty$ ($t=0)$, but less accurate
for finite $t$ and $g$, as we show below.

The article is organized as follows. In Section II, we derive the
momentum average approximation. Its diagrammatic meaning is
discussed in Section III, where we also estimate its corresponding
spectral weight sum rules. We show that MA satisfies exactly the
first 6 sum rules, but more importantly, it remains highly accurate
for higher order sum rules. This is a strong argument in favor of
its accuracy. The accuracy is gauged in more detail in Section IV,
where we compare the MA predictions against the SCBA and generalized
LF predictions, but also against a host of numerical results. This
indeed demonstrates that the MA approximation is remarkably accurate
for all parameter values, especially given its simplicity. In this
section we also present some new results regarding various
properties of the Holstein polaron. Finally, Section V contains our
summary and conclusions.

\section{Calculating the  Green's function}

\subsection{Exact solution}

As is always the case for Green's functions, one can use the
equation of motion technique to generate an infinite hierarchy of
coupled equations for an infinite number of related Green's
functions. We derive it here for the Holstein polaron.

In the frequency domain, this approach is equivalent to using
repeatedly Dyson's identity $\hat{G}(\omega) = \hat{G}_0(\omega) +
\hat{G}(\omega) \hat{V} \hat{G}_0(\omega)$, which holds for any
Green's operators $\hat{G}(\omega)=[\omega - \hat{{\cal
H}}+i\eta]^{-1}$, $\hat{G}_0(\omega)=[\omega - \hat{{\cal
H}}_0+i\eta]^{-1}$ and for any Hamiltonians $\hat{{\cal H}}=
\hat{{\cal H}}_0 + \hat{V}$. As customary, we take $\hat{V}$ to be
the electron-phonon interaction. Applying Dyson's identity once, we
obtain:
\begin{equation} \label{eq:G}
G(\mb{k},\omega) = G_0(\mb{k}, \omega) \left[
1+\frac{g}{\sqrt{N}}\sum_{\mb{q}_1} F_1(\mb{k},\mb{q}_1,\omega)
\right],
\end{equation}
where
\begin{equation}
  G_0(\mb{k}, \omega) = \langle 0| c_{\mb
  k}\hat{G}_0(\omega)c_{\mb k}^\dag |0\rangle = \frac{1}{\omega-
  \varepsilon_{\mb{k}} + i\eta} \label{eq:G0}
\end{equation}
is the free particle Green's function. We made use of the equality
$\hat{V} c_{\mb{k}}^{\dagger} |0 \rangle =
\frac{g}{\sqrt{N}}\sum_{\mb{q}}c_{\mb{k}-\mb{q}}^{\dagger}
b_{\mb{q}}^{\dagger}|0\rangle$ and defined a new Green's function:
$$ F_1(\mb{k},\mb{q}_1,\omega)=\langle0| c_{\mb{k}}
\hat{G}(\omega)c_{\mb{k}-\mb{q}_1}^{\dagger}
b_{\mb{q}_1}^{\dagger}|0 \rangle.
$$ $F_1$ is related to the amplitude of probability to start with the
electron and a phonon at the initial time, and find only the
electron in the system at the final time. Its own equation of motion
relates back to $G(\mb{k}, \omega)$ but also to a new Green's
function with two phonons initially. In general, if we define:
%%%%%%%%%%%%%%%%%%%%%%%%%%%%%% EQUATION %%%%%%%%%%%%%%%%%%%%%%%%%%%%%%
\begin{equation}
\nonumber F_n(\mb{k},\mb{q}_1,\dots,\mb{q}_n,\omega)=\langle0|
c_{\mb{k}} \hat{G}(\omega)c_{\mb{k}-\mb{q}_T}^{\dagger}
b_{\mb{q}_1}^{\dagger}\dots b_{\mb{q}_n}^{\dagger}|0 \rangle,
\end{equation}
%%%%%%%%%%%%%%%%%%%%%%%%%%%%%%%%%%%%%%%%%%%%%%%%%%%%%%%%%%%%%%%%%%%%%%
where $\mb{q}_T = \sum_{i=1}^{n} \mb{q}_i$ is the total momentum of
the $n$ initial phonons, using Dyson's identity we find its equation
of motion to be ($n\ge 1$):
\begin{multline} \label{eq:F}
F_n(\mb{k}, \mb{q}_1, \ldots, \mb{q}_n, \omega) =
\frac{g}{\sqrt{N}} G_0(\mb{k}-\mb{q}_T,\omega-n\Omega) \\
\times \left[ \sum_{i=1}^{n} F_{n-1}(\mb{k}, \mb{q}_1, \ldots,
\mb{q}_{i-1}, \mb{q}_{i+1}, \ldots, \mb{q}_n, \omega) \right. \\ +
\left. \sum_{\mb{q}_{n+1}} F_{n+1}(\mb{k}, \mb{q}_1, \ldots,
\mb{q}_{n+1}, \omega) \right],
\end{multline}
i.e. related to the Green's functions with $n-1$ and $n+1$ initial
phonons. Eqs. (\ref{eq:G}) and (\ref{eq:F}) form the exact infinite
hierarchy of coupled equations whose solution is the Holstein
polaron Green's function $G(\mb{k}, \omega)=F_0(\mb{k},\omega)$.

Obviously, this system of coupled equations can be solved trivially
in the limit $\lambda=g=0$, in which case $G(\mb{k},
\omega)=G_0(\mb{k}, \omega)$ directly from Eq. (\ref{eq:G}). An
exact solution equal to the Lang-Firsov result must also exist if $
t=0$. Indeed, in this limit all Green's functions become independent
of all momenta, and Eqs. (\ref{eq:G}) and (\ref{eq:F}) simplify to:
\begin{multline}
G(\omega) = G_0(\omega) \left[ 1+g\sqrt{N} F_1(\omega) \right]\\
F_n(\omega) = gG_0(\omega-n\Omega)\left[
\frac{n}{\sqrt{N}}F_{n-1}(\omega) +\sqrt{N}F_{n+1}(\omega) \right],
\nonumber
\end{multline}
where $G_0(\omega) = (\omega+i\eta)^{-1}$.  These recurrence
equations can be solved in terms of continued fractions. We briefly
review the solution here. We suppress the functional notation and
rewrite $F_n=\alpha_nF_{n-1}+\beta_nF_{n+1}$, where
$\alpha_n\equiv\frac{ng}{\sqrt{N}}G_0(\omega-n\Omega)$,
$\beta_n\equiv g\sqrt{N}G_0(\omega-n\Omega)$.  On physical grounds
we expect that $F_{n+1}$ becomes vanishingly small for a
sufficiently large $n$, since it describes physical processes which are less
and less likely. This allows one to solve these equations
iterationally starting from this sufficiently large $n$: $F_{n}
\approx \alpha_n F_{n-1}$, to find after solving for $F_{n-1}$,
$F_{n-2}$, etc., that:
\begin{equation}
F_1=\cfrac{\alpha_1}{1-\cfrac{\alpha_2\beta_1}
{1-\cfrac{\alpha_3\beta_{2}}{1-\cdots}}} F_0.
\end{equation}
Allowing the continued fraction to be infinite instead of truncated
after $n$ steps gives the exact solution.  Recalling that $F_0\equiv
G=G_0(1+g\sqrt{N}F_1)$, one can now solve for $G$. With the original
notation, we find:
\begin{equation}
\label{t0} G(\omega)=\cfrac{G_0(\omega)}{1-\cfrac{g^2
G_0(\omega)G_0(\omega - \Omega)}{1 - \cfrac{2g^2 G_0(\omega -
\Omega)G_0(\omega - 2\Omega)}{1 - \cdots}}}.
\end{equation}
After some further work, this can indeed be shown to equal the
Lang-Firsov expression of Eq. (\ref{eq:green_LF_zerohopping}).

\begin{figure}[t]
\includegraphics[width=0.90\columnwidth]{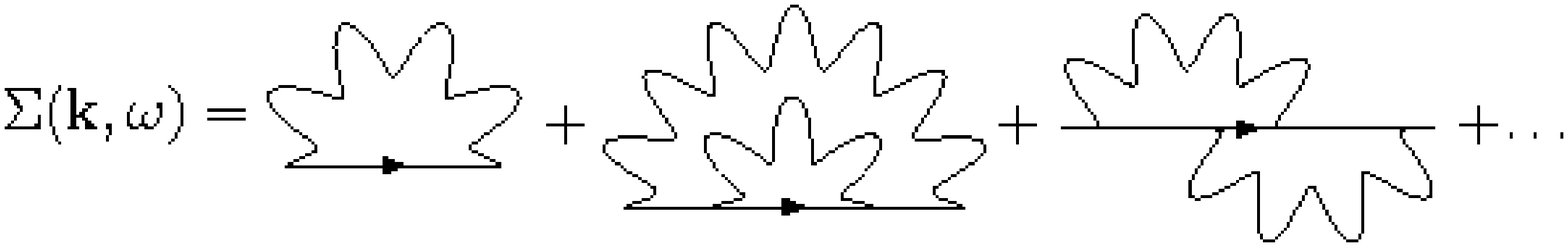}
\caption{The diagrammatic expansion for the self-energy.}
\label{fig:self_energy}
\end{figure}

This exact hierarchy of coupled equations [Eqs. (\ref{eq:G}) and
(\ref{eq:F})] can also be solved in the general case of finite $g$
and finite $t$ by iteratively solving for $F_1$, then $F_2$, etc.,
and removing them from this coupled system. It is straightforward to
verify that the solution obtained in this case is the diagrammatic
expansion, which can be partially resummed to the expected form:
\begin{equation}
G(\mb{k},\omega)=\frac{1}{[G_0(\mb{k},\omega)]^{-1}
  -\Sigma(\mb{k},\omega)},
\end{equation}
where the self-energy $\Sigma(\mb{k},\omega)$ is the sum of all
proper self-energy diagrams, the first few of which are shown in
Fig. \ref{fig:self_energy}. While this solution as a sum of an
infinite number of diagrams is exact, it is clearly not very useful
if the sum cannot be performed. One typical strategy in such cases
is to sum only a subset of these diagrams ({\em e.g.} the
non-crossed ones, in the Self-Consistent Born Approximation). This
is reasonable when one can argue that the diagrams kept contribute
much more than the neglected diagrams, which is not the case for
SCBA in this problem (see below). We propose a new strategy,
explained in the next subsection, to find an approximative solution
of these equations in the case of finite $t$ and finite $g$.

\subsection{The Momentum Average Approximation}

To obtain an approximate solution in the case of finite $t$ and $g$,
we proceed as follows. We first note that $G(\mb k, \omega)$ depends
on $F_1$ only through its average over the Brillouin zone $f_1(\mb
k,\omega) = \sum_{\mb q_1}^{} F_1(\mb k, \mb{q}_1, \omega)$,
\begin{equation}
\label{a1} G(\mb{k},\omega)=G_0(\mb{k},\omega)\left[ 1+g\sqrt{N}
f_1(\mb{k},\omega) \right].
\end{equation}
We define the \emph{momentum averaged} Green's functions:
\begin{equation} \label{eq:averaged}
f_n(\mb{k},\omega)=\frac{1}{N^n}\sum_{\mb{q}_1,
\ldots,\mb{q}_n}F_n(\mb{k}, \mb{q}_1,\ldots,\mb{q}_n,\omega),
\end{equation}
where all momenta sums run over the BZ, and attempt to express Eqs.
(\ref{eq:F}) in terms of these simpler quantities, by performing the
corresponding momenta averages on both sides.  While the first term
on the right-hand side can be written in terms of $f_{n-1}$ exactly,
the second term requires an approximation, which we choose to be:
%%%%%%%%%%%%%%%%%%%%%%%%%%%%%% EQUATION %%%%%%%%%%%%%%%%%%%%%%%%%%%%%%
\begin{eqnarray}
\EqLabel{app} \nonumber \sum_{\mb{q}_1,\ldots,\mb{q}_{n+1}}
F_{n+1}(\mb{k},\mb{q}_1,\ldots, \mb{q}_{n+1},\omega)
G_0(\mb{k}-\mb{q}_T,\omega-n\Omega) & &\\ \approx N
\bar{g}_0(\omega-n\Omega)f_{n+1}(\mb{k},\omega),\hspace{10mm} & &
\end{eqnarray}
%%%%%%%%%%%%%%%%%%%%%%%%%%%%%%%%%%%%%%%%%%%%%%%%%%%%%%%%%%%%%%%%%%%%%%
where
\begin{equation}
\label{eq:g0} \bar{g}_0(\omega)=\frac{1}{N}\sum_{\mb{k}}
G_0(\mb{k},\omega)
\end{equation}
is the free propagator momentum-averaged over the BZ.  One
justification for replacing $G_0(\mb{k}-\mb{q}_T,\omega-n\Omega)$ by
its momentum average $\bar{g}_0(\omega-n\Omega)$ in Eq. (\ref{app})
is that $\mb{q}_T=\sum_i^n\mb{q}_i$ takes, with equal probability,
any value in the BZ.  Moreover, in the impurity limit $t=0$ where
all Green's functions are momentum independent, Eq. (\ref{app}) is
exact. This suggests that our approach should be reasonable at least
in the strong-coupling limit $t\ll g$. In a more practical sense,
this approximation allows us to write $f_n(\mb{k},\omega)$ in terms
of $f_{n-1}(\mb{k},\omega)$ and $f_{n+1}(\mb{k},\omega)$ only, which
is what we require to be able to obtain an analytical expression for
$G(\mb{k},\omega)$. We discuss the meaning and consequences of this
approximation in more detail below.

After the approximation of Eq. (\ref{app}), Eqs.  (\ref{eq:F})
become:
$$
f_n(\mb{k},\omega)=\frac{g\bar{g}_0(\omega-n\Omega)}{\sqrt{N}}\left[
nf_{n-1}(\mb{k},\omega) +N f_{n+1}(\mb{k},\omega) \right].
$$ Together with Eq. (\ref{a1}) these simplified recurrence relations
can be solved similarly to the $t=0$ case. We find:
\begin{equation}
\label{eq:green_MA}
G(\mb{k},\omega)=\frac{1}{\omega-\varepsilon_{\mb{k}}-
\Sigma_{\textrm{MA}}(\omega)+i\eta},
\end{equation}
where the self-energy is, within the MA approximation,
\begin{equation} \label{eq:sigma_MA}
\Sigma_{\textrm{MA}}(\omega) = \cfrac{g^2\bar{g}_0(\omega-\Omega)}{1
- \cfrac{2g^2 \bar{g}_0(\omega-\Omega)
\bar{g}_0(\omega-2\Omega)}{1-\cfrac{3g^2 \bar{g}_0(\omega-2\Omega)
\bar{g}_0(\omega-3\Omega)}{1-\cdots}}}.
\end{equation}
This is the main new result of this work.  As discussed, the MA
approximation becomes exact in the limit of zero hopping ($t=0$),
where $\bar{g}_0(\omega) \to G_0(\omega)=(\omega+i\eta)^{-1}$, but
also for zero coupling, $g=0$.  In the following sections we show
that the range of validity of this approximation extends well beyond
these asymptotic limits, and that in fact the MA expression is
reasonably accurate over the entire parameter space.

Note that although the MA self-energy looks similar to the DMFT
self-energy,\cite{ciuchi:1997} this is in fact a very different
approximation. This issue is discussed in Appendix \ref{dmft}.

\section{Meaning of the MA Approximation}
\subsection{Diagrammatics}

To understand the diagrammatical meaning of the MA approximation we
expand Eq. (\ref{eq:sigma_MA}) in powers of $g^2$:
\begin{multline}
\Sigma_{\textrm{MA}}(\omega) = g^2 \bar{g}_0(\omega-\Omega) + g^4
\left[ 2 \bar{g}_0^2(\omega-\Omega) \bar{g}_0(\omega-2\Omega)
\right]
\\ + g^6 \left[ 4 \bar{g}_0^3(\omega-\Omega)
\bar{g}_0^2(\omega-2\Omega) + 6 \bar{g}_0^2(\omega-\Omega) \right. \\
\left. \times \bar{g}_0^2(\omega-2\Omega)
\bar{g}_0(\omega-3\Omega)\right] + \mathcal{O}(g^8).
\end{multline}
and analyze the various terms.  First, one can verify that the MA
approximation generates the correct number of proper self-energy
diagrams to all orders. Indeed, there is one term of order $g^2$,
two terms of order $g^4$, 4+6=10 terms of order $g^6$, and so on
(see Table \ref{tab1}). Moreover, to each of these terms we can
associate an MA diagram. These have the same topology as the exact
proper self-energy diagrams.  The difference is that each free
propagator $G_0(\mb{k},\omega)$ in the exact self-energy diagrams is
replaced with a momentum averaged free propagator
$\bar{g}_0(\omega)$ (with the correct frequency) in each MA diagram.

Using Eq. (\ref{eq:g0}), the first-order self-energy diagram is (see
Fig. \ref{fig:self_energy}):
$$ \Sigma^{(1)}(\mb{k}, \omega)=\frac{g^2}{N} \sum_{\mb{q}}^{}
G_0(\mb{k}-\mb{q},\omega-\Omega) = g^2 \bar{g}_0(\omega-\Omega),
$$ i.e. $\Sigma^{(1)}(\mb{k}, \omega)=\Sigma^{(1)}_{\textrm{MA}}(\omega)$, and
thus MA is exact to first order. Differences appear from the second
order diagrams, where the two exact contributions (see Fig.
\ref{fig:self_energy}):
\begin{multline} \label{eq:second_exact}
\frac{g^4}{N^2} \sum_{\mb{q}_1,\mb{q}_2}
G_0(\mb{k}-\mb{q}_1,\omega-\Omega) G_0(\mb{k} - \mb{q}_1 -
\mb{q}_2,\omega - 2\Omega) \\ \times \left[
G_0(\mb{k}-\mb{q}_1,\omega-\Omega) +
G_0(\mb{k}-\mb{q}_2,\omega-\Omega) \right]
\end{multline}
are replaced, within MA, by two equal contributions:
\begin{multline} \label{eq:second_MA}
2 g^4 \bar{g}_0^2(\omega-\Omega) \bar{g}_0(\omega-2\Omega) =
\frac{2g^4}{N^3}\sum_{\mb{q}_1,\mb{q}_2,\mb{q}_3}
G_0(\mb{q}_1,\omega-\Omega)\\ \times G_0(\mb{q}_2, \omega
-2\Omega)G_0(\mb{q}_3,\omega-\Omega) .
\end{multline}
Comparing Eq. (\ref{eq:second_exact}) to Eq. (\ref{eq:second_MA}),
we see that the MA self-energy diagrams have the correct number of
free propagators with the correct frequencies, however the momenta
of the free propagators are un-correlated and individually averaged
over. It is as if there is no connection between the momentum
carried by a cloud phonon when it is emitted and when it is
re-absorbed by the electron. Precisely the same holds for all higher
order self-energy diagrams.

To gain a better understanding of the difference between the exact
and the MA diagrams, let us further analyze the dependence on $t$ of
Eqs. (\ref{eq:second_exact}) and (\ref{eq:second_MA}). For $t=0$ the
expressions are identical, because the free propagators become
independent of momenta and, as already discussed, MA becomes exact.
For finite $t$, we expand each free propagator as:
$$ G_0(\mb{k},\omega) =G_0(\omega) \left[ 1+
\varepsilon_{\mb{k}}G_0(\omega)+ \left(
\varepsilon_{\mb{k}}G_0(\omega) \right)^2 + \cdots \right],
$$ where $G_0(\omega)=(\omega+i\eta)^{-1}$.  Inserting this expansion
into Eqs. (\ref{eq:second_exact}) and (\ref{eq:second_MA}) and
collecting powers of $t$ yields the following. All $\mathcal{O}(t)$
terms vanish in both the exact and the MA diagrams because they are
proportional to a $\sum_{\mb{q}}\varepsilon_{\mb{q}}=0$.  In fact,
all odd-order powers in $t$ vanish because
$\sum_{\mb{q}}\varepsilon_{\mb{q}}^{2n+1}=0$. Next, consider terms
of order $t^2$. Such terms arise either from expanding one free
propagator to $\mathcal{O}(t^2)$, or from expanding two different
free propagators to $\mathcal{O}(t)$. The former case leads to the
same result for both the exact and the MA diagrams, and we obtain 6
contributions proportional to $ {1\over N} \sum_{\mb{q}}^{}
\epsilon^2_{\mb{q}}=2dt^2$.  The latter case, however, reveals a
difference. Five of the six $\mathcal{O}(t^2)$ such contributions
from the exact diagrams vanish because they involve propagators
carrying different momenta, and, for example,
$\sum_{\mb{q}_1,\mb{q}_2}\varepsilon_{\mb{k}-
\mb{q}_1}\varepsilon_{\mb{k}-\mb{q}_1-\mb{q}_2}=0$. The exception
comes from the two outside free propagators of the non-crossed
diagram, which carry the same momentum and result into another
${1\over N} \sum_{\mb{q}_1} \varepsilon^2_{\mb{k}- \mb{q}_1}=2dt^2$
contribution. This is absent in the MA approximation, where
different propagators always carry different momenta. It follows
that the MA second order self-energy diagrams capture 6 out of the 7 finite
$\mathcal{O}(t^2)$ contributions correctly. Similar considerations
apply for higher order $t$ powers and for higher order diagrams,
differences between the MA and the exact self-energy diagrams coming
only from terms involving free propagators carrying equal momenta in
non-crossed diagrams.  However, the error from such missed terms
becomes smaller and smaller as one goes to higher order diagrams
because the percentage of self-energy diagrams with one or more
pairs of free propagators of equal momenta decreases exponentially.

We conclude that the MA approximation captures most of the $t$
dependence of each self-energy diagram, while summing over all
diagrams. This analysis suggests that MA should be quite accurate
for any finite $g$ and $t$ values. In the next section, we reinforce
this conclusion by analyzing the sum rules of the spectral weight.

\subsection{Sum Rules}

For an even better idea of the accuracy of the MA approximation, we
consider the sum rules for the spectral weight
$A(\mb{k},\omega)=-{1\over \pi}\textrm{Im} G(\mb{k},\omega)$:
\begin{equation} \label{eq:sum_rule}
M_n(\mb{k})=\int_{-\infty}^{\infty} d\omega \, \omega^n
A(\mb{k},\omega).
\end{equation}
For a problem of this type (single dressed particle), the sum rules
can be calculated exactly to arbitrary order.  The usual approach is
based on the equation of motion technique.\cite{kornilovitch:2002}
We review it briefly here in order to make a few useful
observations. The key step is to rewrite $\omega^n = \left. \left(i
{d\over d\tau}\right)^n e^{-i\omega \tau}\right|_{\tau=0}$, so that
we have:
$$ M_n(\mb{k})=-{1\over \pi}\textrm{Im} \left.\left(i {d\over
d\tau}\right)^n \int_{-\infty}^{\infty} d\omega \, e^{-i\omega
\tau}G(\mb{k},\omega) \right|_{\tau=0}.
$$ The integral is now simply $G(\mb{k}, \tau \rightarrow 0^+)$. Using
the definition of Eq. (\ref{eq:green_simplified}), we find, for any
$\tau >0$:
$$ \left(i {d\over d\tau}\right)^n G(\mb{k}, \tau) = -i \Theta(\tau)
\langle 0 | c_{\mb k}{\cal H}^n e^{-i{\cal H} \tau} c^\dag_{\mb
k}|0\rangle,
$$ so that the sum rules simplify to:
%%%%%%%%%%%%%%%%%%%%%%%%%%%%%% EQUATION %%%%%%%%%%%%%%%%%%%%%%%%%%%%%%
\begin{equation}
\EqLabel{sr} M_n(\mb{k})=\langle 0 | c_{\mb k}{\cal H}^n c^\dag_{\mb
k}|0\rangle.
\end{equation}
%%%%%%%%%%%%%%%%%%%%%%%%%%%%%%%%%%%%%%%%%%%%%%%%%%%%%%%%%%%%%%%%%%%%%%
These vacuum expectation values can be evaluated directly with some
effort. We find $M_0(\mb{k}) =1$, $M_1(\mb{k}) =\epsilon_{\mb k}$,
$M_2(\mb{k}) =\epsilon^2_{\mb k}+ g^2$,
$M_3(\mb{k})=\varepsilon_{\mb{k}}^3 + 2g^2 \varepsilon_{\mb{k}} +
g^2 \Omega$, etc.

One very important conclusion that can be drawn from this derivation
is that these sum rules have the same functional dependence on the
energy scales $t, \Omega, g$ anywhere in the parameter space. Of
course, in various asymptotic regimes, different terms dominate the
overall value ({\em e.g.} $M_2(\mb{k})\approx g^2$ if $g\gg t$ while
$M_2(\mb{k})\approx \epsilon^2_{\mb k}$ if $g\ll t$). However, this
shows that if one can evaluate the sum rules {\em exactly} in any
asymptotic regime, for instance by using perturbation theory, the
results hold true {\em everywhere} in the parameter space, even
where perturbation fails.

The second important conclusion one can draw from Eq. (\ref{sr}) is
that each term in $M_n(\mb{k})$ is proportional to $t^p \Omega^m
g^{n-m-p}$, where $0\le p, m, n-m-p \le n$ are integers, i.e. the
sum rule $M_n$ is a polynomial of total order $n$ in the energy
scales of the problem. More complicated dependence on $t, \Omega,
g$, for example through $\exp(-g^2/\Omega^2)$, simply cannot appear
(see below).

The first conclusion suggests an alternative derivation of the sum
rules, which can also be used for the MA and SCBA sum rules. Namely,
we use the diagrammatic perturbational expansion of the Green's
function valid for $g\ll t$ to evaluate directly the integrals
$\int_{-\infty}^{\infty} d\omega \, \omega^n G(\mb{k},\omega)$ and
retain the imaginary part. In this case, $G(\mb{k},\omega)
=\sum_{n=0}^{\infty}\sum_{i=1}^{s_n} D_{n,i}(\mb{k},\omega)$, where
$D_{n,i}(\mb{k},\omega), i=1,s_n$ are all Green's function diagrams
of order $n$, i.e. containing $n$ phonon lines. The multiplicity
$s_n=(2n-1)!!=1\cdot3\dots(2n-1)$. Each diagram
$D_{n,i}(\mb{k},\omega)$ is a product of $2n+1$ free propagators,
summed over internal phonon momenta. Since for large frequency each
$G_0(\mb{k},\omega)\rightarrow (\omega+i\eta)^{-1}$, it follows that
for $|\omega|\rightarrow \infty$, each
$D_{n,i}(\mb{k},\omega)\rightarrow g^{2n}/(\omega+i\eta)^{2n+1}$.
Since any integrand that decreases faster than $1/\omega^2$ has a
vanishing contribution to Eq. (\ref{eq:sum_rule}), it follows that
the diagrams of order $n$ only contribute to the sum rules
$M_p(\mb{k})$ with $p\ge 2n$. Thus, even though $G(\mb{k},\omega)$
is the sum of an infinite number of diagrams, only a finite number
of them, of low order, contribute to any given sum rule and the
calculation can be done.  The same holds true for the MA sum rules,
the only difference being that the self-energy parts in the Green's
functions diagrams are replaced with the corresponding MA
self-energy parts.

Let us analyze the differences between contributions of the exact
and of the MA diagrams to the sum rules. It is straightforward to
verify that:
$$ -{1\over \pi}\textrm{Im}\int_{-\infty}^{\infty} d\omega \, \omega^{2n}
g^{2n} \prod_{i=1}^{2n+1} G_0(\mb{q}_i, \omega-\Omega_i) = g^{2n},
$$
\begin{multline}
\nonumber -{1\over \pi}\textrm{Im}\int_{-\infty}^{\infty} d\omega \,
\omega^{2n+1} g^{2n} \prod_{i=1}^{2n+1} G_0(\mb{q}_i,
\omega-\Omega_i)
\\ = g^{2n}\sum_{i=1}^{2n+1} (\Omega_i+\epsilon_{\mb q_i}),
\end{multline}
and
\begin{multline}
\nonumber -{1\over \pi}\textrm{Im}\int_{-\infty}^{\infty} d\omega \,
\omega^{2n+2} g^{2n} \prod_{i=1}^{2n+1} G_0(\mb{q}_i,
\omega-\Omega_i)
\\ = g^{2n}\left[\sum_{i=1}^{2n+1}\left(\Omega_i+\epsilon_{\mb
    q_i}\right)^2 +\sum_{i<j}^{}
  \left(\Omega_i+\epsilon_{\mb{q}_i}\right)\left(\Omega_j +
  \epsilon_{\mb{q}_j}\right)\right].
\end{multline}
Both the exact and the MA diagrams of order $n$ contain products of
the general form $g^{2n} \prod_{i=1}^{2n+1} G_0(\mb{q}_i,
\omega-\Omega_i)$. Some of the free propagators are actually
$G_0(\mb{k},\omega)$ (always the first and the last one, but there
can also be intermediary ones connecting proper self-energy parts).
All other free propagators have momenta dependent on the phonon
momenta, which are summed over (in the exact diagrams), or are
individually averaged over (in the MA diagrams). Since there is
one-to-one correspondence between the number of exact vs. MA
diagrams and their topologies, and since $\sum_{\mb
{q}}^{}\epsilon_{\mb{q}}=0$, it follows that the exact and the MA
diagrams of order $n$ give precisely the same contributions to
$M_{2n}(\mb{k})$ and $M_{2n+1}(\mb{k})$. Differences appear in the
contribution to $M_{2n+2}(\mb{k})$ if there is at least one pair of
propagators in any of the self-energy parts of the exact diagram
that carry the same momenta. In this case, the corresponding
$\epsilon_{\mb{q}_i}\epsilon_{\mb{q}_j}$ averages to $2dt^2$ when
the sums over phonon momenta are carried for the exact diagrams,
whereas these terms always average to zero for the MA diagrams.
Since most free propagators in the self-energy parts have different
momenta, such differences are quantitatively small. This is
especially true for large phonon frequencies $\Omega$, where the
contributions proportional to $\Omega$ captured correctly by MA
scale like some power of $2n+1$.  This analysis can be continued for
higher sum rules, with similar conclusions.

We can now summarize our findings. Only the $0^{th}$ order Green's
function diagram (the free propagator $G_0({\mb k},\omega)$)
contributes to $M_0(\mb{k})$ and $M_1(\mb{k})$. Since this is
included correctly in the MA and the SCBA cases, both give the
correct $M_0(\mb{k})$ and $M_1(\mb{k})$. In fact, this diagram
contributes an $\epsilon_{\mb{k}}^n$ to $M_n(\mb{k})$, which is
always the leading power in $t$ contribution. The 1$^{st}$ order
diagram is also exact in both the MA and SCBA cases, therefore both
give the correct $M_2(\mb{k})$ and $M_3(\mb{k})$ sum rules as well.
Differences appear from $M_4$ onwards. Because SCBA only keeps 2 out
the 3 second order Green's function diagrams, and 5 out of 15 third
order diagrams, etc., the leading terms in $g$ are $2g^4$ instead of
$3g^4$ in $M_4(\mb{k})$, $5g^6$ instead of $15 g^6$ in
$M_6(\mb{k})$, etc. This shows that SCBA fails all sum rules with
$n>3$ significantly in the strong coupling regime where the term
proportional to $g^{2n}$ gives the most significant contribution to
$M_{2n}(\mb{k})$ (similar conclusions hold for odd sum rules). So
even though SCBA always satisfies exactly the first 4 sum rules, it
is a bad approximation for large $g$, where big discrepancies appear
for $n>3$.

\begin{figure}[t]
\includegraphics[width=0.95\columnwidth]{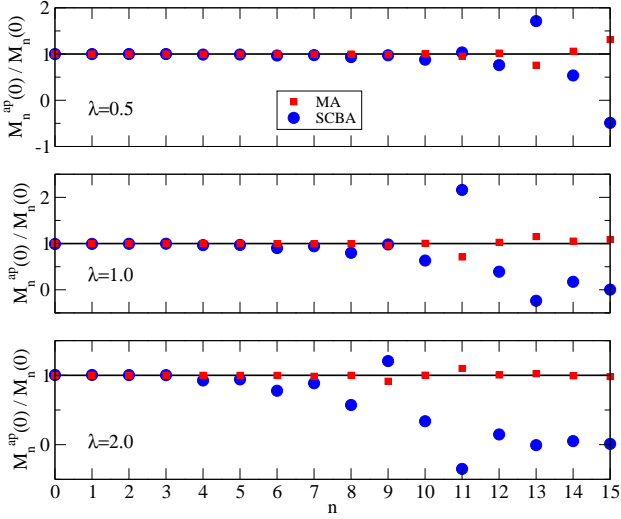}
\caption{(color online) Ratio of MA (squares), respectively SCBA (circles)
sum rules and the exact sum rules, vs. order $n$. Results
are for 1D and $k=0$, $\Omega=0.5t$ and $\lambda=0.5, 1$, 2.}
\label{sumrules0}
\end{figure}

\begin{figure}[t]
\includegraphics[width=0.95\columnwidth]{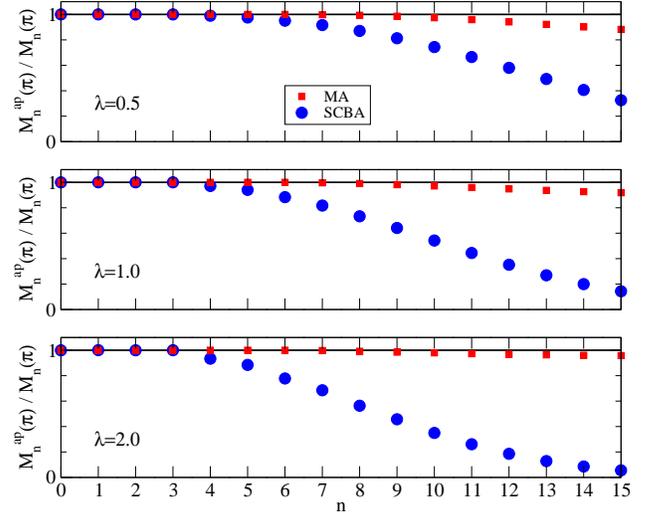}
\caption{(color online) Ratio of MA (squares), respectively SCBA (circles)
sum rules and the exact sum rules, vs. order $n$. Results
are for 1D and $k=\pi$, $\Omega=0.5t$ and $\lambda=0.5, 1$, 2.}
\label{sumrulespi}
\end{figure}

MA satisfies $M_4(\mb{k})$ and $M_5(\mb{k})$ exactly as well,
because these only depend on having the correct number and topology
for the 2nd order diagrams. MA fails from $M_6(\mb{k})$ onwards,
however in a very different manner than SCBA. The leading term in
$g^6$ has the correct prefactor, because MA has the correct number
of 3rd order diagrams. The error comes from the 2nd order diagram
containing the non-crossed self-energy diagram, as discussed.
Indeed, instead of the exact sum rule:
\begin{multline}
M_6(\mb{k}) = \varepsilon_{\mb{k}}^6 + g^2 \left[ 5
\varepsilon_{\mb{k}}^4 + 6t^2(2d^2-d) + 4 \varepsilon_{\mb{k}}^3
\Omega + 3 \varepsilon_{\mb{k}}^2 \Omega^2 \right. \\ + \left. 6dt^2
(\varepsilon_{\mb{k}}^2 + \varepsilon_{\mb{k}} \Omega + 2 \Omega^2)
+ 2 \varepsilon_{\mb{k}} \Omega^3 + \Omega^4 \right] \\ + g^4
(18dt^2 +12 \varepsilon_{\mb{k}}^2 + 22 \varepsilon_{\mb{k}}\Omega +
25 \Omega^2) + 15g^6,
\end{multline}
MA finds a sum rule
$M_6^{\textrm{MA}}(\mb{k})=M_6(\mb{k})-2dt^2g^4$. The leading terms
in the $g\ll t$ and $g \gg t$ limits are always exact (as expected,
since MA becomes exact in these limits), and this is true for all
orders $n$. For $n\ge 6$ some of the cross terms are missing, but
these are a minority related to non-crossed diagrams, as explained.
We therefore expect the MA sum rules to remain highly accurate for
larger $n$ values.  That this is indeed true for higher sum rules is
shown numerically in Figs. \ref{sumrules0} and \ref{sumrulespi},
where we plot the ratio of the MA respectively SCBA sum rules, and
the exact sum rules of same order $n$. The results shown are for 1D
and $k=0, \pi$, but similar trends are found in the other cases. For
$k=0$ all the spectral weight is at negative frequencies, therefore
$M_n(0)$ alternate signs for even/odd $n$, and this is reflected in
the non-monotonic behavior with $n$. For $k=\pi$, most of the weight
is at positive frequencies and sum rules are always positive. The
magnitude of the exact sum rules increases roughly exponentially
with $n$, for instance for $\lambda=2$ and $\Omega=0.5t$,
$M_{14}(0)=119,516,000$. For $k=0$ and $\lambda=0.5$, both MA and
SCBA are reasonably accurate, with a slight edge for MA at higher
$n$. However, MA is clearly much more accurate for all the other
cases shown, and its accuracy is expected to improve even more as
one moves further into the asymptotic regions of weak or strong
coupling.

Before ending this section, one more issue needs to be addressed. It
is obvious that the MA sum rules must capture the contributions to
$M_n(\mb{k})$ proportional respectively to $t^n$ and $g^n$ exactly,
since MA is exact for both $g=0$ and $t=0$. One may assume that this
alone is sufficient for a good interpolation at finite $t$ and $g$.
That this is not so is shown by the generalized LF approximation,
which is also exact for $t=0$ or $g=0$. However, in this
approximation one finds $M_0^{\textrm{LF}}(\mb{k})=1,
M_1^{\textrm{LF}}(\mb{k})=\varepsilon_{\mb{k}}
\exp\left(-{g^2\over\Omega^2}\right)$, etc. $M_1$ and all higher sum
rules have unacceptable dependence on the energy scales $g$ and $
\Omega$ (see second observation above), even though they become
exact asymptotically. As shown in the next section, this
approximation indeed performs rather poorly for finite $t$ and $g$.

We conclude that while MA satisfies exactly the first 6 sum rules,
it remains accurate for all higher order sum rules, and is
asymptotically exact. This is another argument in favor of the
accuracy of this approximation over the whole parameter space. In
the next section, we compare the MA predictions to those of existing
numerical simulations to further support this claim.

\section{Results}

We first list the explicit expressions of the momentum averaged
Green's function $\bar{g}_0(\omega)$ of Eq.  (\ref{eq:g0}). For
nearest-neighbor hoping, it is straightforward to derive:
$$ \bar{g}_0^{1\textrm{D}}(\omega)
=\frac{\textrm{sgn}(\omega)}{\sqrt{(\omega+i\eta)^2-4t^2}},
$$
\begin{equation}
\nonumber \bar{g}_0^{2\textrm{D}}(\omega) =
\frac{2}{\pi(\omega+i\eta)} \, \mathcal{K} \!
\left(\frac{4t}{\omega+i\eta}\right)
\end{equation}
and
\begin{equation}
\nonumber \bar{g}_0^{3\textrm{D}}(\omega) = \frac{1}{2\pi^2 t}
\int_0^{\pi} dk_z \, \textrm{sgn} \nu |\nu| \mathcal{K}(\nu)
\end{equation}
respectively, where
\begin{equation}\nonumber
\nu=\frac{4t}{\omega+2t\cos(k_za) + i\eta},
\end{equation}
and
\begin{equation}\nonumber
\mathcal{K}(\nu)=\int_0^{\pi/2} \frac{d \phi}{\sqrt{1-\nu^2
\cos^2\phi}}
\end{equation}
is the complete elliptical function of the first
kind.\cite{abramowitz:1964} These integrals can be performed
numerically very efficiently. More generally, for any free electron
dispersion $\epsilon_{\mb{k}}$ to which corresponds the free
electron density of states (DOS) $\rho_0(\epsilon)={1\over
N}\sum_{\mb{k}}^{}\delta(\epsilon - \epsilon_{\mb{k}})$, we have
$\bar{g}_0(\omega) =\int_{-\infty}^{\infty}d\epsilon
\rho_0(\epsilon) (\omega -\epsilon +i\eta)^{-1}$ (also see Appendix
\ref{dmft}).

The self-energy $\Sigma_{\textrm{MA}}(\omega)$ is then calculated
easily from Eq. (\ref{eq:sigma_MA}) by truncating the continued
fraction to a high-enough level. For an error of order $\epsilon$ it
is necessary to go to a level with $n$ such that $n g^2
\bar{g}_0(\omega-n\Omega)\bar{g}_0(\omega-(n+1)\Omega) < \epsilon$.
Using the fact that for large enough $n$ we can approximate
$\bar{g}_0(\omega-n\Omega)\approx
\bar{g}_0(\omega-(n+1)\Omega)\approx -1/(n\Omega)$, it follows that
we must have
$$ n>{1\over \epsilon}{g^2\over \Omega^2}.
$$ This result is expected, since $g^2\over \Omega^2$ is roughly the
average number of phonons in the polaron cloud (see below). This
condition shows that all diagrams with at least that many phonons
have to be included. In practice, we always use $n$ large enough so
that the change in $\Sigma_{\textrm{MA}}(\omega)$ after doubling $n$
is below a threshold much smaller than $\eta$. All MA error bars in
the figures we show are less than the thickness of lines/symbols
used for the plots.

With an explicit form for $\Sigma_{\textrm{MA}}(\omega)$ we are now
in a position to calculate the Green's function
$G_{\textrm{MA}}(\mb{k},\omega)$ 
and extract various polaron properties.

\subsection{Polaron Ground State Properties}

We begin by discussing polaron ground-state properties. Most of
these are already known from numerical studies, but they give us an
opportunity to further test the accuracy of the MA approximation.
Given the simplicity and efficiency of the MA approximation, we can
also investigate higher dimensionality and larger parameter ranges
than typical numerically intensive approaches. In this section, we
use for comparison 1D and 2D numerical results obtained from
diagrammatic Quantum Monte Carlo (QMC)
simulations.\cite{macridin:2003} 

For
$k=0$, we track the energy and weight of the lowest pole in the
spectral weight, which give  the ground-state energy $E_0$ and the
ground-state quasi-particle weight $Z_0= |\langle GS |
c_{\mb{k=0}}^{\dagger} | 0 \rangle|^2.$ Using the Hellmann-Feynman
theorem, \cite{feynman:1939} we then find the average number of phonons
in the ground-state to be
\begin{equation} \label{eq:phonons}
N_{\textrm{ph}} \equiv \langle GS| \sum_{\mb{q}}
b_{\mb{q}}^{\dagger} b_{\mb{q}} | GS \rangle = \frac{\partial
E_0}{\partial \Omega}.
\end{equation}
Note that one can also calculate the correlation function:
\begin{equation} \label{eq:corr}
\langle GS| \sum_{i}c^\dag_i c_i \left( b_{i}^{\dagger}+ b_{i}
\right)| GS \rangle = \frac{\partial E_0}{\partial g}
\end{equation}
just as easily. We do not show it here because we do not have the
corresponding numerical data for the comparison, however some
typical results for this quantity are shown at the end of this
section. Also note that all these quantities can be calculated
similarly for other eigenstates. We will show such results in other
sections.

We also show the effective mass, $m^*$. Because the MA self-energy
is momentum independent for this simple Holstein model, one
has:\cite{slezak:2006}
\begin{equation} \label{eq:effective}
\frac{m^*}{m}= \frac{1}{Z_0} = 1-\left.\frac{d
\Sigma_{\textrm{MA}}(\omega)}{d \omega}\right|_{\omega=E_0}.
\end{equation}
This result also gives us a consistency check on our calculations.
For MA we generally show effective mass results obtained from the
first equality.

\begin{figure}[t]
\includegraphics[width=0.76\columnwidth]{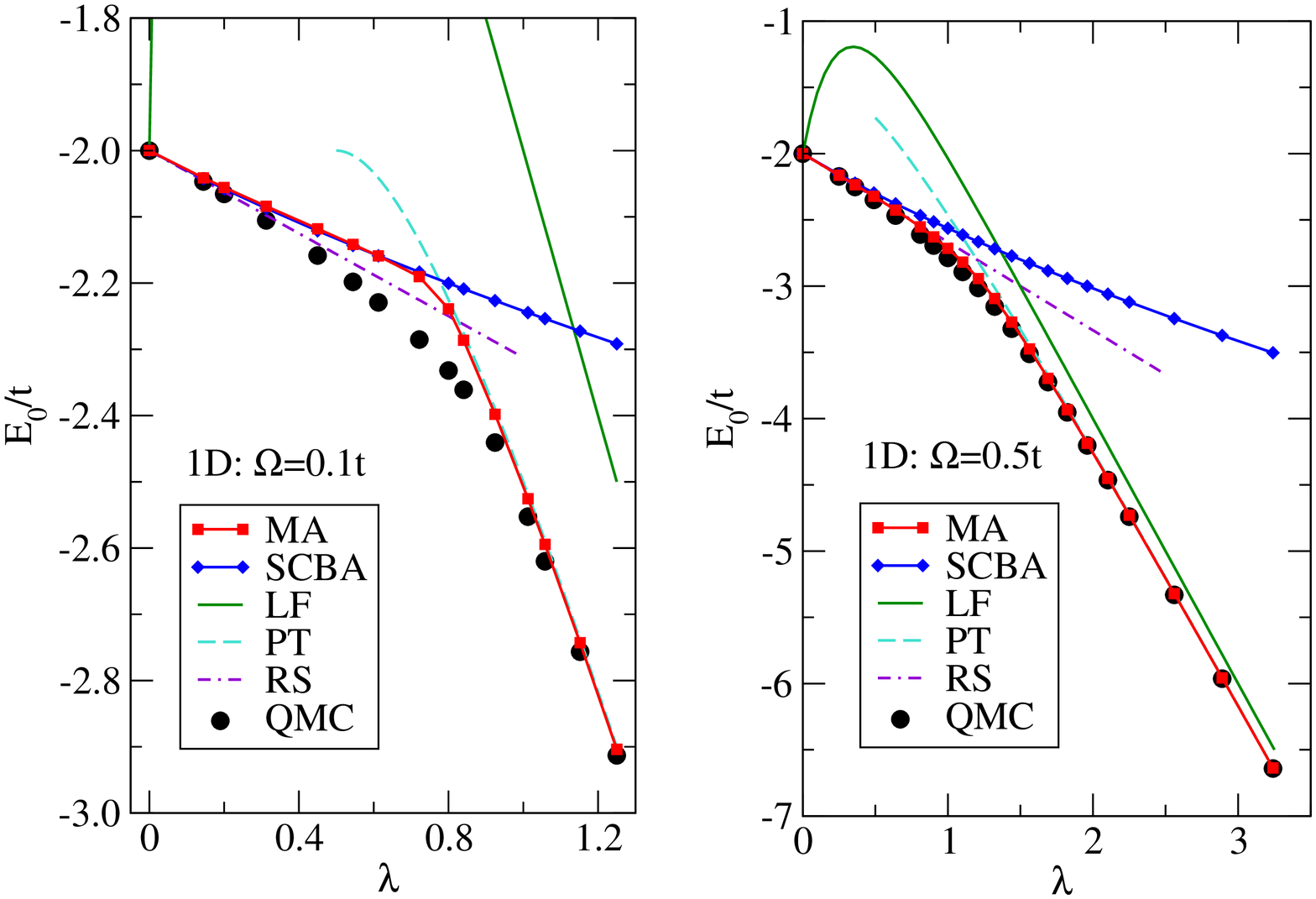}
\includegraphics[width=0.76\columnwidth]{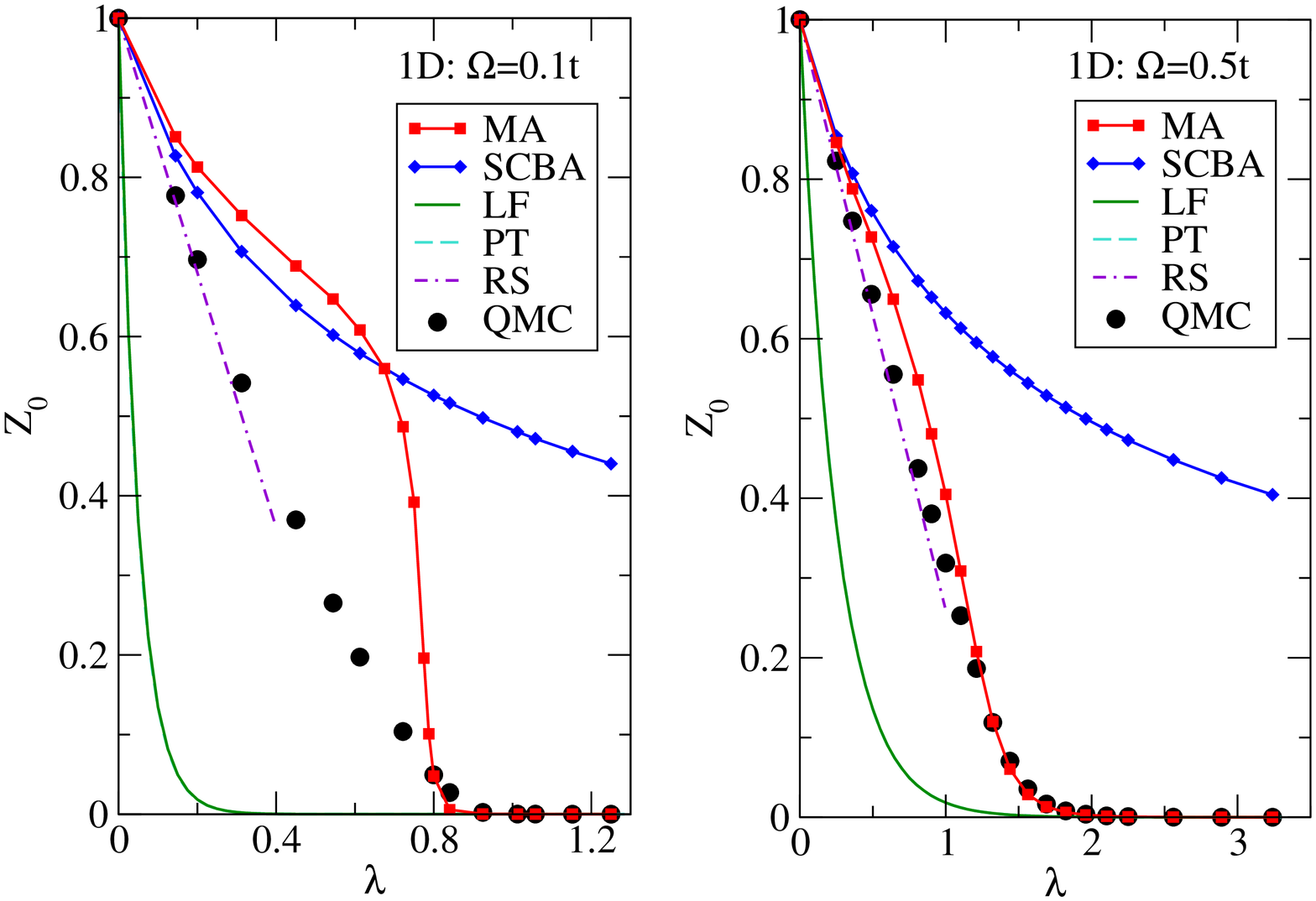}
\includegraphics[width=0.76\columnwidth]{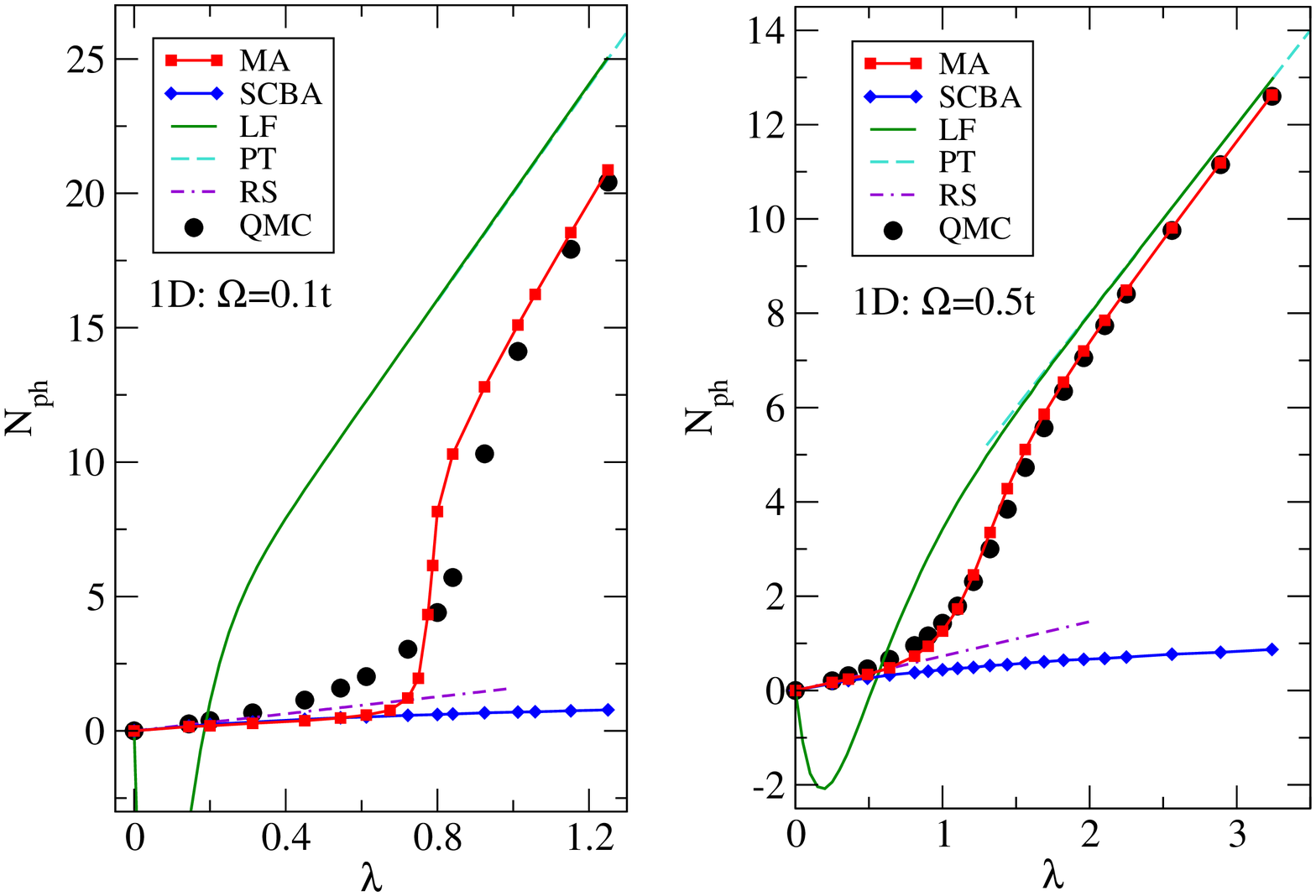}
\includegraphics[width=0.76\columnwidth]{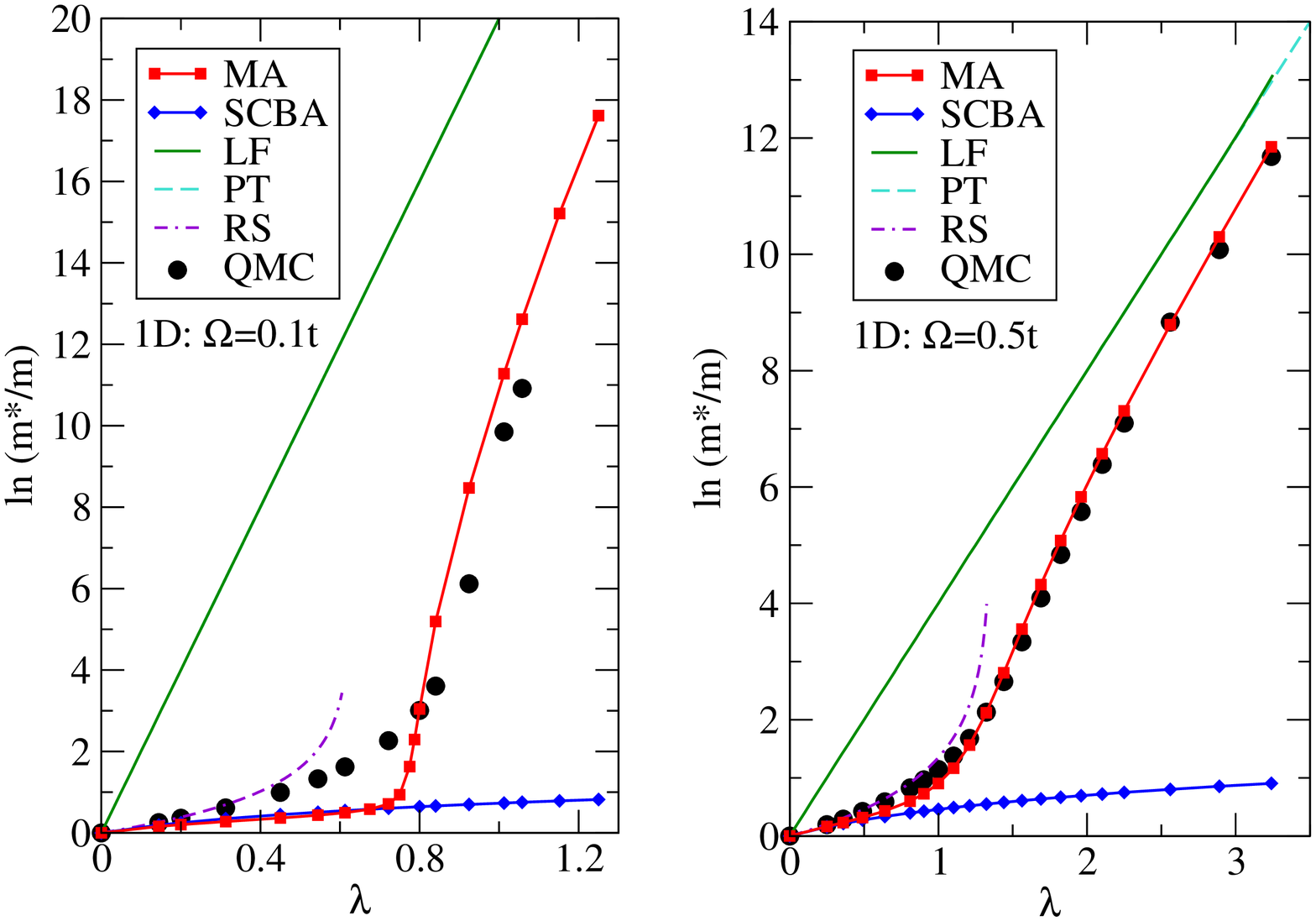}
\caption{(color online) Ground state results in 1D. Shown as a
  function of the coupling $\lambda$ are the ground state energy
  $E_0$; the $qp$ weight $Z_0$; the average number of phonons $N_{\textrm{ph}}$
  and the effective mass $m^*$ on a logarithmic scale. The left panels
  correspond to $\Omega/t = 0.1$ and the right ones to
  $\Omega/t=0.5$.}
\label{fig:1D_E}
\end{figure}

\begin{figure}[t]
\includegraphics[width=0.75\columnwidth]{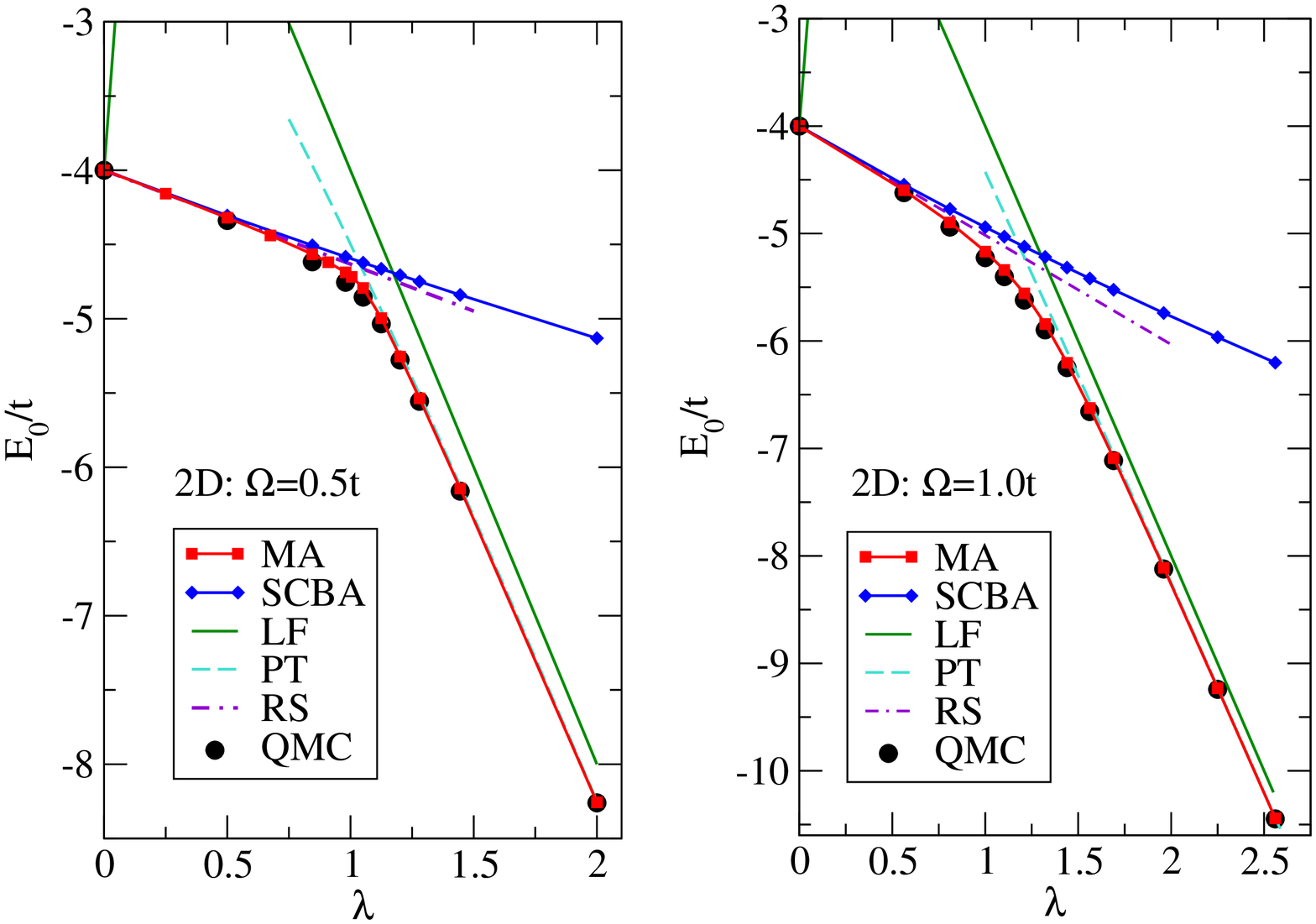}
\includegraphics[width=0.75\columnwidth]{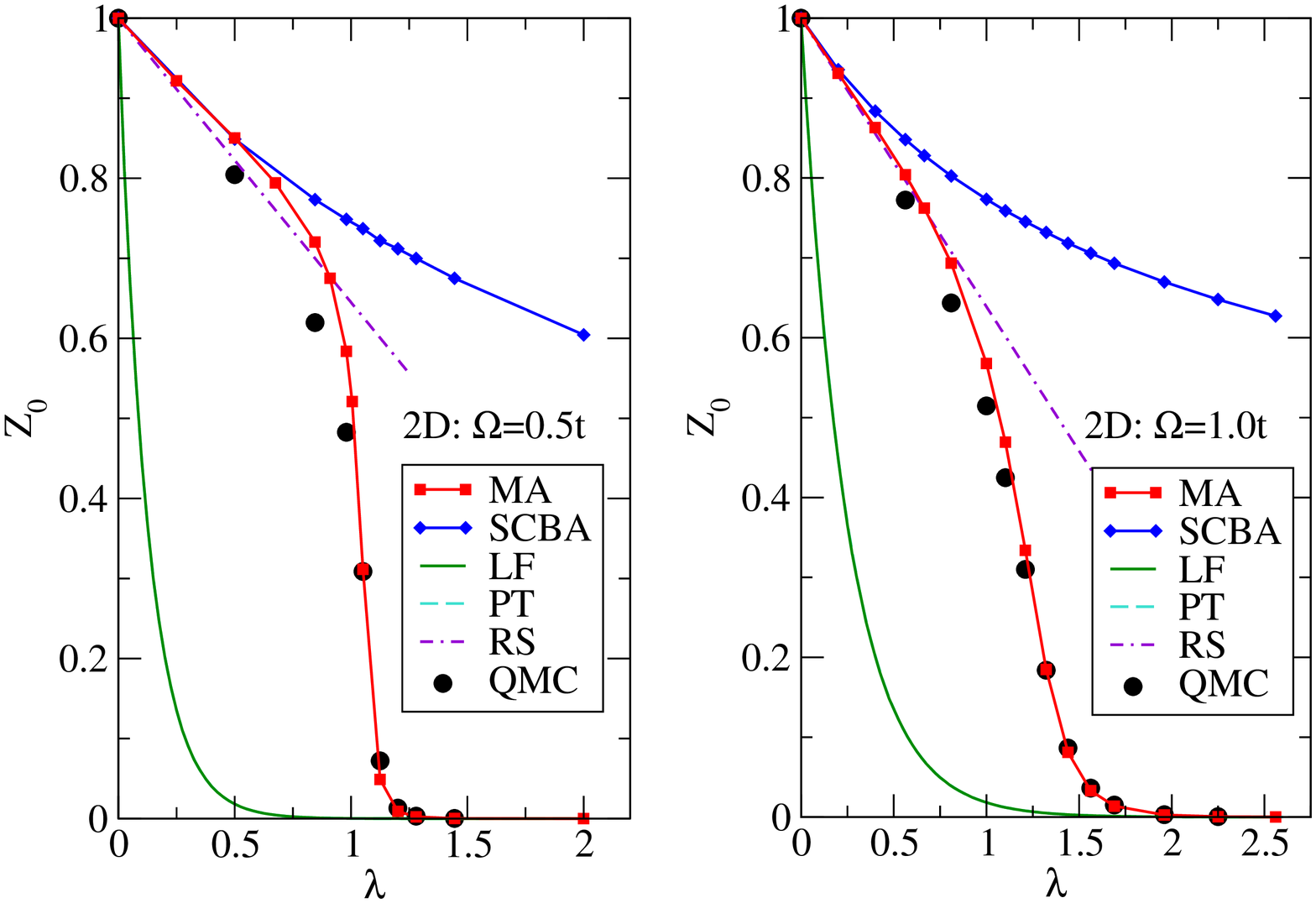}
\includegraphics[width=0.75\columnwidth]{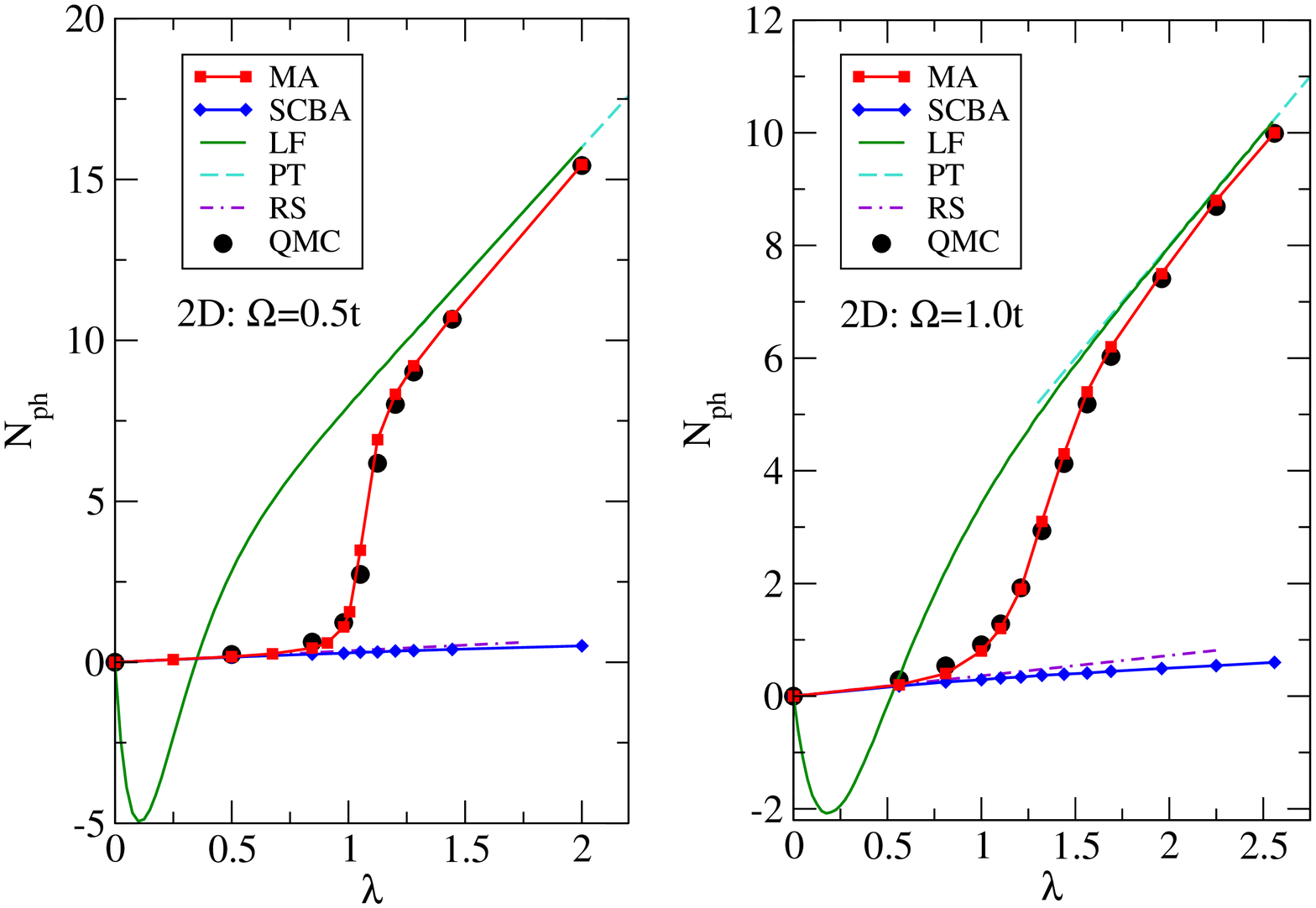}
\includegraphics[width=0.75\columnwidth]{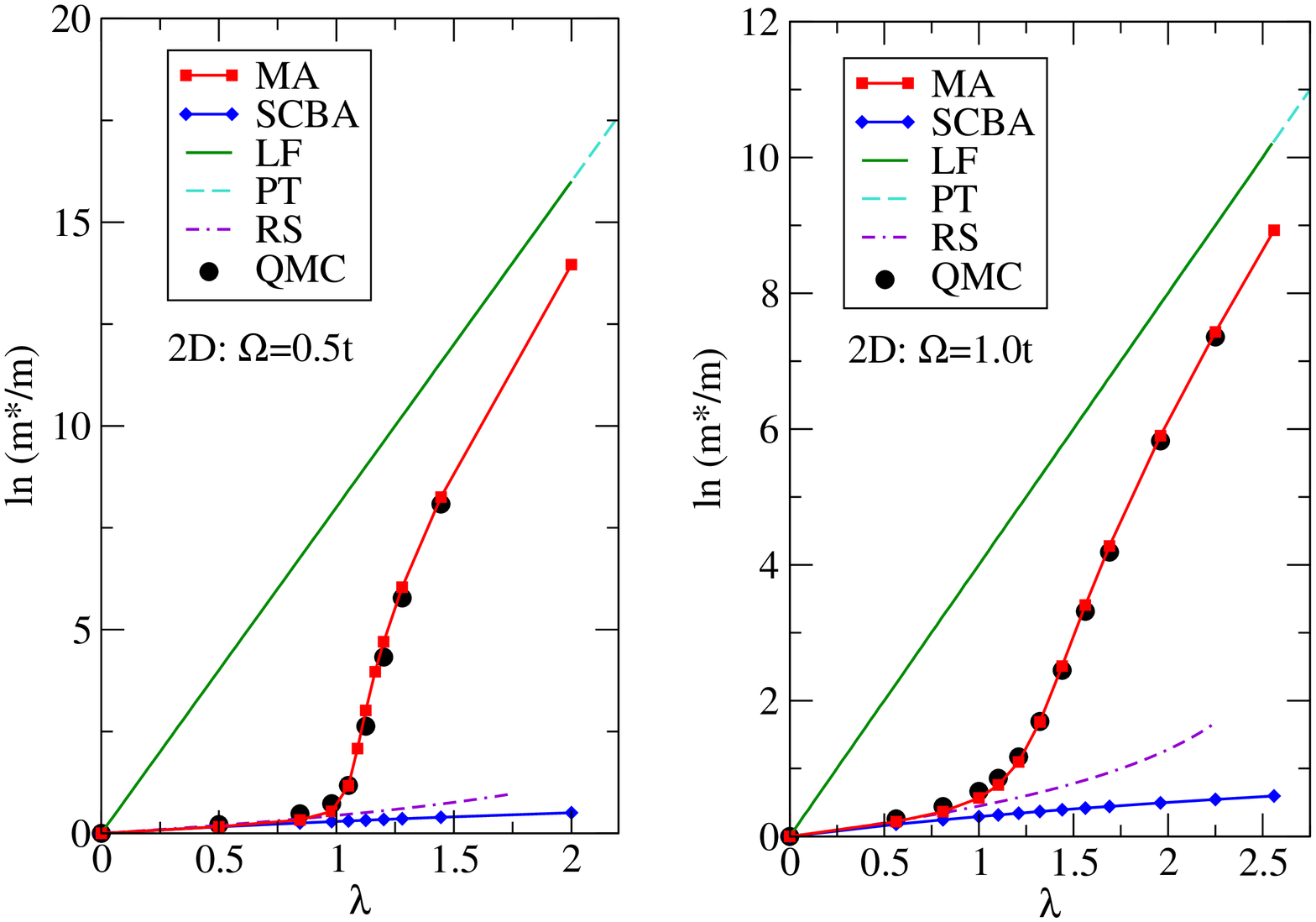}
\caption{(color online) Ground state results in 2D. Shown as a
  function of the coupling $\lambda$ are the ground state energy
  $E_0$; the $qp$ weight $Z_0$; the average number of phonons $N_{\textrm{ph}}$
  and the effective mass $m^*$ on a logarithmic scale. The left panels
  correspond to $\Omega/t = 0.1$ and the right ones to
  $\Omega/t=0.5$.}
\label{fig:2D_E}
\end{figure}

A comparison of the results for these four quantities as obtained
with
 QMC (black circles) and the different approximations is shown in
 Fig. \ref{fig:1D_E}. As expected, SCBA (blue line) fares well at
 small couplings $\lambda$ but very poorly at strong couplings. The
 generalized LF (green line) is asymptotically exact, but quite wrong
 at  finite $\lambda$. Because these are GS properties, we can also
 use perturbation theory in the two asymptotic limits to estimate
 them. At weak coupling we use the Rayleigh - Schr\"odinger
 perturbation theory (RS, violet line) which gives the lowest energy
 for a state with total momentum $\mb{k}$ as:\cite{negele:1988}
\begin{equation} \label{eq:E_RS}
E_{\mb{k}} = \varepsilon_{\mb{k}} - \frac{1}{N} \sum_{\mb{q}}
\frac{g^2}{\Omega + \varepsilon_{\mb{k}-\mb{q}} -
\varepsilon_{\mb{k}}}.
\end{equation}
At strong couplings we use the second order perturbation theory
result (PT, cyan line):\cite{lang:1963,marsiglio:1995}
\begin{equation} \label{eq:E_PT}
E_{\mb{k}} = -\frac{g^2}{\Omega} +\epsilon_{\mb{k}}e^{-{g^2\over\Omega^2}}
-\frac{d\Omega t^2}{g^2}.
\end{equation}
The GS energy is simply $E_0=E_{\mb{k}=0}$.  $N_{\textrm{ph}}$ is
obtained from $E_0$ as before; $m^*$ can be evaluated from the
second derivative of $E_{\mb{k}}$ with respect to $\mb{k}$, and
$Z_0$ is extracted from the effective mass [Eq.
(\ref{eq:effective})].

Fig. \ref{fig:1D_E} shows that one or the other of these
perturbational values describe the GS energy $E_0$ quite well,
especially for the larger $\Omega$ value. However, the agreement for
the other quantities is somewhat poorer, especially at strong
couplings (PT and LF give identical results for $Z_0$).  Part of the
reason is that the $t^2$ term in Eq. (\ref{eq:E_PT}) has in fact a
more complicated dependence of $g$ and $ \Omega$, which is only
asymptotically equal to the one used here.\cite{marsiglio:1995} More
importantly, neither perturbational theory describes well the
transition area, or can be easily applied to high energy states.

\begin{figure}[t]
\includegraphics[width=0.82\columnwidth]{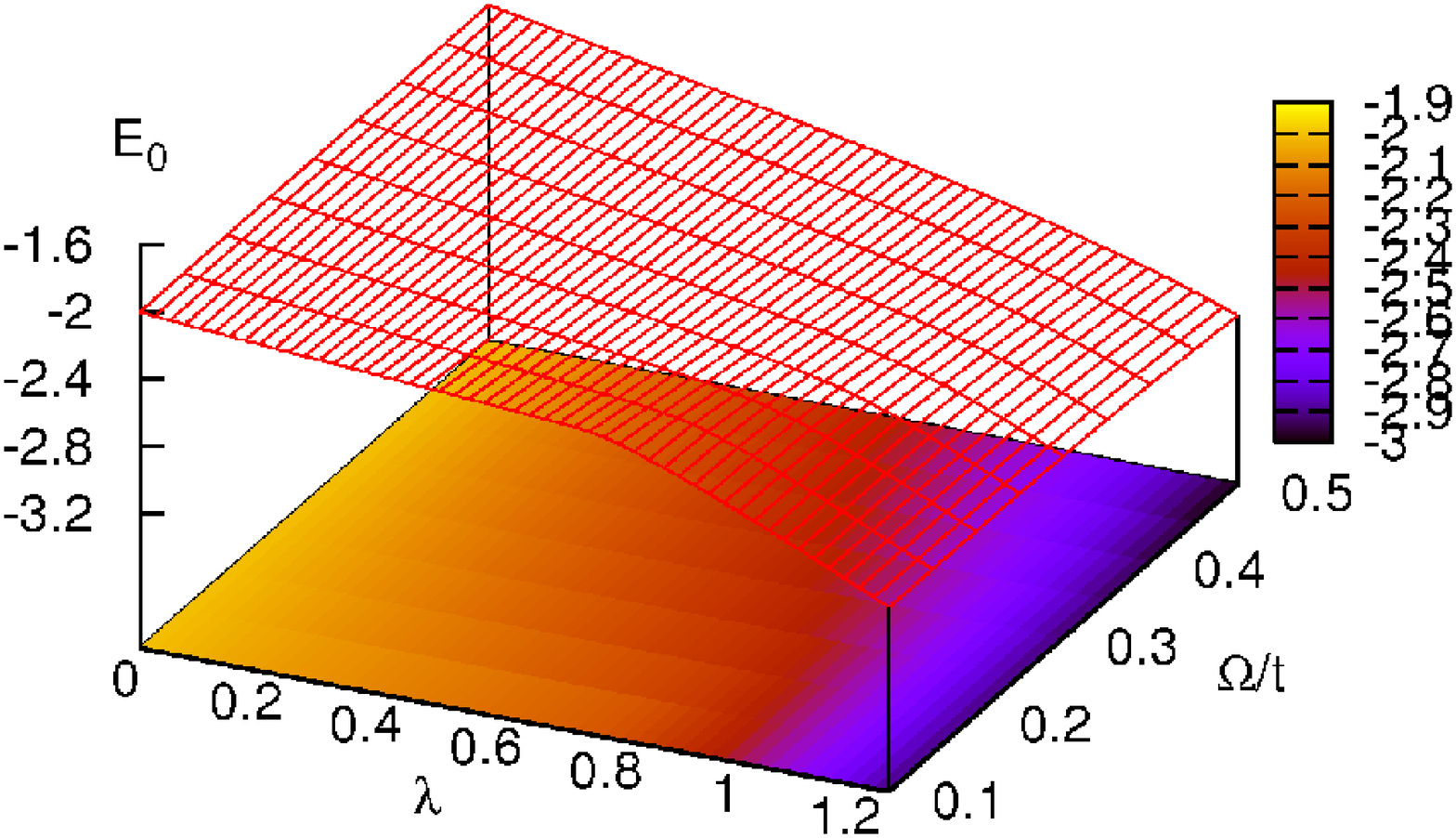}
\includegraphics[width=0.82\columnwidth]{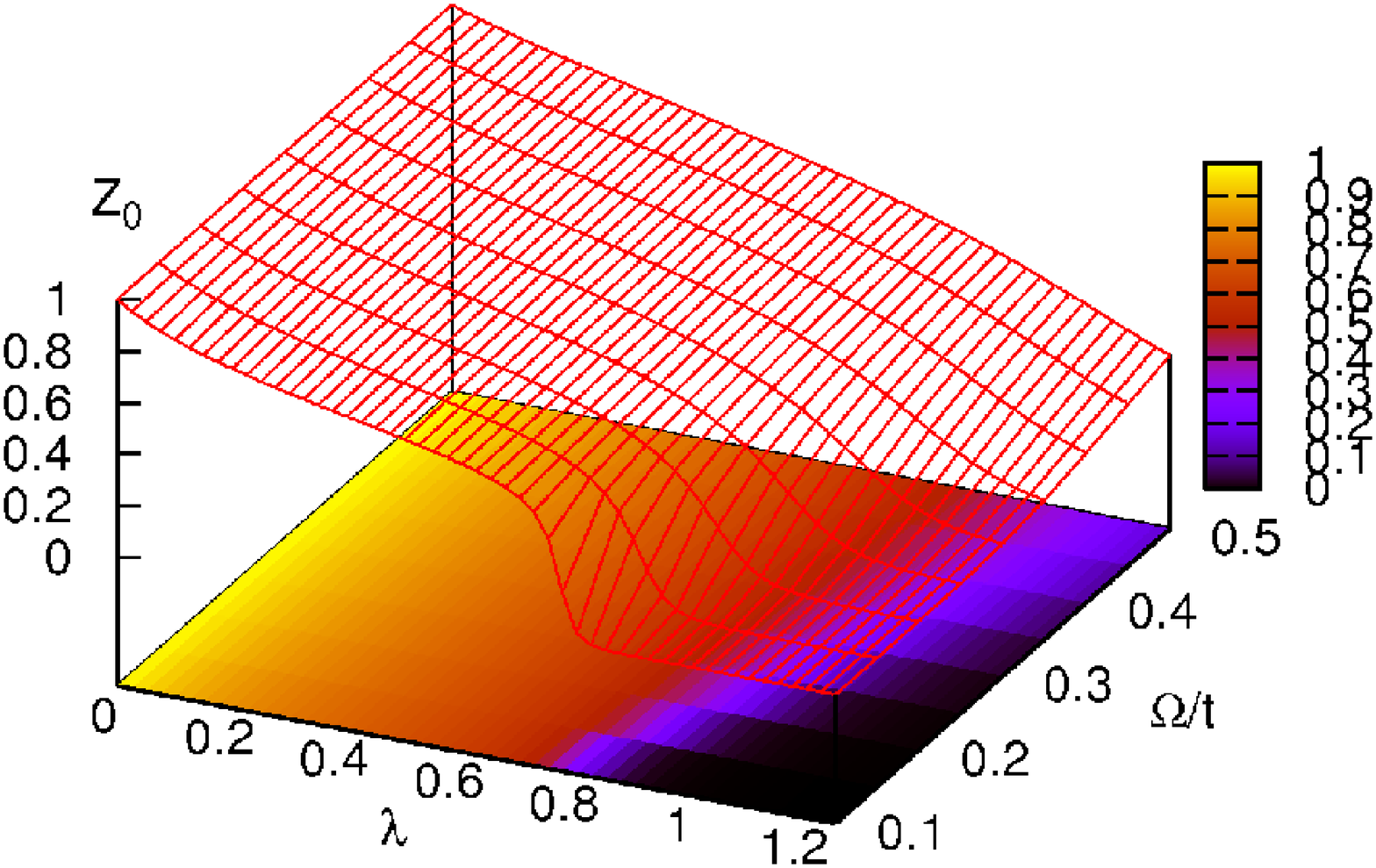}
\includegraphics[width=0.82\columnwidth]{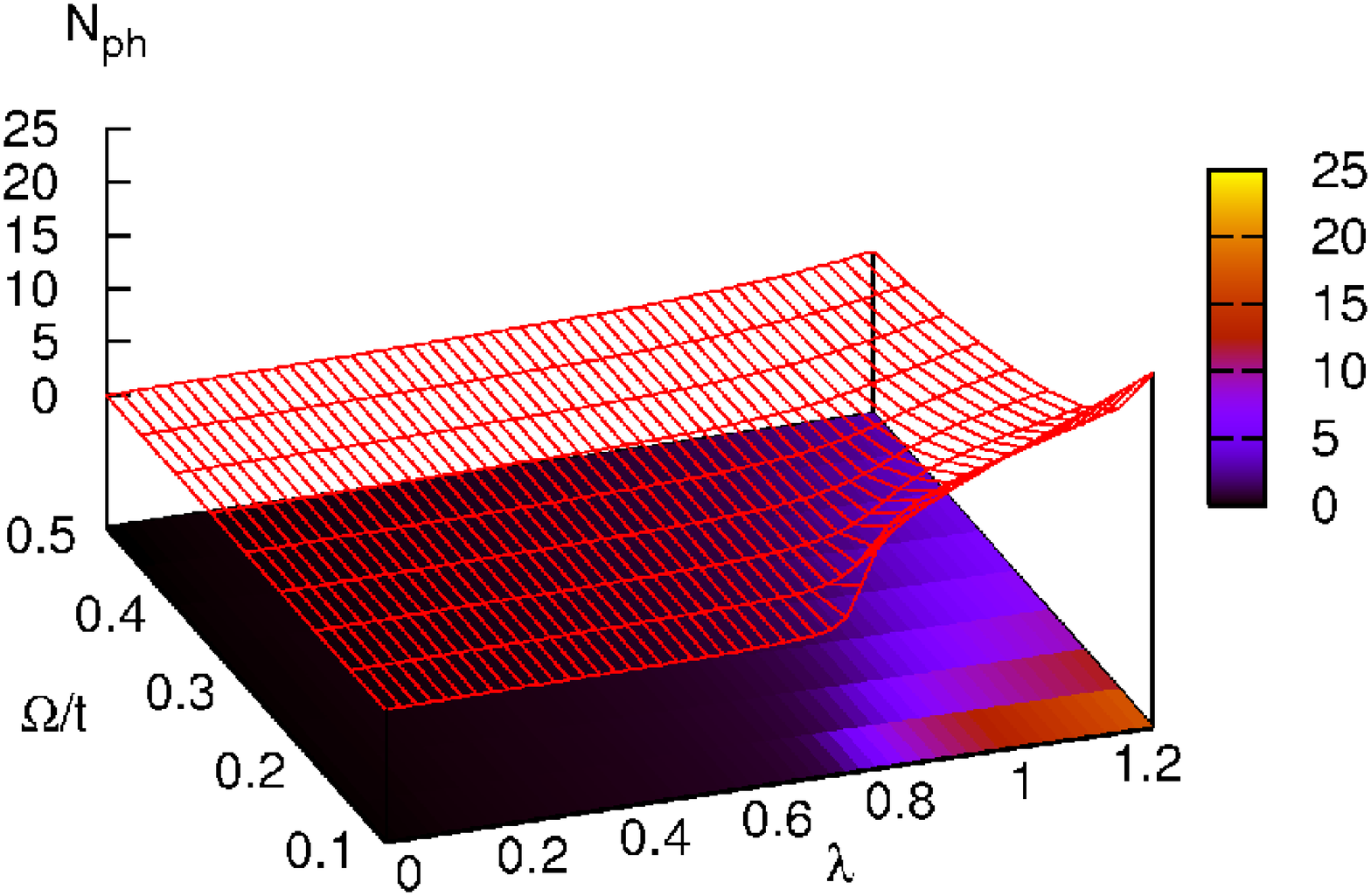}
\includegraphics[width=0.82\columnwidth]{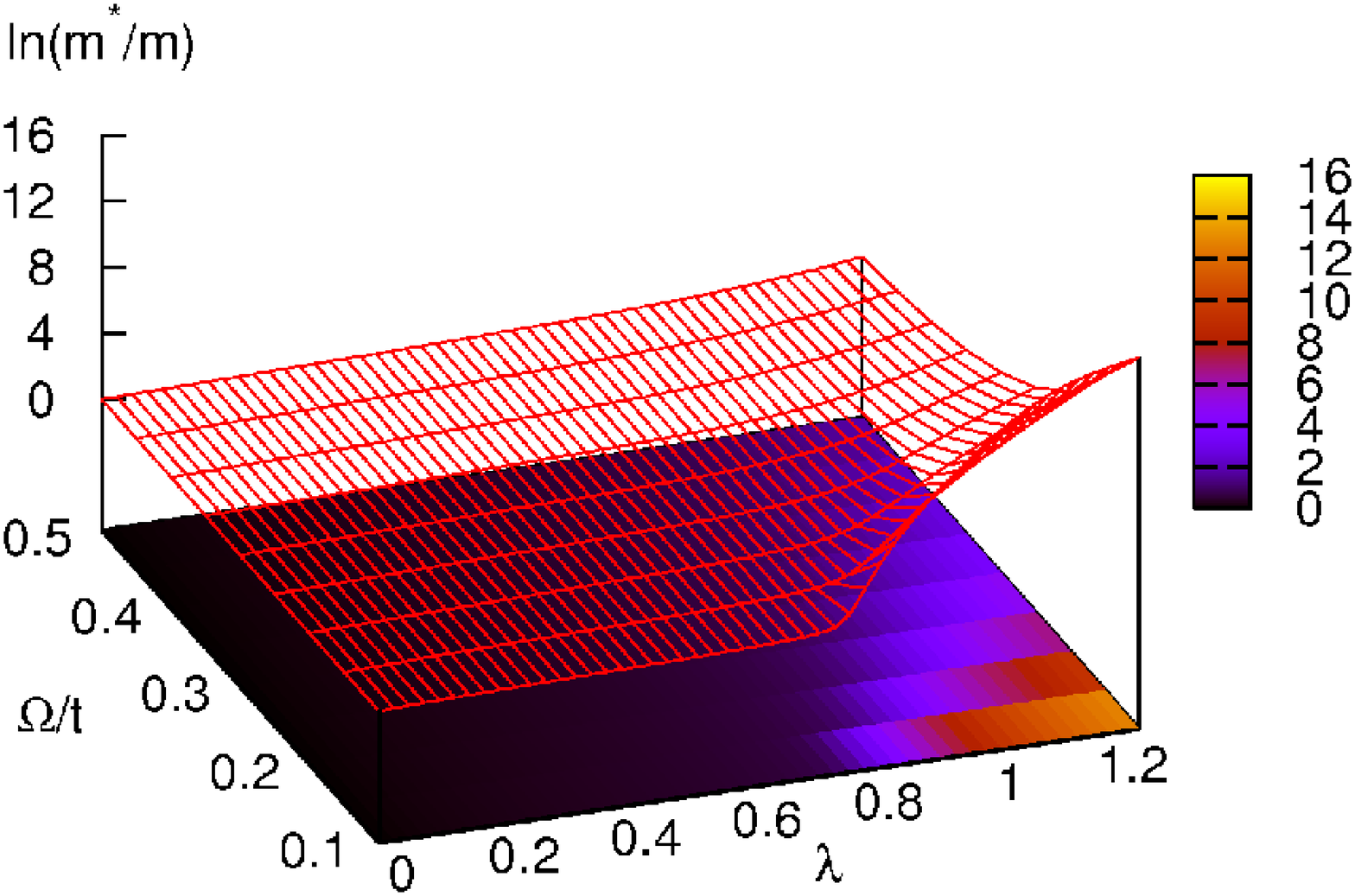}
\caption{(color online) Ground state energy $E_0$; the $qp$ weight
  $Z_0$; the average number of phonons $N_{\textrm{ph}}$ and the effective mass
  $m^*$ as a function of $\lambda$ and of $\Omega/t$, for $d=1$.}
\label{fig:1D_c}
\end{figure}

\begin{figure}[t]
\includegraphics[width=0.8\columnwidth]{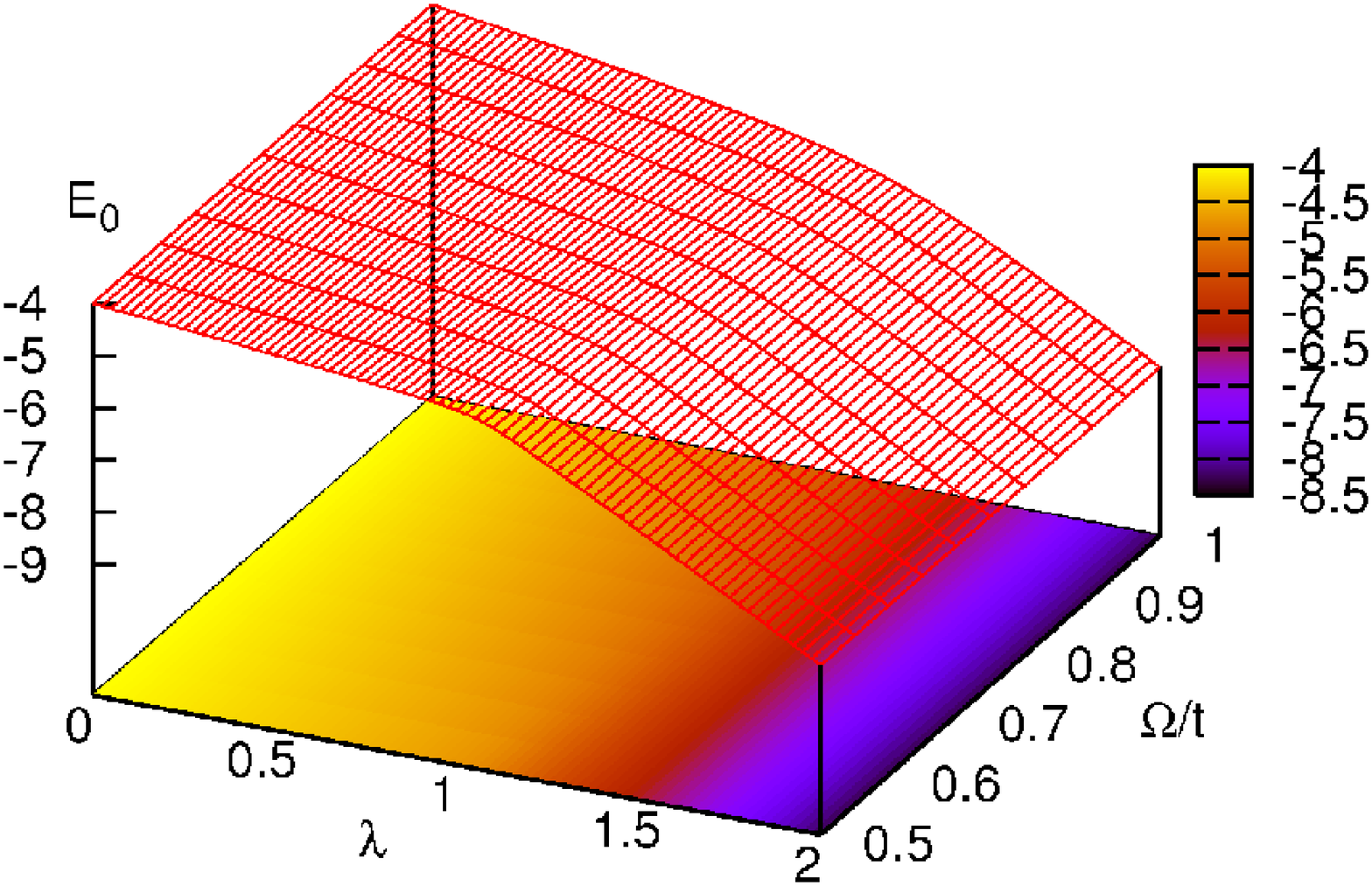}
\includegraphics[width=0.8\columnwidth]{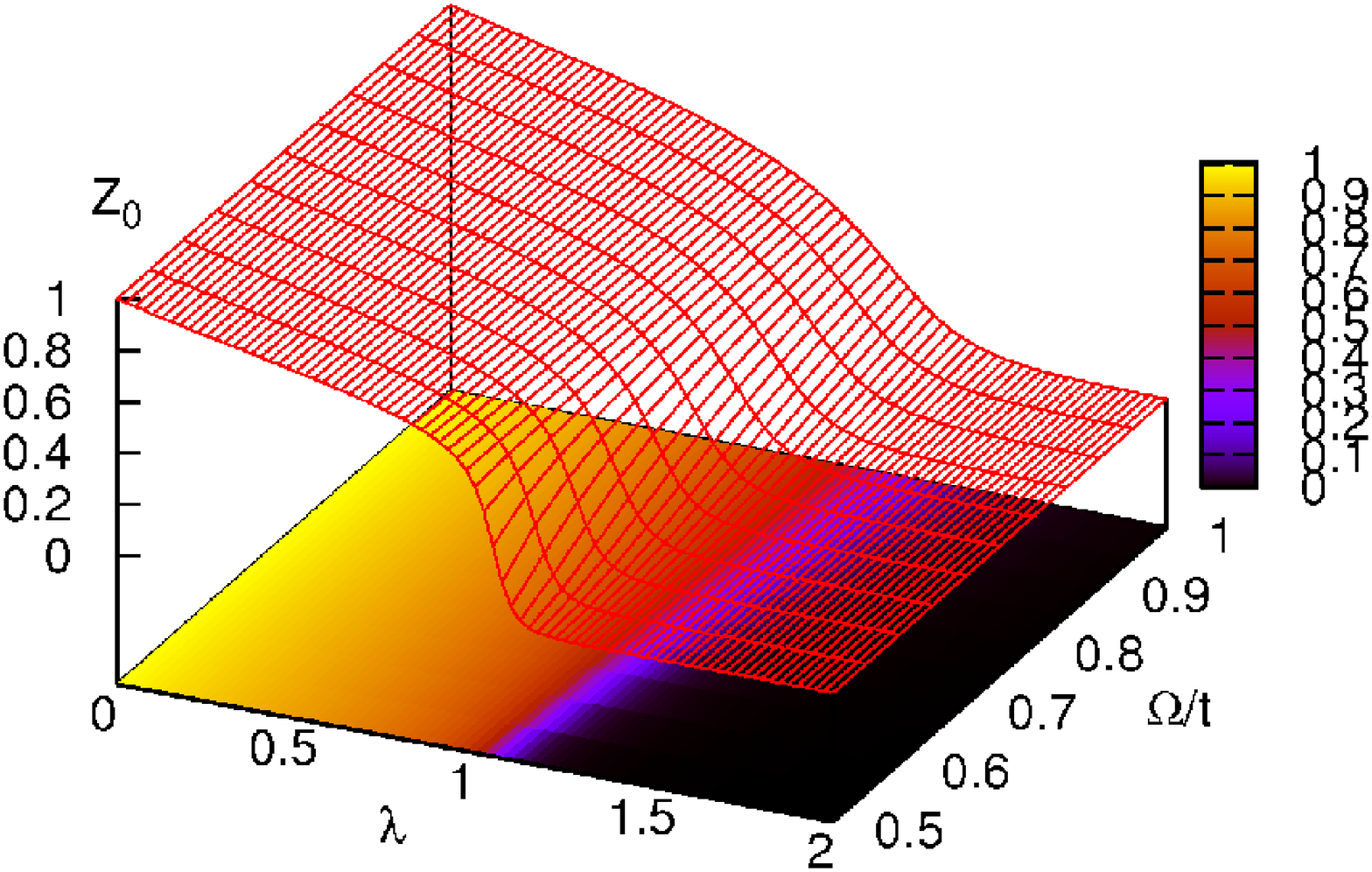}
\includegraphics[width=0.8\columnwidth]{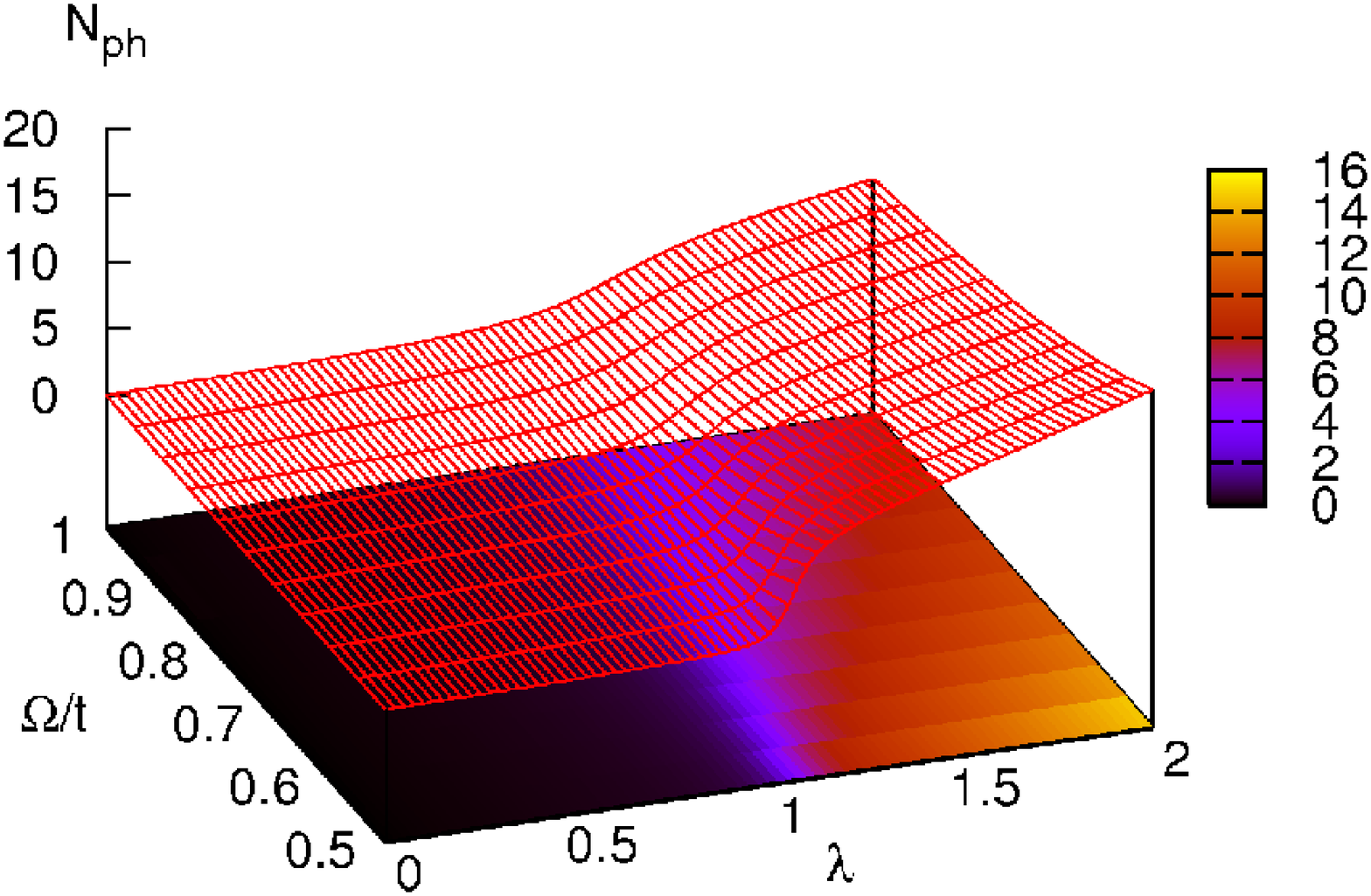}
\includegraphics[width=0.8\columnwidth]{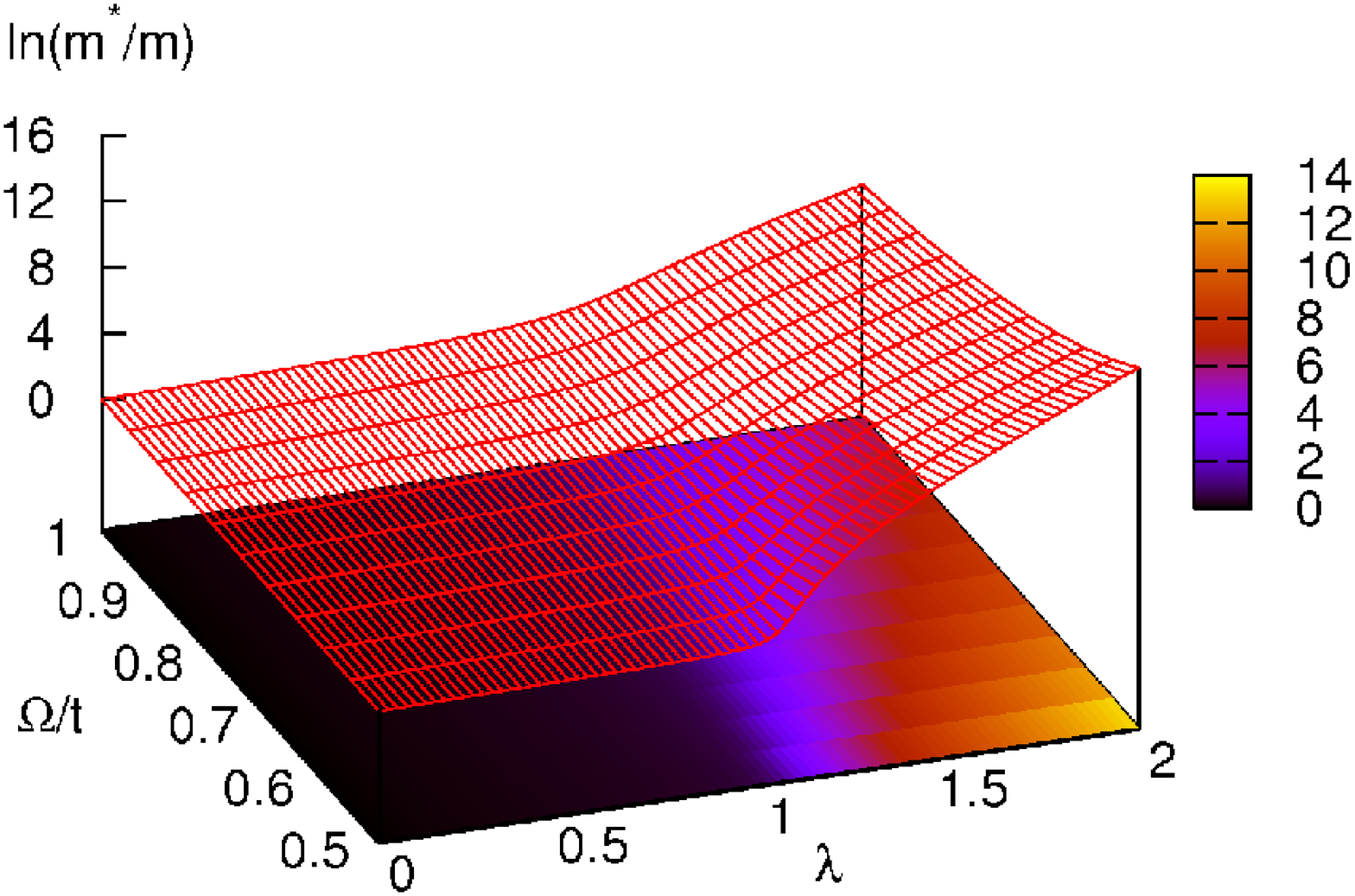}
\caption{(color online) Ground state energy $E_0$; the $qp$ weight
  $Z_0$; the average number of phonons $N_{\textrm{ph}}$ and the effective mass
  $m^*$ as a function of $\lambda$ and of $\Omega/t$, for $d=2$.}
\label{fig:2D_c}
\end{figure}

Clearly, MA (red line with red symbols, for easier comparison with
QMC
 data) has the best agreement with the QMC data. As expected from the
 sum rule analysis and the discussion on the convergence of
 $\Sigma_{\textrm{MA}}$, MA improves for larger $\Omega$. The worst
 disagreements we ever found are the ones shown in the intermediary
 $\lambda$ regime for $\Omega/t=0.1$. Even there, the error in the GS
 energy is always below $5\%$. The $qp$ weight has a more significant
 disagreement, however note that it indeed becomes asymptotically
 correct for $\lambda < 0.2$ and $\lambda > 0.8$. The second claim is
 supported by the $m^*$ data, which indeed shows convergence towards
 the expected PT values. Most importantly, even though it is
 quantitatively somewhat wrong in this intermediary regime for small
 $\Omega$, the MA approximation is the only one that clearly captures
 the crossover from the large to the small polaron, which is
 accompanied by the collapse of the $qp$ weight and the increase in
 the number of trapped phonons and thus of the effective mass.

The agreement with the QMC data is significantly improved for 2D
polarons, as shown in Fig. \ref{fig:2D_E}. Here MA gives excellent
agreement at all couplings $\lambda$. This is all the more
remarkable when considering what a numerically trivial task it is to
evaluate the MA results compared to the QMC simulations.  The
physics is similar to that seen in 1D, however the cross-over from
the large (light) polaron at weak couplings, to the small (heavy)
polaron at strong couplings, becomes somewhat sharper, especially
for smaller $\Omega/t$ values.

Given the simplicity of MA, we can also generate contour plots of
these quantities as a function of both $\lambda$ and $\Omega/t$, and
investigate the entire parameter space. Such results are shown in
Figs. \ref{fig:1D_c} and \ref{fig:2D_c} for $d=1$ and 2,
respectively. The cross-over from the large to the small polaron can
now be tracked (for instance, from the collapse of $Z_0$) and one
can quantitatively trust the results to a high degree. The
transition occurs for $\lambda \approx 1$, although this shifts to
lower values as $\Omega$ decreases.

\begin{figure}[t]
\includegraphics[width=0.75\columnwidth]{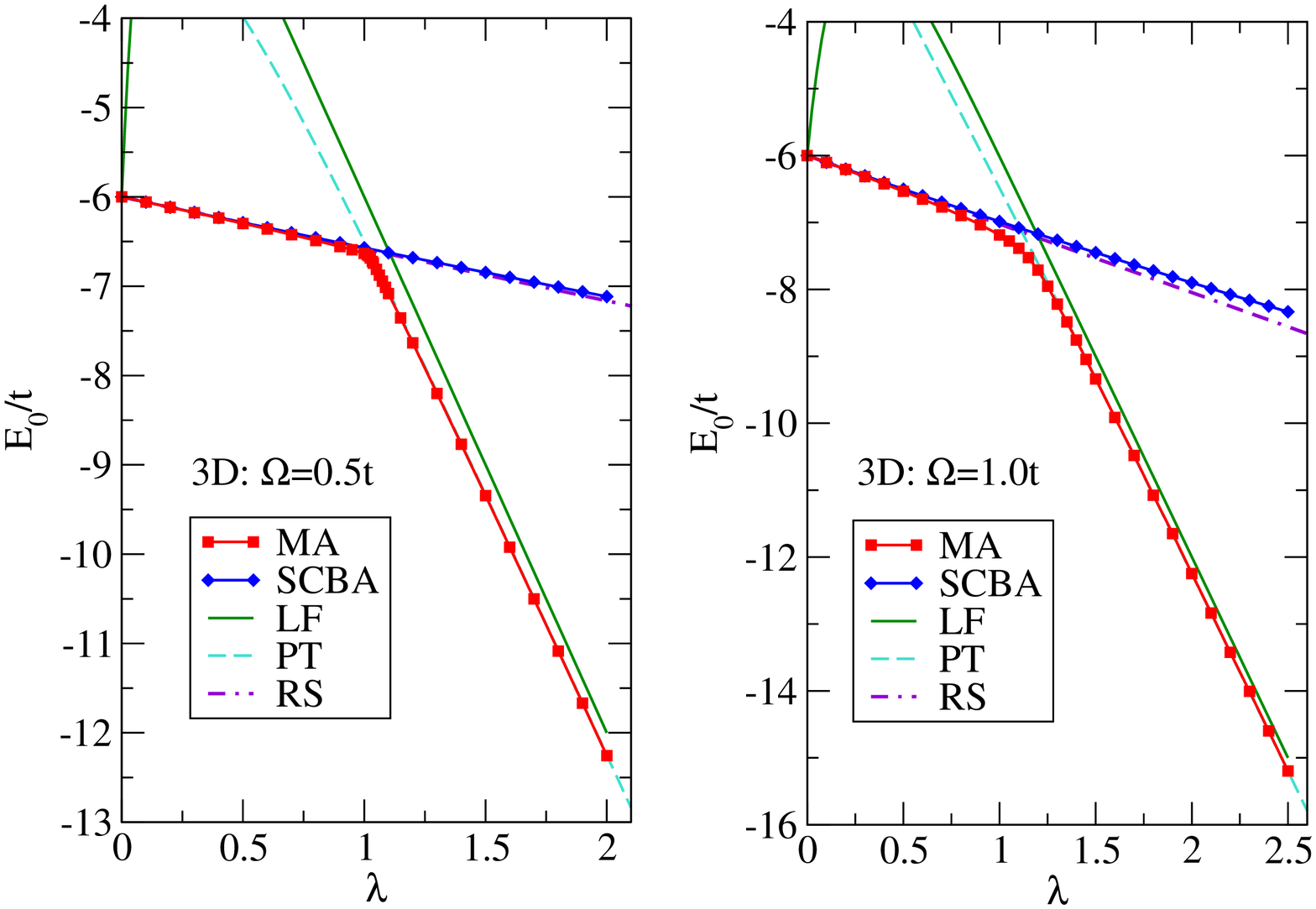}
\includegraphics[width=0.75\columnwidth]{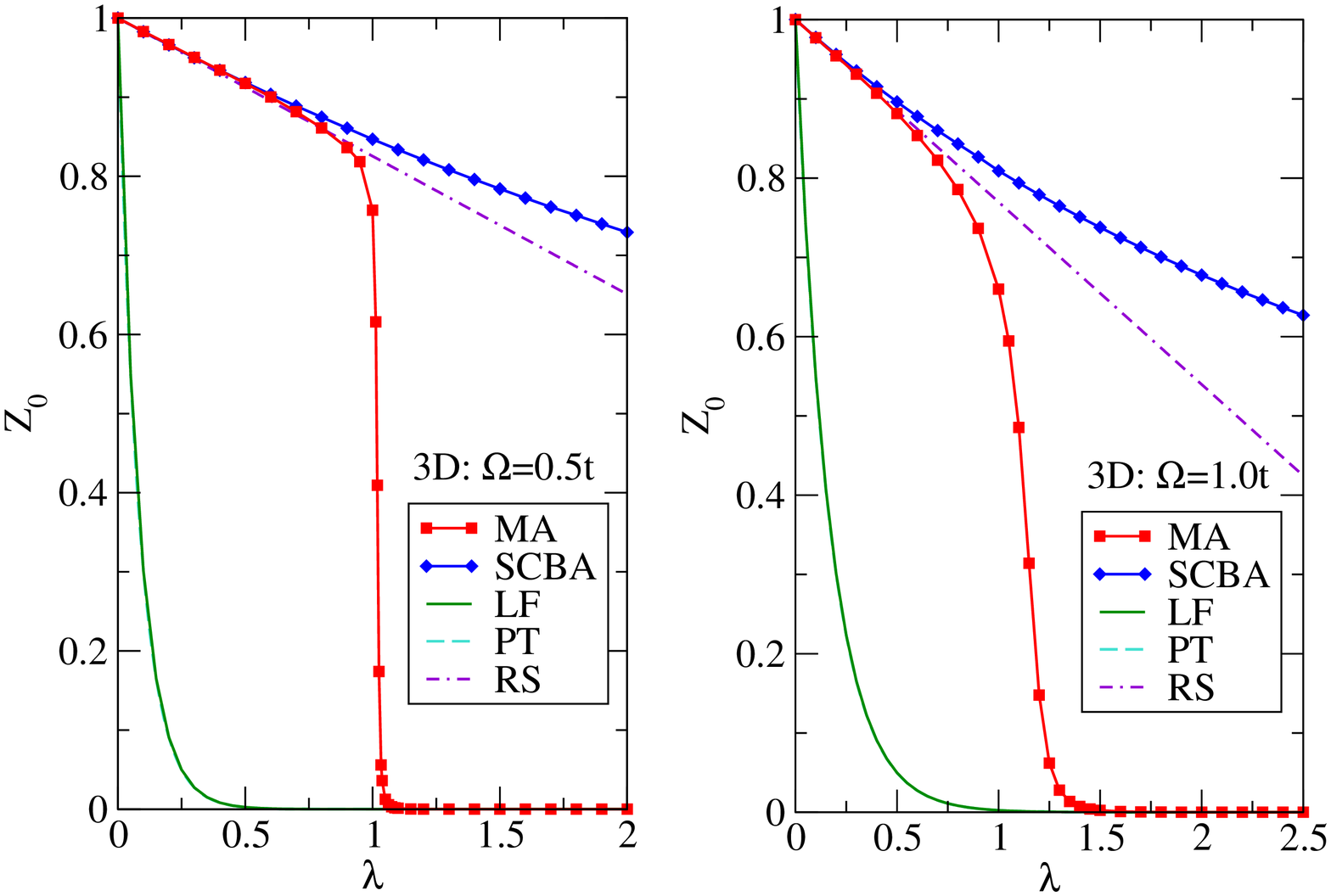}
\includegraphics[width=0.75\columnwidth]{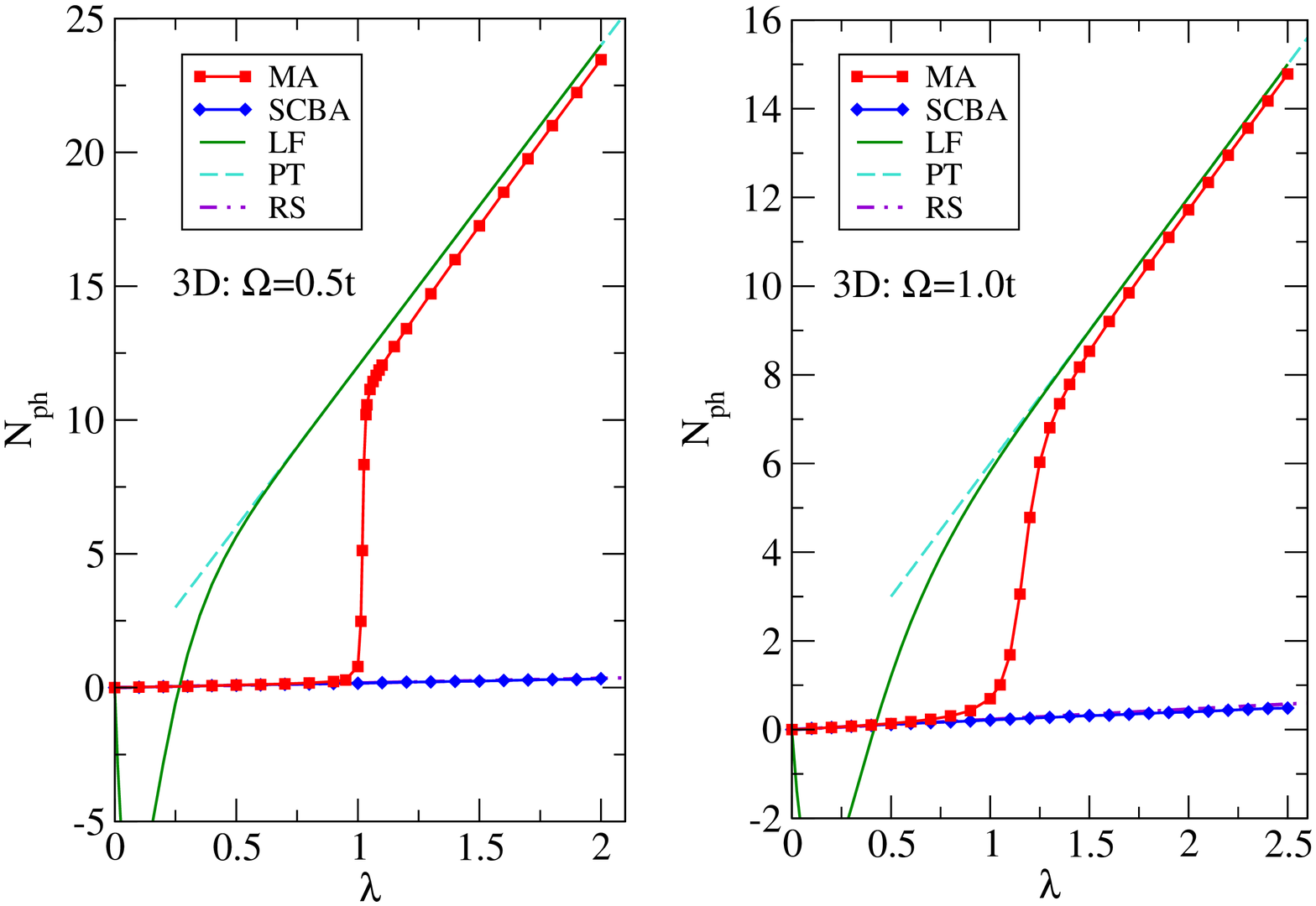}
\includegraphics[width=0.75\columnwidth]{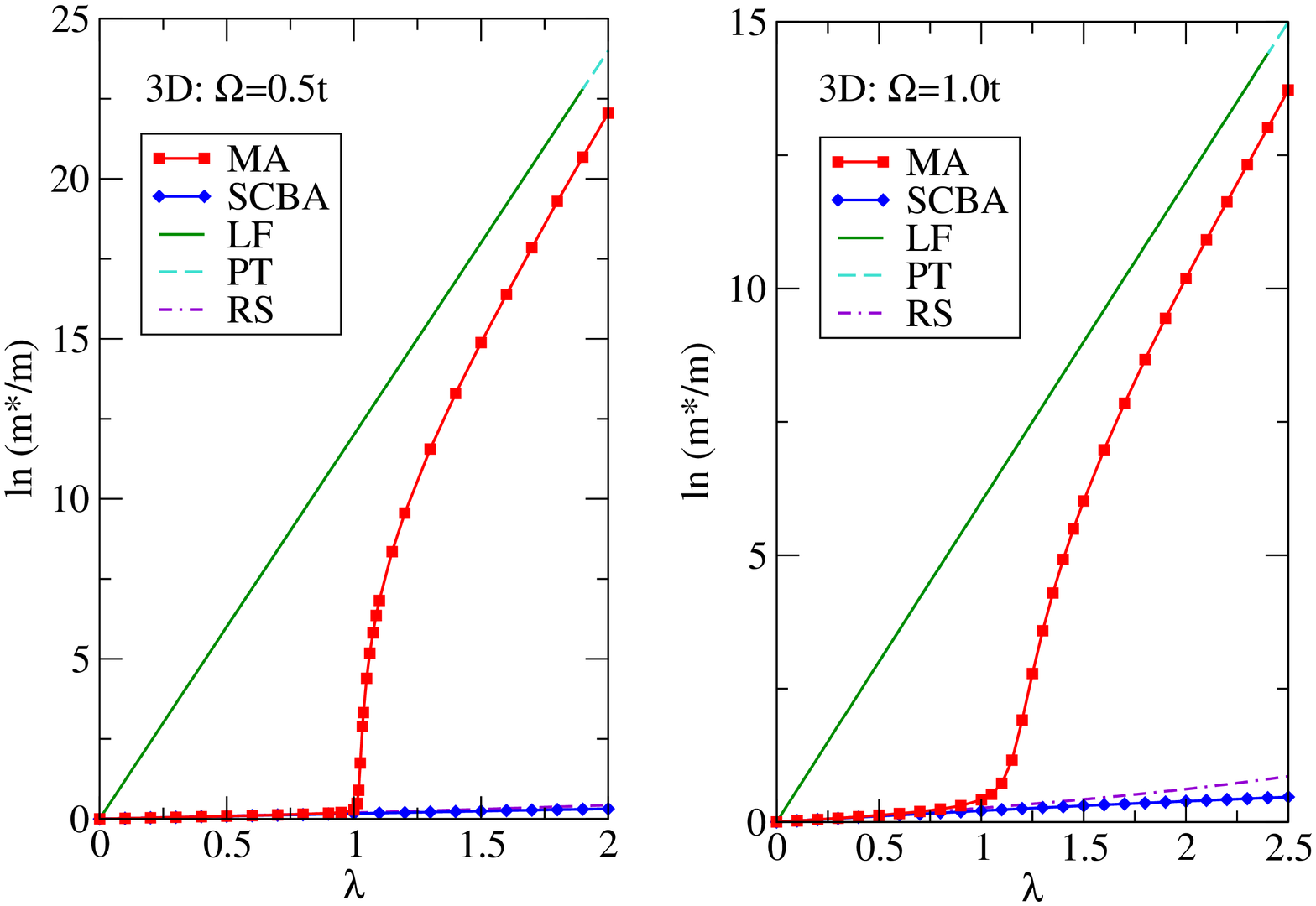}
\caption{(color online) Ground state results in 3D. Shown as a
  function of the coupling $\lambda$ are the ground state energy
  $E_0$; the $qp$ weight $Z_0$; the average number of phonons $N_{\textrm{ph}}$
  and the effective mass $m^*$ on a logarithmic scale. The left panels
  correspond to $\Omega/t = 0.1$ and the right ones to
  $\Omega/t=0.5$.}
\label{fig:3D_E}
\end{figure}

\begin{figure}[t]
\includegraphics[width=0.75\columnwidth]{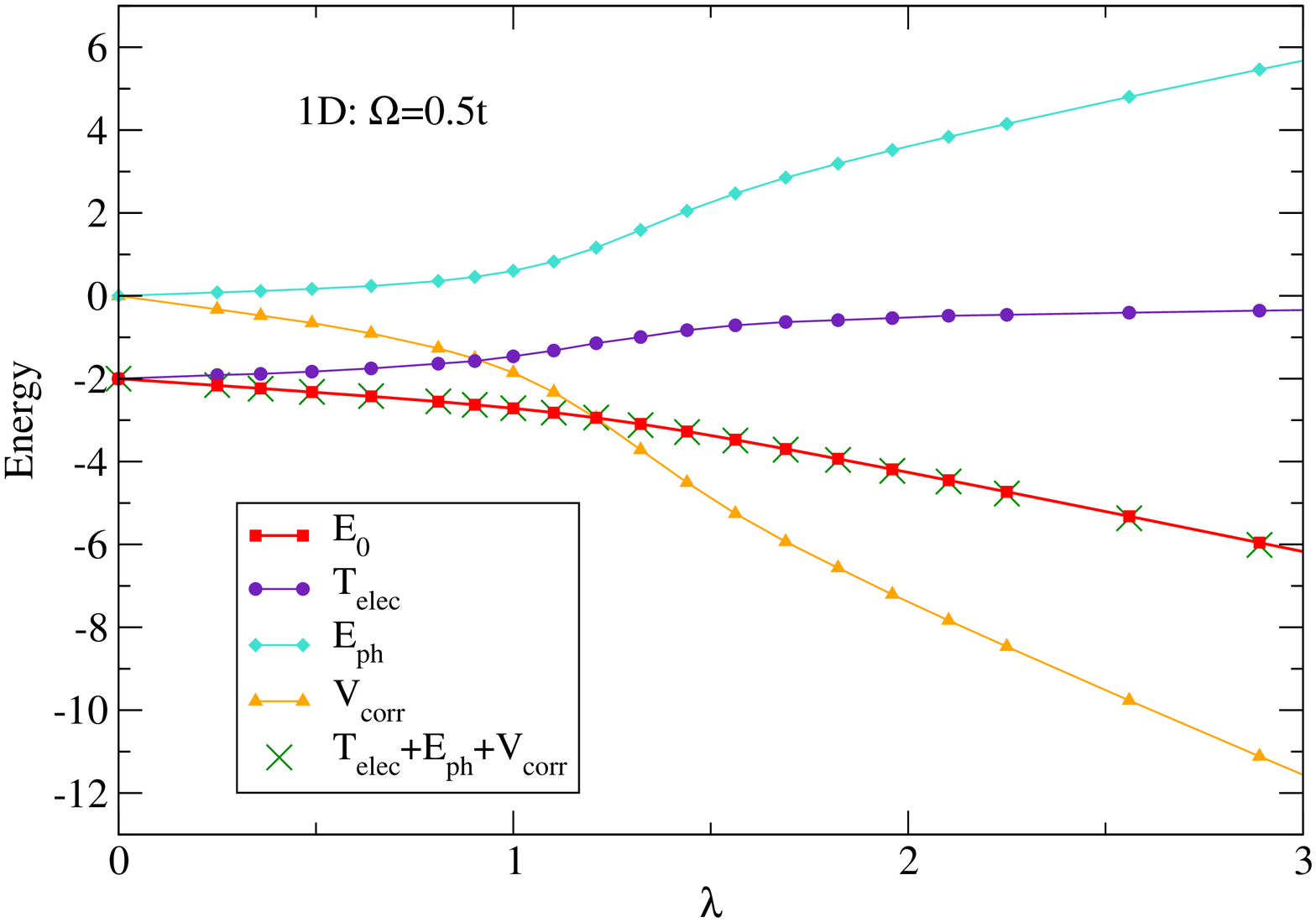}
\includegraphics[width=0.75\columnwidth]{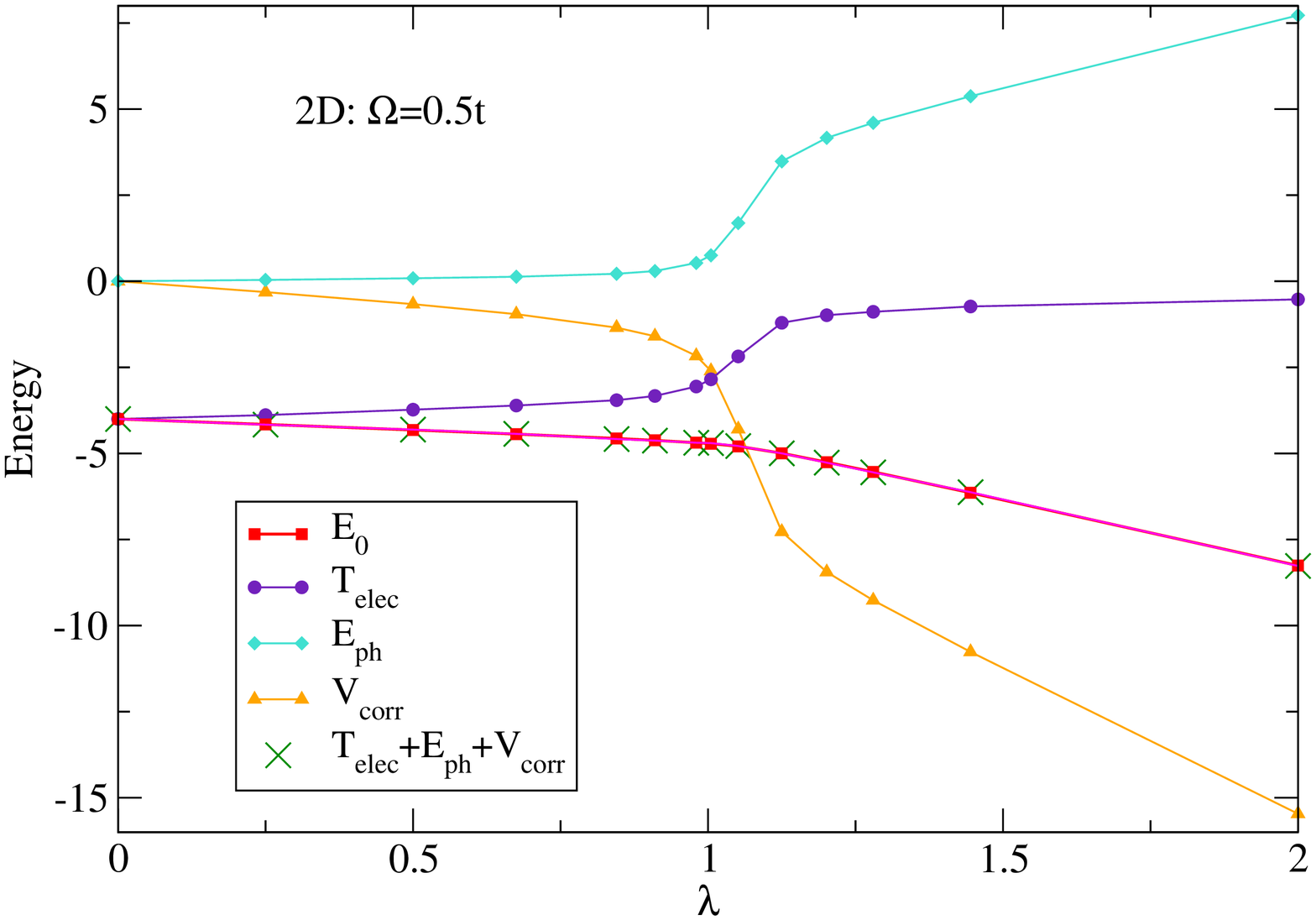}
\includegraphics[width=0.75\columnwidth]{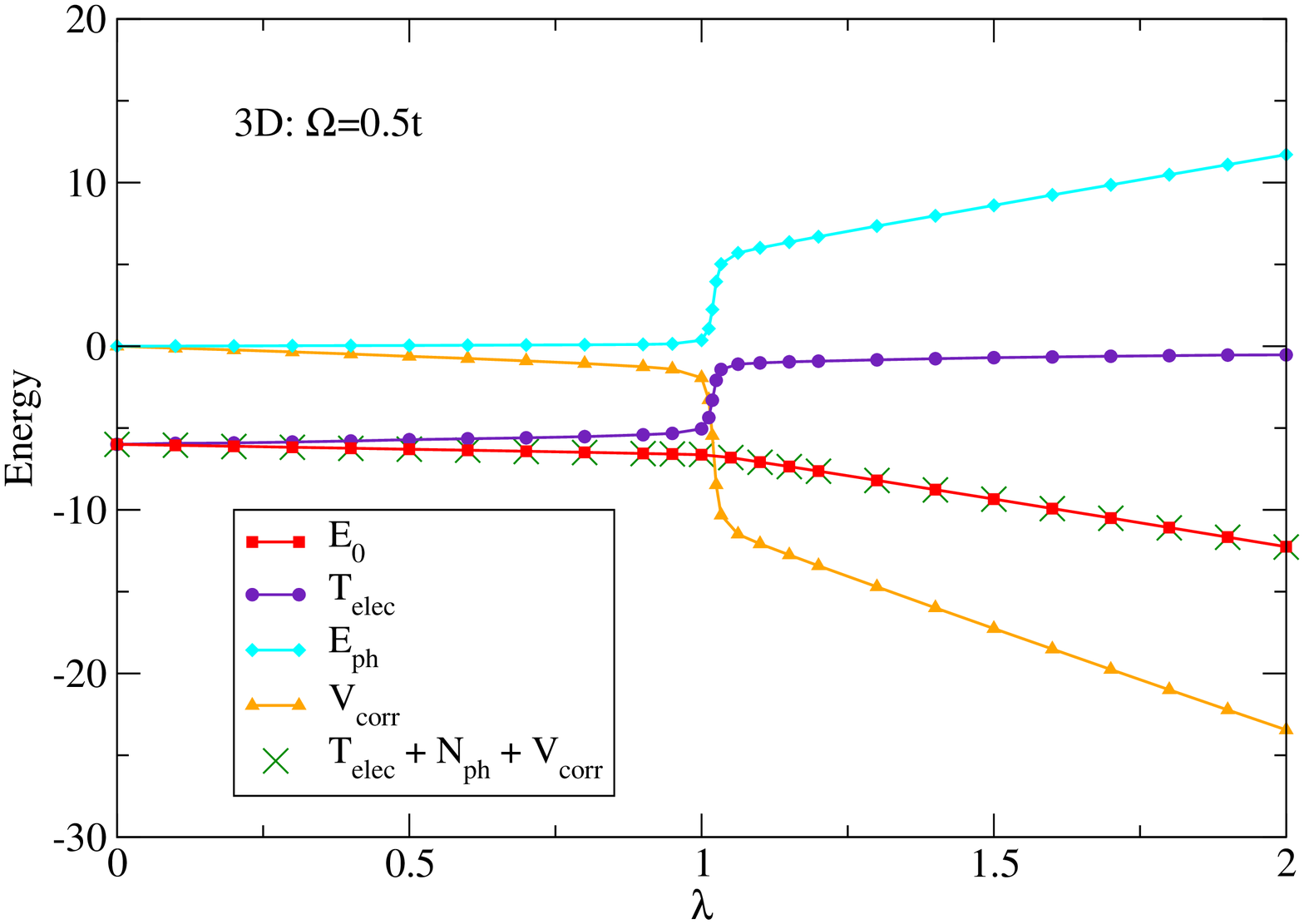}
\caption{(color online) GS expectation values for the electron
kinetic
  energy (violet), the phonon energy (cyan), and the electron-phonon
  interaction (orange), as a function of $\lambda$, for $\Omega=0.5t$
  and $d=1,2$, 3.}
\label{fig:1D_components_0-5}
\end{figure}

We also show MA results in 3D, see Fig. \ref{fig:3D_E}. Here we do
not have QMC results for comparison, however the good agreement with
one or the other perturbational theories for most coupling strengths
suggests that the MA results are probably even more accurate than in
2D. This is consistent with the expectation of improved agreement
in higher dimensions. The crossover from large to small polaron
becomes even sharper, especially for lower $\Omega$. It is still
located in the neighborhood of $\lambda \approx 1$.

To the best of our knowledge, the highly accurate three-dimensional
results shown in Fig. \ref{fig:3D_E} are the first of their kind.
This is likely due to the fact that the numerically intensive
techniques require far too much computational
effort to investigate such
cases.\cite{jeckelmann:1998,romero:1999,cataudella:2000,bonca:1999} 

Using the Hellmann-Feynman theorem like in Eqs. (\ref{eq:phonons})
and (\ref{eq:corr}) also allows us to separate the individual
contributions of
$$ T_{\textrm{elec}} = \langle GS| \sum_{\mb{k}}^{} \epsilon_{\mb k}
c^\dag_{\mb k} c_{\mb k}|GS\rangle,
$$
$$ E_{\textrm{ph}} = \langle GS|\Omega \sum_{\mb{q}}^{} b^\dag_{\mb q} b_{\mb
q}|GS\rangle = \Omega N_{\textrm{ph}}$$ and
$$V_{\textrm{corr}} = \langle GS|g \sum_{i}^{} c^\dag_ic_i\left(b_i^\dag + b_i
\right)|GS\rangle$$ to the total GS energy. Plots of these
individual contributions as a function of coupling strength
$\lambda$ are shown in Fig. \ref{fig:1D_components_0-5} for
$d=1,2,3$. As expected, the kinetic energy is close to $-2dt$ at
weak couplings, but it becomes vanishingly small in the strong
coupling limit, where the polaron becomes very heavy. The phonon
energy $E_{\textrm{ph}}$ increases roughly like $g^2/\Omega$ in the
strong coupling limit, where $N_{\textrm{ph}}\approx g^2/\Omega^2$.
It follows that the decrease in the total GS energy is due to the
interaction term, as expected. Note that this energy is proportional
to the correlation of Eq. (\ref{eq:corr}). Since $E_0 \approx
-g^2/\Omega$ in the strong coupling limit (see agreement with PT
results), it follows that this correlation becomes asymptotically
equal to $-2g/\Omega$ in the strong coupling limit.

\subsection{Low energy states: momentum dependence}

\begin{figure}[t]
\includegraphics[width=0.8\columnwidth]{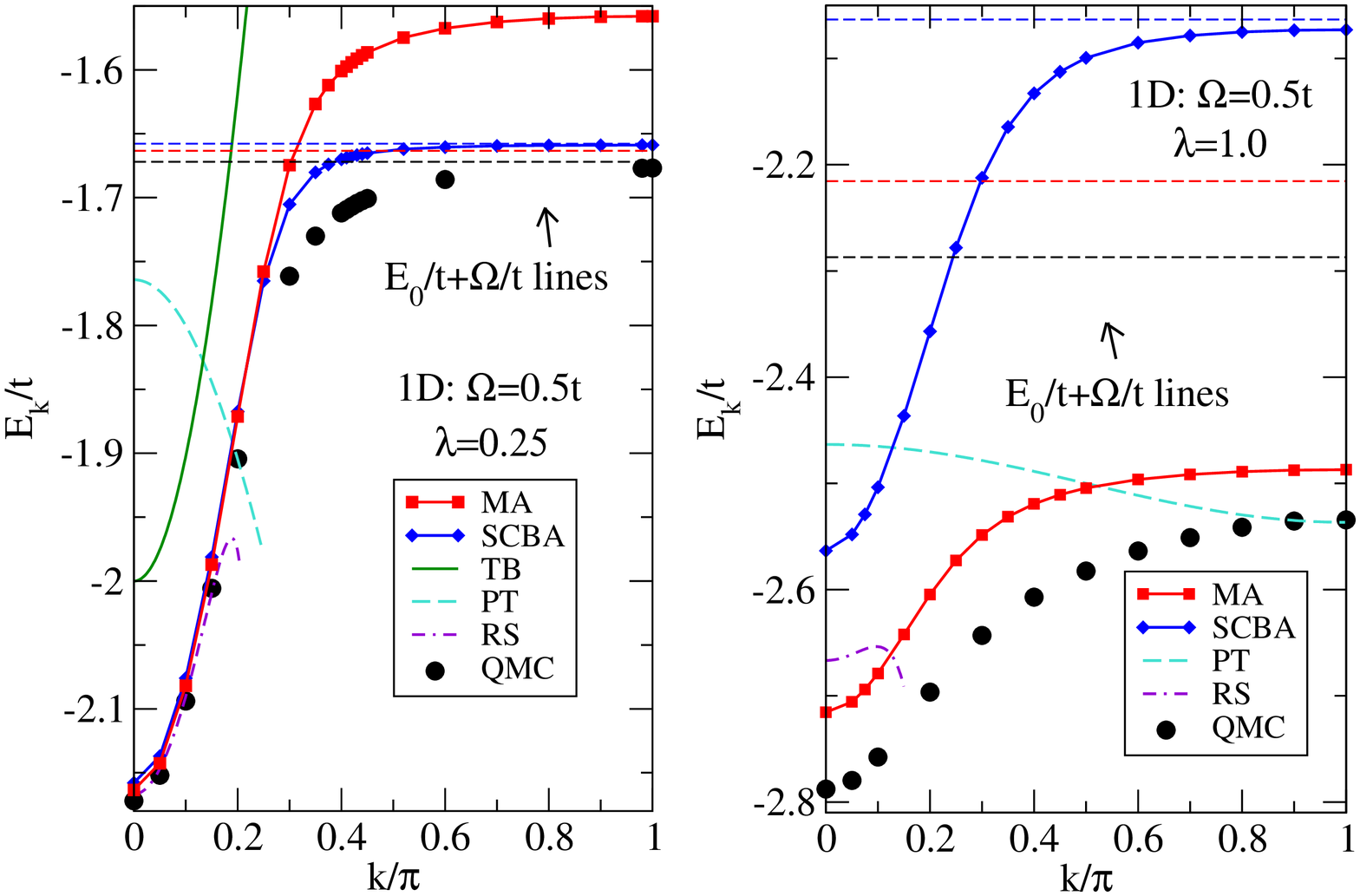}
\includegraphics[width=0.8\columnwidth]{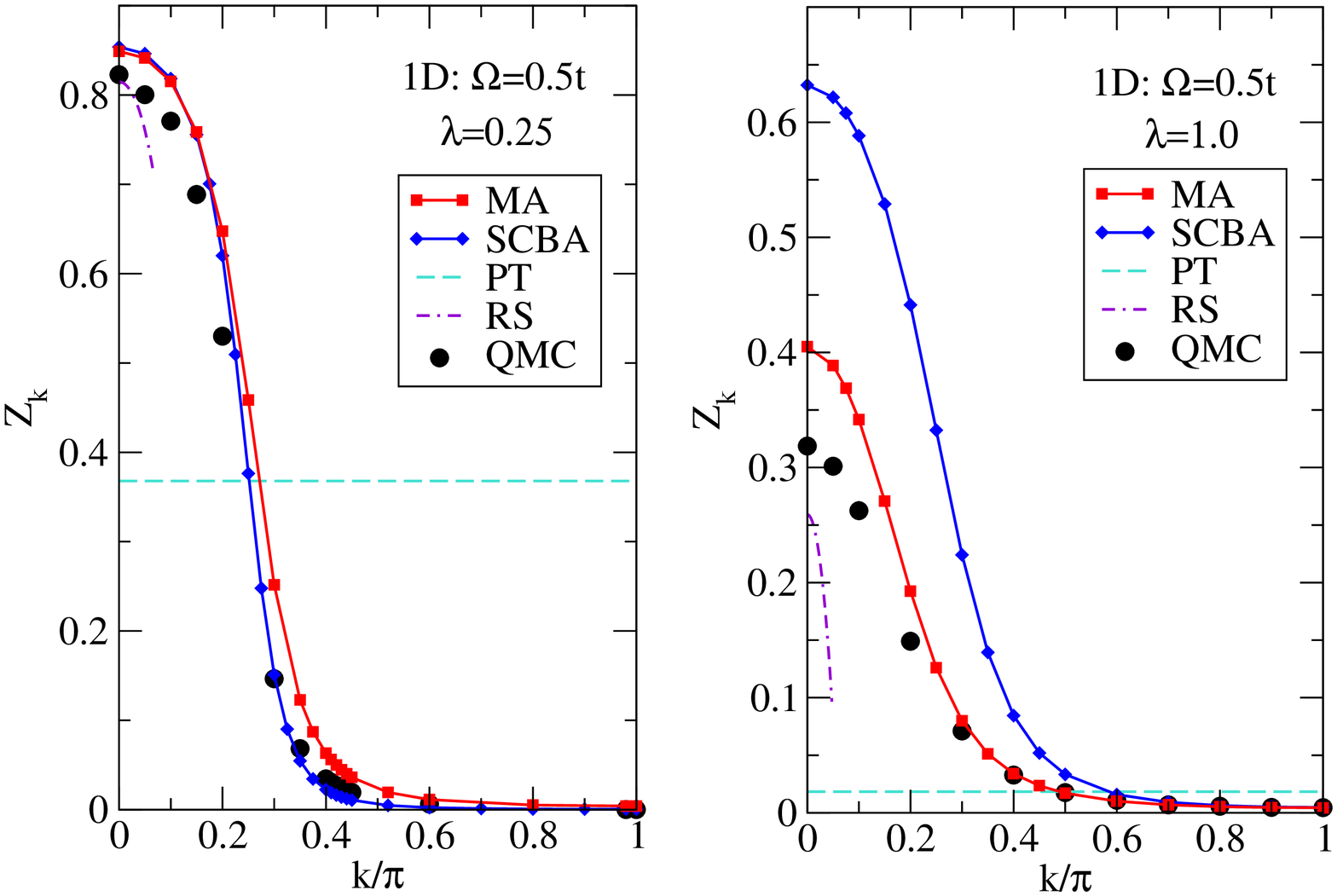}
\includegraphics[width=0.8\columnwidth]{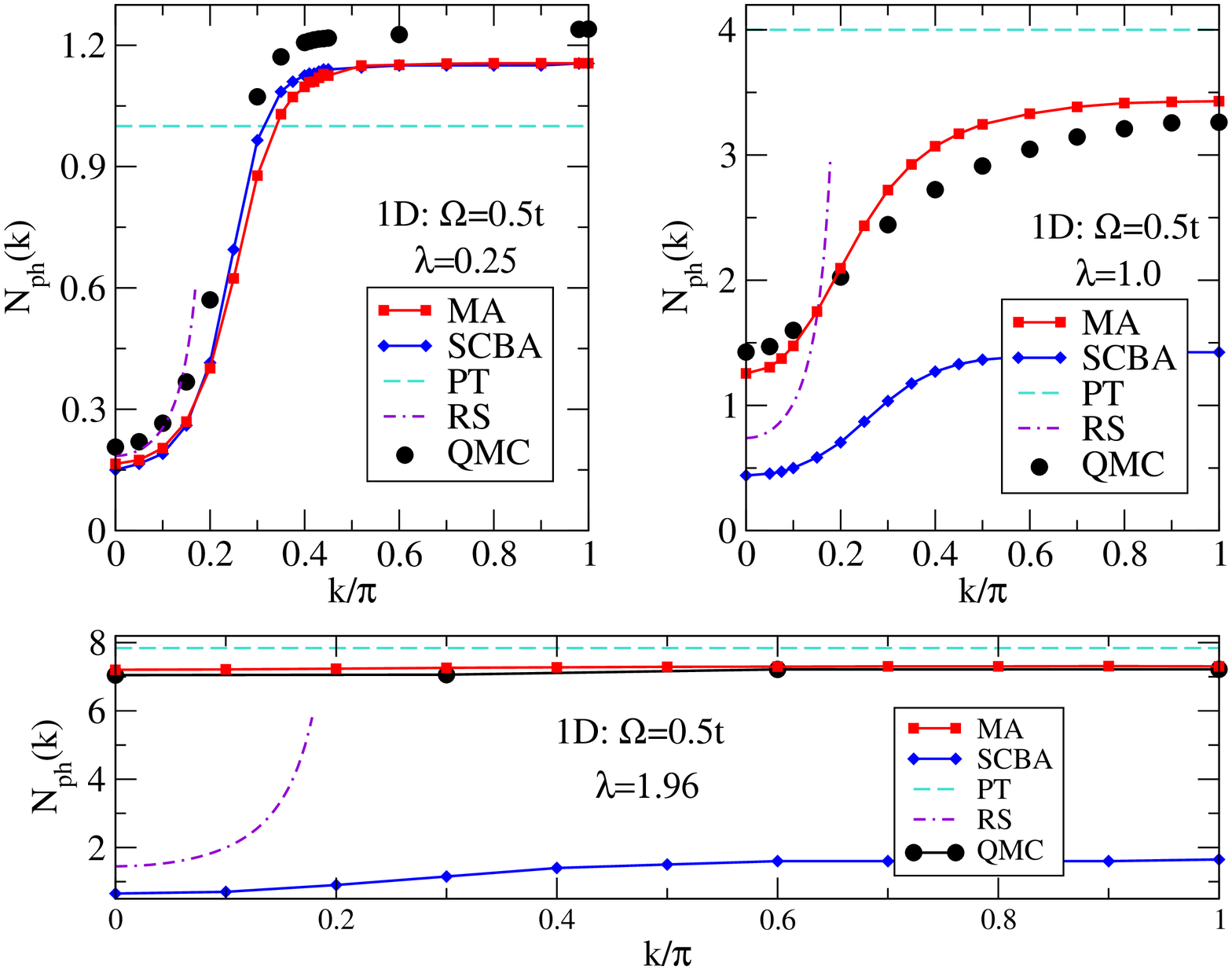}
\caption{(color online) Lowest polaron eigenenergy for a given $k$,
  $E_k$, and the corresponding $qp$ weight $Z_k$ and average phonon
  number $N_{\textrm{ph}}(k)$. Results are for $1D$, $\Omega/t=0.5$,
  $\lambda=0.25$ and 1 (for $E_k$ and $Z_k$) as well as 1.96, for
  $N_{\textrm{ph}}(k)$. Only half of the BZ is shown.  }
\label{fig:1D_Ek}
\end{figure}

We can also calculate the same properties for the lowest energy state
corresponding to each given momentum $\mb{k}\ne 0$, to find the
low-energy behavior of the polarons. In this section we present
comparisons with available QMC results\cite{macridin:2003} for 1D and
2D systems.

In Fig. \ref{fig:1D_Ek} we show 1D results for the polaron
dispersion $E_k$, the associated $qp$ weight $Z_k$ and average
phonon number $N_{\textrm{ph}}(k)$ for two couplings. For the very
weak $\lambda=0.25$, we see that MA and SCBA are equally good at
small $k$, however for large $k$ MA overestimates the energy, such
that the continuum which is expected to appear at a distance
$\Omega$ above the GS energy $E_0$, is in this case pushed somewhat
higher. We will return to a discussion of this discrepancy later on.
As expected, the $qp$ weight is large for small $k$, where the main
contribution to the eigenstate comes from the free electron state
$c^\dag_k|0\rangle$. However, $Z_k$ goes to zero for larger $k$,
since these are primarily linear combinations of states of type
$c^\dag_{k-q}b^\dag_{q}|0\rangle$, as confirmed also by the the
average phonon number of about unity. Note that the RS perturbation
also works well for small $k$, as expected. It however breaks done
at a finite $k$ where $\epsilon_k \approx -2t + \Omega$, i.e. the
free electron dispersion crosses into the continuum of electron plus
one phonon states. Here, RS predicts an unphysical peak in the
polaron dispersion (see the denominator of (\ref{eq:E_RS})) and it
fails for larger $k$.

\begin{figure}[t]
\includegraphics[width=0.8\columnwidth]{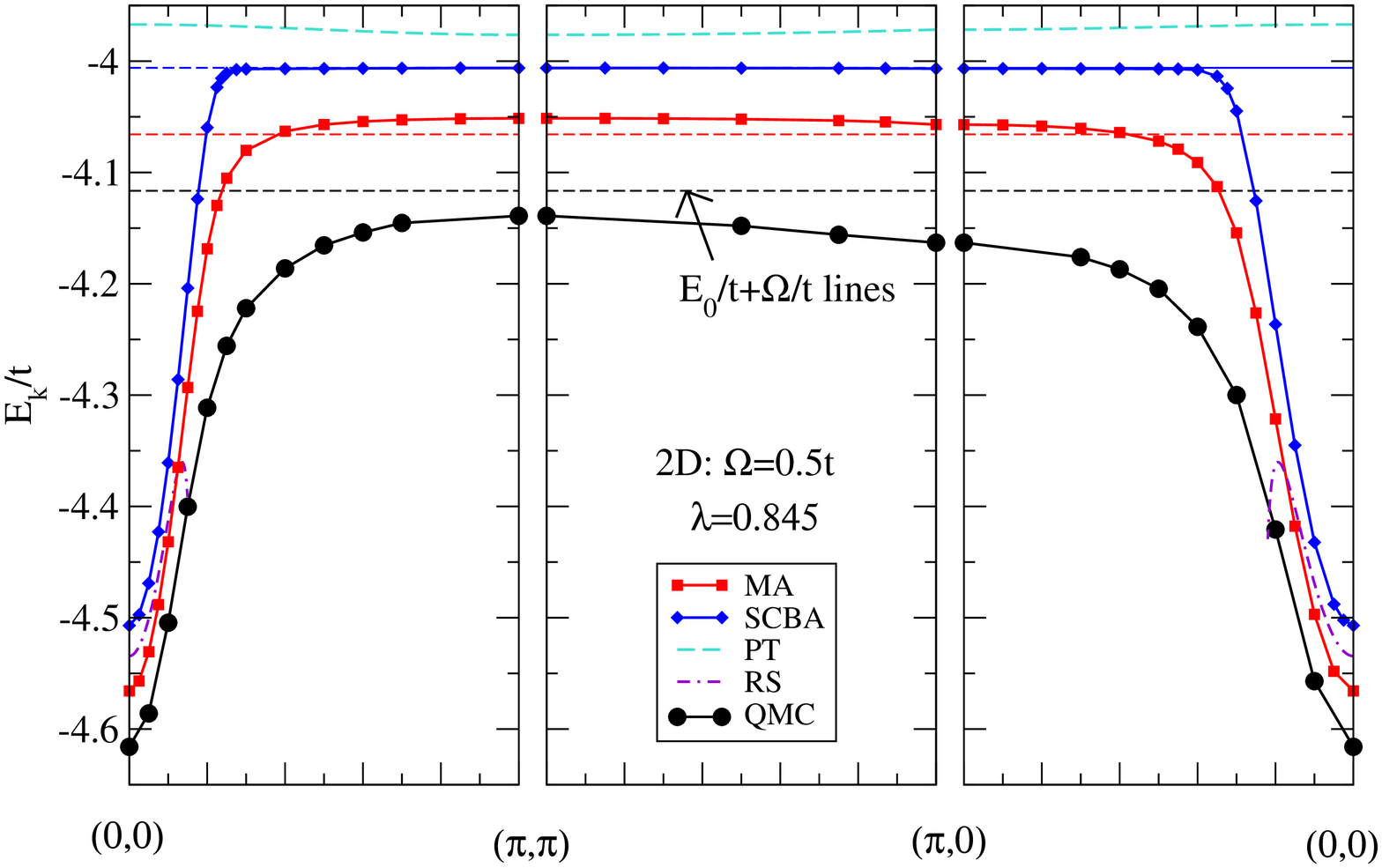}
\includegraphics[width=0.8\columnwidth]{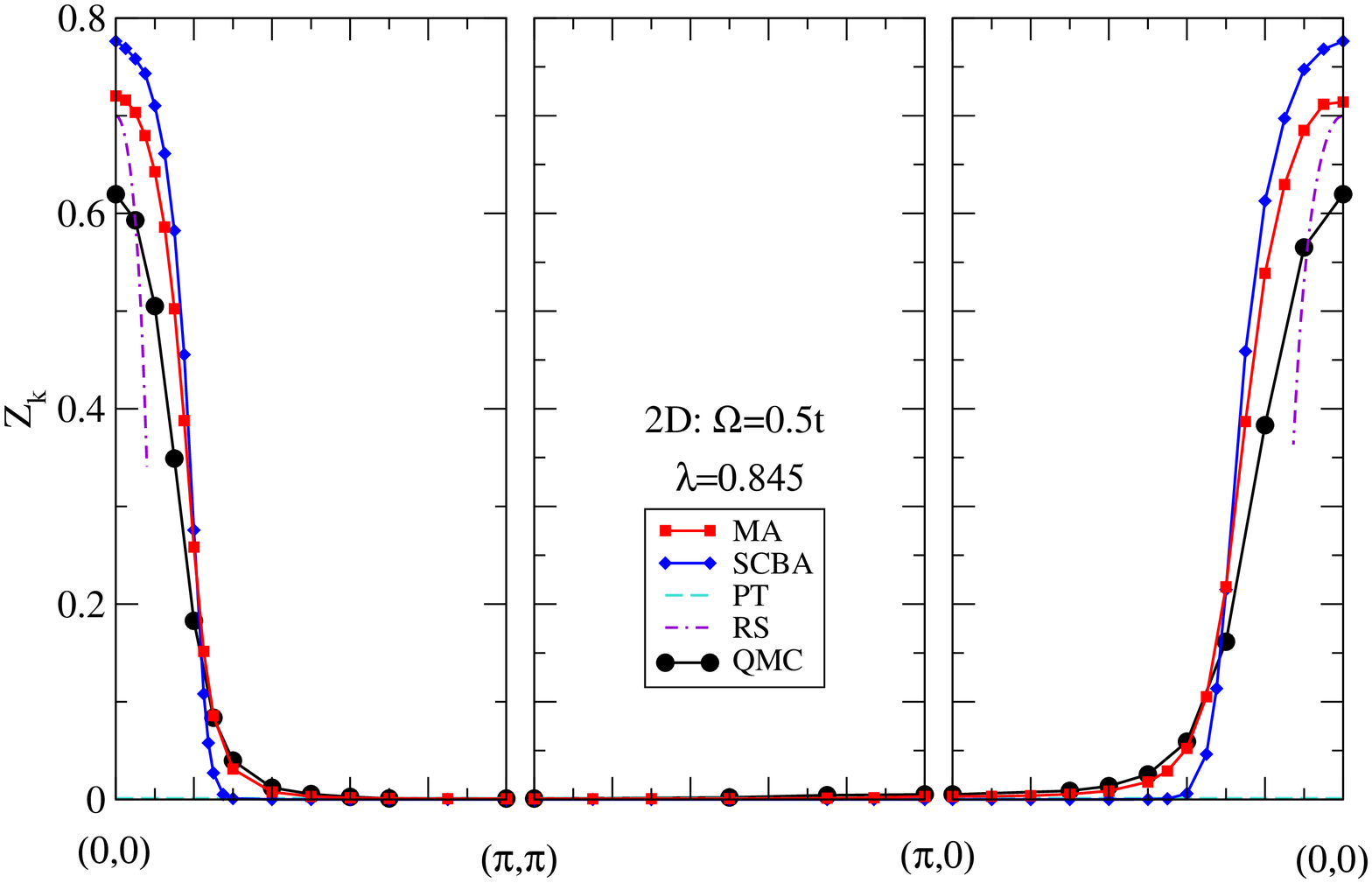}
\includegraphics[width=0.8\columnwidth]{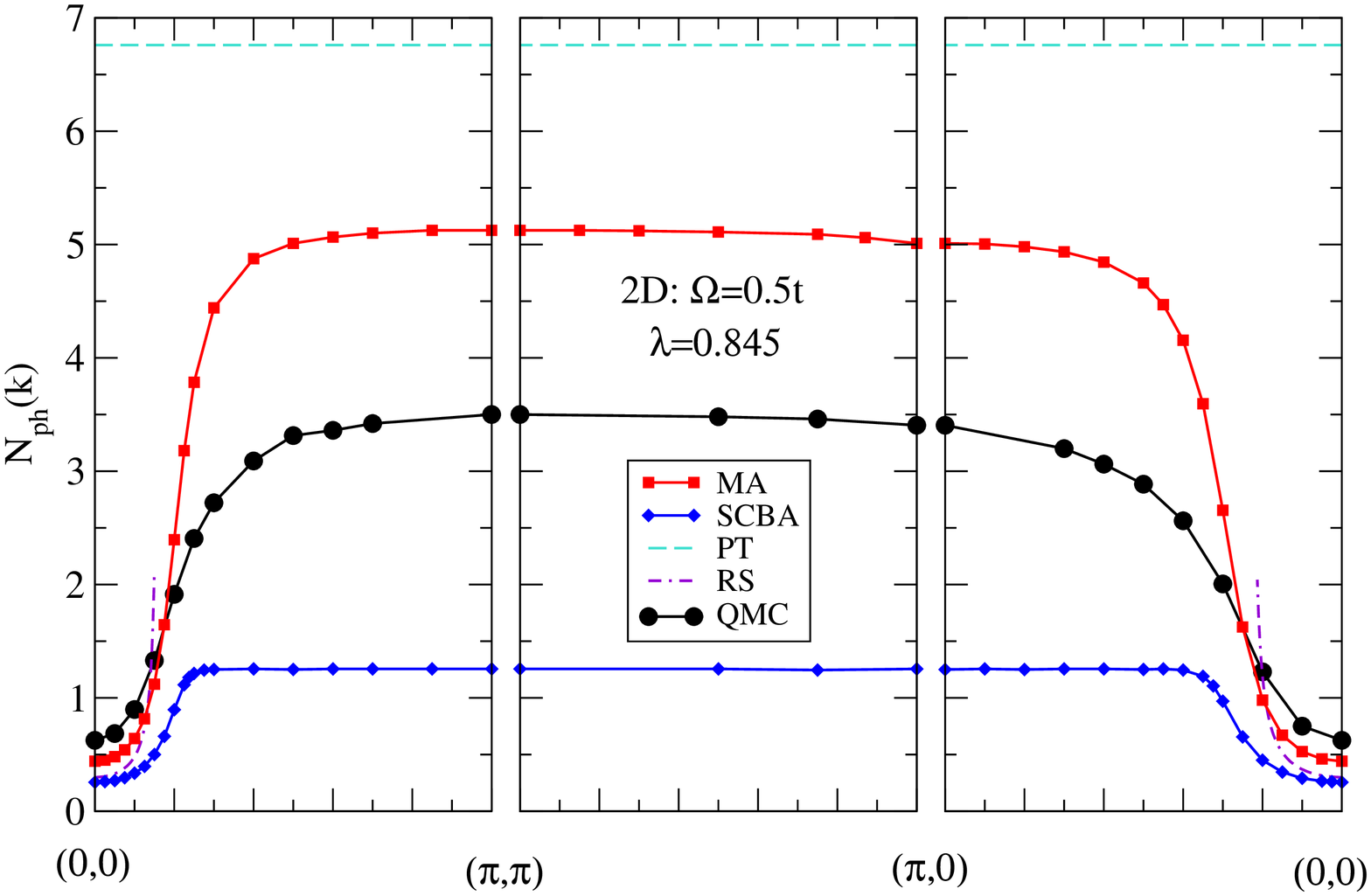}
\caption{(color online) Lowest polaron energy for a given $\mb{k}$,
  $E_{\mb k}$, and 
  the corresponding $qp$ weight $Z_{\mb k}$ and average phonon number
  $N_{\textrm{ph}}(\mb{k})$. Results are for $2D$, $\Omega/t=0.5$,
  $\lambda=0.845$. Three high-symmetry cuts in the Brillouin zone are
  shown.  }
\label{fig:2D_Ek}
\end{figure}

The second coupling $\lambda=1$ is roughly in the cross-over regime,
see also Fig. \ref{fig:1D_E}.  Here MA gives a much better agreement
with QMC than SCBA or the perturbational theories. The polaron
bandwidth is already renormalized and well below the weak coupling
value of $ \Omega$. In fact, as we show later, there is another
bound state between these states and the continuum. This is not
captured in SCBA, which always predicts a polaron bandwidth of
$\Omega$ with a continuum above, and roughly between zero and one
average number of phonons as $k$ increases from 0 to $\pi$. For the
strong coupling $\lambda=1.96$, low-energy properties become almost
$k$-independent, as expected since the Lang-Firsov impurity limit is
being approached.

A second such comparison is possible for the 2D case with
$\Omega=0.5t$ and $\lambda=0.845$, where QMC data is available, see
Fig. \ref{fig:2D_Ek}. This coupling is on the weak side of the
crossover, with the GS $qp$ weight still large, $Z_0\approx 0.6$.
Here MA is already doing better than SCBA. As in the 1D
weak-coupling case, one can again see that the MA polaron bandwidth
is slightly larger than $\Omega$. MA also somewhat overestimates the
average number of phonons for large $\mb{k}$ values. Overall, given
that all this data is in the crossover regime where the MA is at its
worst, one can conclude that MA is also reasonably accurate in
capturing low-energy polaron behavior.

\subsection{High-energy states}

The main motivation in trying to find an approximation for the
Green's function is that this quantity gives not only low-energy
information, but the whole spectrum. Here we compare MA predictions
with various high-energy results available in the literature.
Unfortunately there are much fewer of these, because the
computational effort to obtain the whole spectrum through the usual
numerical approaches is generally forbidding.

\begin{figure}[t]
\includegraphics[width=0.95\columnwidth]{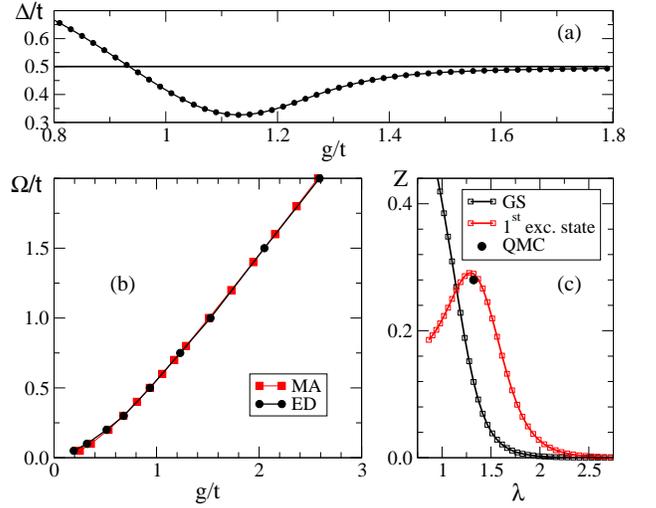}
\caption{(color online) (a) Energy gap $\Delta = E_1- E_0$ between GS and first
  excited $k=0$ state. A second bound state is stable if $\Delta <
  \Omega$ (here $\Omega=0.5 t$); (b) Line below which a second bound
  state appears: ED\cite{bonca:1999} (circles) and MA
  (squares); (c) MA $qp$ weight of the second bound state when
  stable (red squares), and that of the GS (black squares), for
  $\Omega/t=0.5$. The circle is the one QMC result available for the
  $qp$ weight of the second bound state.\cite{macridin:2003} These
  results are for 1D.  }
\label{fig:2Dpeak}
\end{figure}

\begin{figure}[b]
\includegraphics[width=0.95\columnwidth]{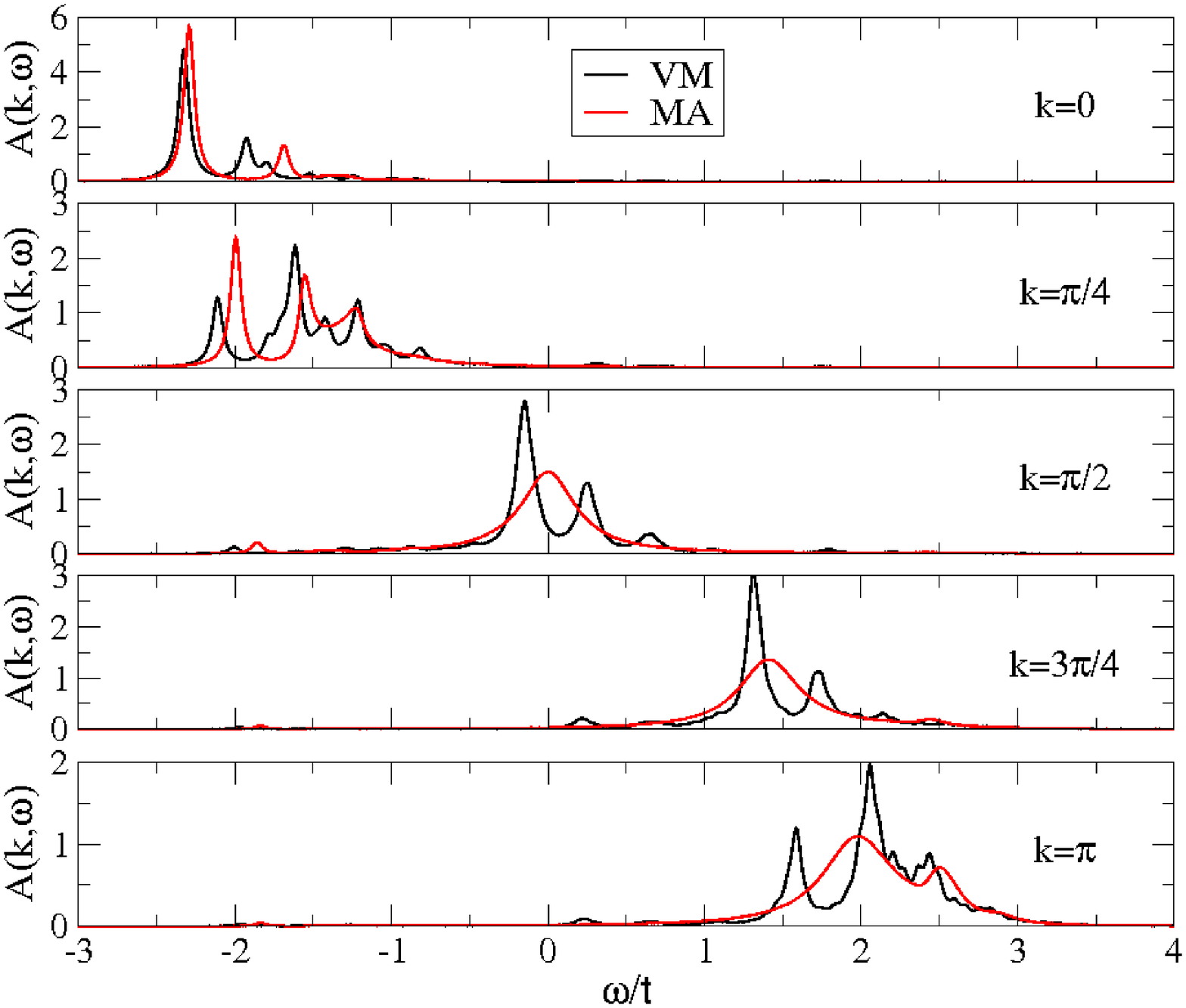}
\caption{(color online) 1D spectral weight $A(k,\omega)$ vs.
$\omega$,
  for $k=0, {\pi\over 4}, {\pi\over2}, {3\pi\over4}$ and $\pi$. MA
  results (red line) vs. data from Ref. \onlinecite{filippis:2005}
  (black line). Parameters are $\Omega=0.4t$, $\lambda=0.5$,
  $\eta=0.1\Omega$.  }
\label{Cat05}
\end{figure}

We begin with a comparison against exact diagonalization (ED) 1D
data, from Ref. [\onlinecite{bonca:1999}]. The results are shown in
Fig. \ref{fig:2Dpeak}. As already discussed, for weak coupling there
is a continuum starting at $E_0+\Omega$, but for stronger coupling a
second so-called bound state appears below the continuum. In Fig.
\ref{fig:2Dpeak}(a) we track the energy $E_1$ of the second $k=0$
state.  For small couplings, the data actually shows the maximum DOS
in the continuum, not its edge (the maximum is generally located
close to the lower edge. This data shows again that MA somewhat
overestimates this energy, which should be $\approx \Omega$).  When
$E_1< E_0+\Omega$, there is a true discrete state. Note that panel
$(a)$ is in very good quantitative agreement with similar data shown
in Fig. 8 of Ref. \onlinecite{bonca:1999}. The only difference is
for strong coupling, where the ED data shows $E_1 > E_0 + \Omega$
again, however, with significant finite-size dependence on the chosen
Hilbert space cutoff.

We can thus find the coupling $g/t$ where the second bound state
appears, for different values of $\Omega/t$. This line is shown in
panel (b), together with the ED results. The agreement between the
two data sets is excellent, even at small $\Omega/t$ values where we
expect MA to be less accurate. We also show in panel (c) the $qp$
weight of this second bound state, where stable. This data is not
given in Ref. \onlinecite{bonca:1999}, however one QMC point is
available in Ref. \onlinecite{macridin:2003}, in good agreement with
the MA prediction.

\begin{figure}[t]
\includegraphics[width=0.95\columnwidth]{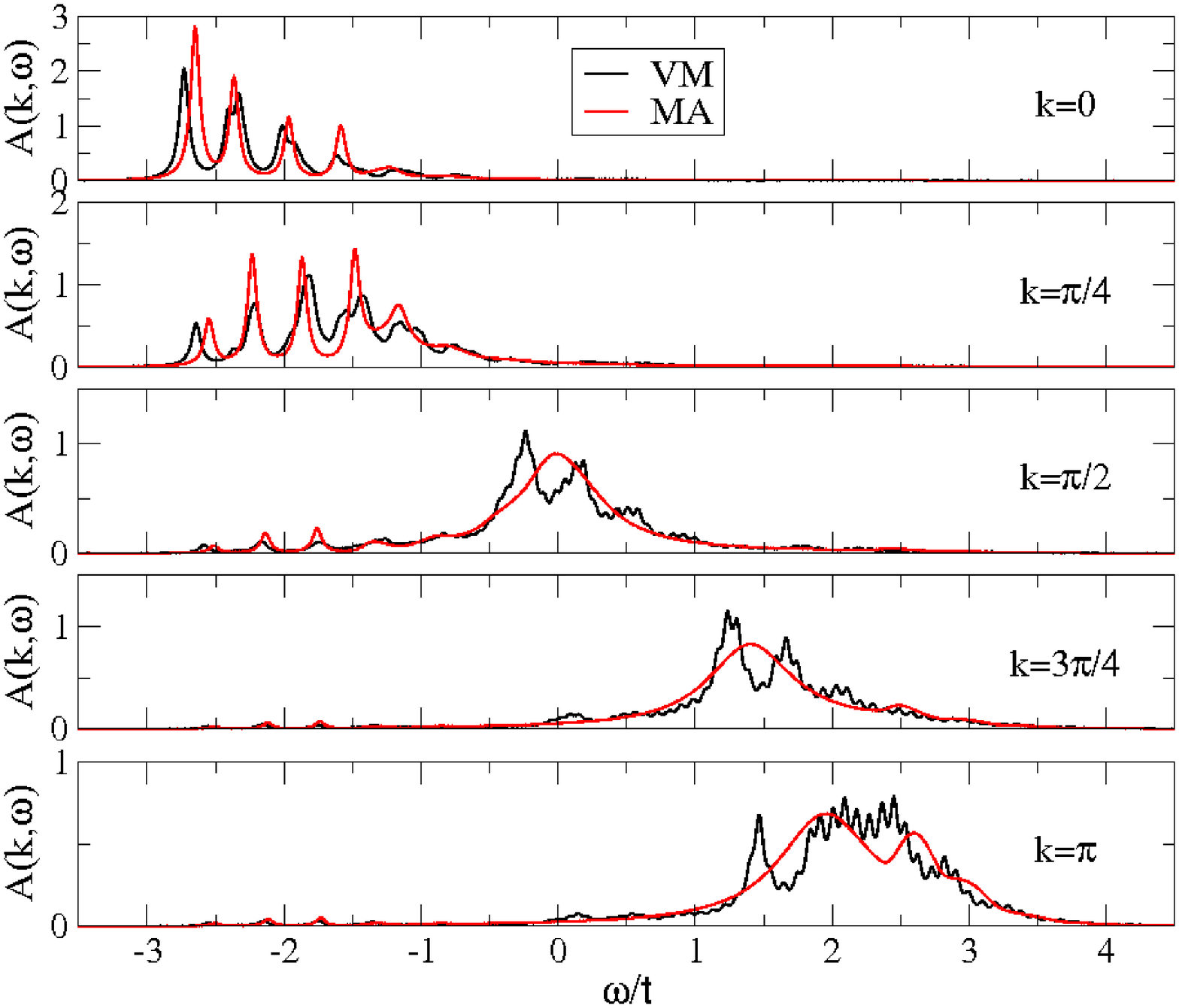}
\caption{(color online) 1D spectral weight $A(k,\omega)$ vs.
$\omega$,
  for $k=0, {\pi\over 4}, {\pi\over2}, {3\pi\over4}$ and $\pi$. MA
  results (red line) vs. data from Ref. \onlinecite{filippis:2005}
  (black line). Parameters are $\Omega=0.4t$, $\lambda=1$,
  $\eta=0.1\Omega$.  }
\label{Cat1}
\end{figure}

\begin{figure}[b]
\includegraphics[width=0.95\columnwidth]{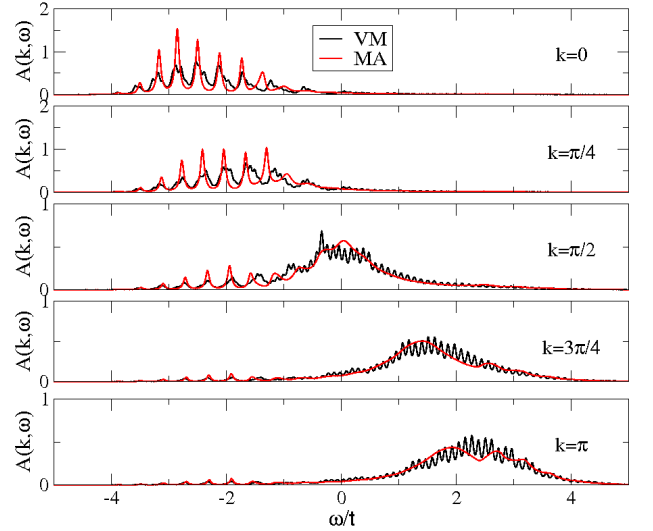}
\caption{(color online) 1D spectral weight $A(k,\omega)$ vs.
$\omega$,
  for $k=0, {\pi\over 4}, {\pi\over2}, {3\pi\over4}$ and $\pi$. MA
  results (red line) vs. data from Ref. \onlinecite{filippis:2005}
  (black line). Parameters are $\Omega=0.4t$, $\lambda=2$,
  $\eta=0.1\Omega$.  }
\label{Cat2}
\end{figure}

We now move to comparisons for the entire spectral weight
$A(k,\omega)$. In Figs. \ref{Cat05}, \ref{Cat1} and \ref{Cat2} we
show comparisons for a 1D system with $\Omega=0.4t$ and three
different coupling strengths $\lambda=0.5, 1$ and 2, respectively.
In each case, data for 5 values of $k$, namely $0, {\pi\over 4},
{\pi\over2}, {3\pi\over4}$ and $\pi$ are shown. The numerical data
(black line) is obtained using a variational method by De Filippis
et al. \cite{filippis:2005} Numerical data obtained through exact
diagonalization of a finite system and by QMC, for the same
parameters but somewhat different $k$ values, is also presented by
Hohenadler et al. in Refs.
\onlinecite{hohenadler:2003,hohenadler:2005}. These sets of
numerical data are in good agreement with one another.

In all three cases the agreement between MA results and the
numerical data is very good. As expected, it is best for the largest
$\lambda$, but even for the smaller $\lambda$ values, which are just
below and within the cross-over region, the agreement is very
satisfactory. For $\lambda = 0.5$ and $k=0$ (upper panel of Fig.
\ref{Cat05}) we see the polaron state as a Lorentzian peak (a
broadening $\eta=0.1\Omega$ was used) which accounts for most of the
weight, and a small continuum at a higher energy. MA overestimates
the gap between the two, which should be $\Omega$. As $k$ increases,
the polaron peak disperses but also looses significant weight, as
discussed in the previous section. Most of the weight is now in the
high energy continuum, located roughly near the corresponding
$\epsilon_k$ value. This simply shows that these higher energy
states are not significantly affected by this rather weak coupling.
The VM data shows somewhat more structure in these continua than the
MA data, but most of the weight occupies similar frequency ranges.

\begin{figure}[t]
\includegraphics[width=0.85\columnwidth]{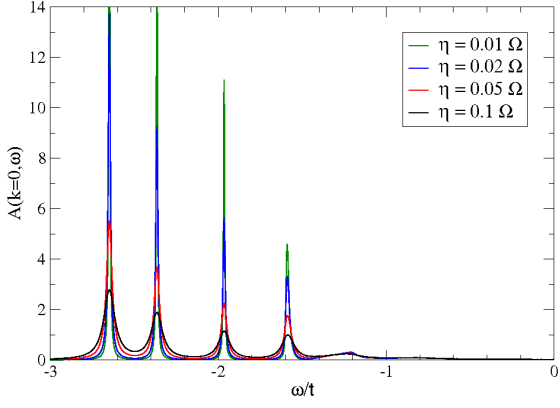}
\caption{(color online) MA 1D spectral weight $A(k=0,\omega)$ vs.
  $\omega$, for $\Omega=0.4t$, $\lambda=1$, and $\eta/\Omega=0.1,
  0.05, 0.02, 0.01$. The first three peaks are discrete states
  (lorentzians) whereas the fourth marks the band-edge singularity of
  the continuum.  }
\label{etacomp}
\end{figure}

For $\lambda=1$ and $k=0$ (upper panel of Fig. \ref{Cat1}), the MA
data shows 3 Lorentzian peaks plus a continuum starting at
$\omega/t=-1.6$. For the rather large $\eta$ used it is hard to
distinguish which peaks come from individual poles, and which are
true continua. This can be easily done by studying their behavior as
a function of the broadening $\eta$, as shown in Fig. \ref{etacomp}.
The height of peaks corresponding to discrete states scales
precisely like $1/\eta$, as expected for lorenzians. The continuum
is affected very little by changes in $\eta$, except the peak near
its lower edge where the finite $\eta$ smoothes out a singularity in
the DOS. Since this singularity is not of the $1/\omega$ type, its
scaling with $\eta$ is different from that of the Lorentzians. The
two lower states are closer to one another than $\Omega$, however
the MA data shows no sign of the continuum that is expected to start
at $E_0 + \Omega$. Note that the numerical data in Fig.
\ref{Cat1}(a) shows more structure, that could be consistent with
this continuum. We will address the issue of this continuum below.
As $k$ is increased (see Fig. \ref{Cat1}) the low-energy peaks show
some dispersion, but with a strongly renormalized bandwidth. At
higher $k$ most weight shifts again at high energies, in a rather
broad continuum. Finally, for $\lambda=2$ and $k=0$, Fig. \ref{Cat2}
shows even more discrete peaks spaced by $\Omega$. The GS is at
$E_0\approx -4.25t$, but its weight is so small that it cannot be
seen on this scale, unless $\eta$ is decreased significantly. A
continuum is seen above $\omega= -1.6t$. As $k$ is now increased,
there is almost no dispersion of the discrete peaks, however the
weight shifts again to an even broader high-energy continuum.

\begin{figure}[t]
\includegraphics[width=0.85\columnwidth]{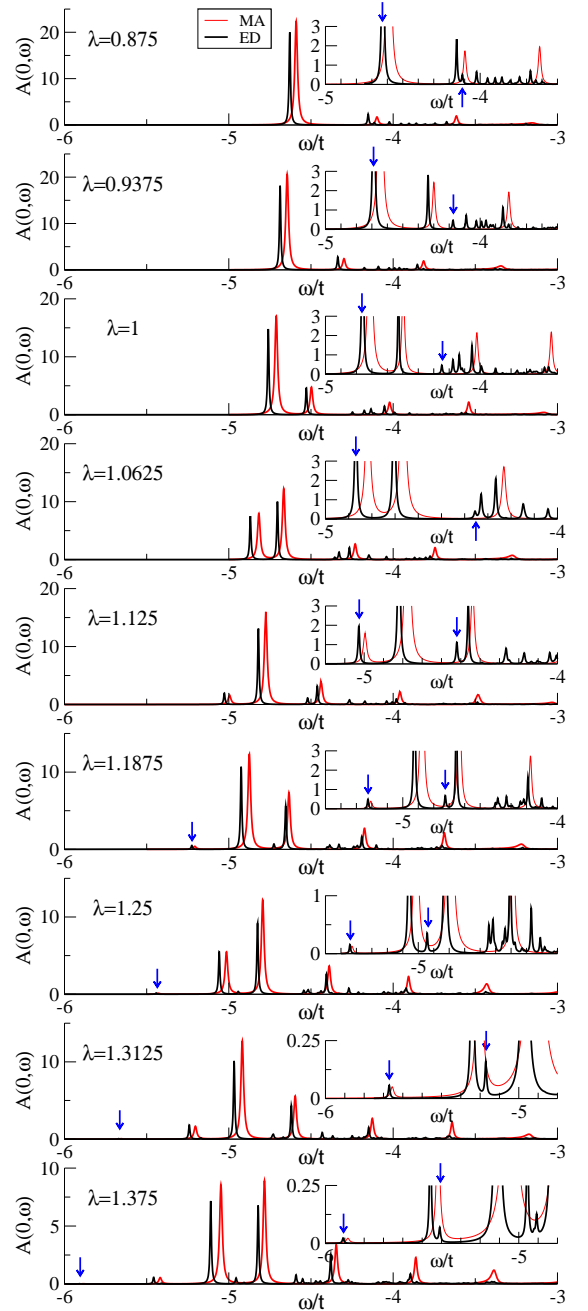}
\caption{(color online) 2D spectral weight $A(k=0,\omega)$ for
  $\Omega=0.5t$, $\eta=0.01t$ and various $\lambda$ values, from exact
  diagonalization\cite{bayo} (black) and MA (red). For $\lambda >
  1.125$, the  arrows point the GS location. The insets show the
  same data on a smaller scale, so that low-weight states are more
  visible. The arrows show the GS and the state appearing at precisely
  $\Omega$ above GS, in the ED results.  }
\label{bayo}
\end{figure}

The issue of the continuum at $E_0+\Omega$ in the exact case, and of
its absence in the MA approximation for moderate and large couplings
can be understood from Fig. \ref{bayo}. Here we show a comparison of
$A(k=0,\omega)$ in 2D obtained from exact
diagonalization,\cite{bayo} vs. MA results, for various couplings
$\lambda$. The first remark is that MA captures quite well all the
large-weight features, both as far as their energy and their weight
are concerned. This is expected, given the good sum rules agreement
demonstrated previously. However, the ED results clearly show more
states than MA predicts. There is always a low-weight peak at
precisely $E_0 +\Omega$ (the GS and this state are marked by arrows
in the insets). For couplings of up to $\lambda\approx 1$, this peak
is followed by several nearby peaks with comparably low-weight,
which can be argued to be part of the expected continuum. For larger
$\lambda$, however, only the state at $E_0 +\Omega$ can still be
seen, although more states suggesting more continua are seen between
the large higher-energy  peaks. The gradual disappearance of the
first continuum is not surprising, since one expects its width to
narrow exponentially as the coupling increases. Moreover, one
expects that the largest contribution to this continuum is from
states with one or more phonons, explaining their low $qp$ weight.

As far as MA is concerned, Fig. \ref{bayo} suggests that for
couplings where there is more than one discrete state, the very
little weight in the $E_0 +\Omega$ and similar higher-energy
continua is absorbed in the discrete states predicted by MA. This is
consistent with the systematic up-shift of the MA peaks compared to
the ED data.

In fact, it is straightfoward to see that the MA approximation can
only predict a continuum starting at $-2dt+\Omega$. A continuum is
signaled by a finite imaginary part of
$\Sigma_{\textrm{MA}}(\omega)$, and the lowest frequency where this
can occur is that for which $\bar{g}_0(\omega-\Omega)$ acquires a finite
imaginary part [see Eq. (\ref{eq:sigma_MA})]. However, the imaginary
part of $\bar{g}_0(\omega)$ is proportional to the total density of
states of the free model, i.e. it is finite for $\omega \in [-2dt,
2dt]$ for nearest-neighbor hopping. It follows that the MA continuum
always starts precisely at $-2dt +\Omega$. This explains why for
small $\lambda$, where there is only one peak below this continuum,
the gap between the two is somewhat larger that the expected
$\Omega$ value: the GS energy decreases below $-2dt$ with increasing
$\lambda$, whereas the continuum edge is pinned at $-2dt + \Omega$,
in the MA approximation. As the coupling increases, bound states
start to split from this continuum, and spectral weight is shifted
to lower energies, in good agreement with the sum-rule predictions
of the exact solution. These new bound states have to account for
the (small) weight that is present in lower energy continua, in the
exact solution, and this is precisely what Fig. \ref{bayo} shows.
Clearly, a self-energy that would account for these continua as well
would have to be a lot more complicated than that of Eq.
(\ref{eq:sigma_MA}).

\begin{figure}[t]
\includegraphics[width=0.8\columnwidth]{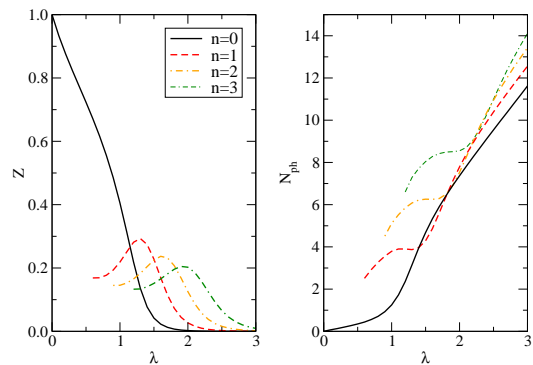}
\caption{(color online) $qp$ weight and average number of phonons in
  the GS (black line) and the next three higher $k=0$ bound states,
  when they become stable according to MA. Results are for $1D$,
  $\Omega=0.5t$.  }
\label{peaks}
\end{figure}

This shows again that MA is remarkably successful in predicting the
main features of the Green's function, given its simplicity and
trivial numerical cost. This should make it (and generalizations of
it to other models) of large interest for comparison against
experiments.

In the following, we use MA to investigate more properties of the
Green's function. To our knowledge, there are few or no similar
results with similar accuracy published in the literature.  We begin
with Fig. \ref{peaks}, where we plot the $qp$ weight and average
phonon numbers for a few of the higher-energy peaks, once they
appear below the continuum. For comparison, we also show the already
discussed GS results (black line). Results in higher dimension are
qualitatively similar and we do not show them here.  Unlike for
the GS, both these quantities are non-monotonic functions of
$\lambda$ for all higher-energy bound states. Each of these states
disperses with $k$, like in Figs. \ref{Cat1} and \ref{Cat2} (more
data for this is shown below), so an effective mass can be
associated with each such band. This effective mass satisfies
$m^*/m=1/Z$, and therefore also shows non-monotonic behavior, first
decreasing and then increasing as $\lambda$ is increased. The
average phonon number in the $n$th state must approach $g^2/\Omega^2
+ n$ asymptotically, as can be verified in the Lang-Firsov limit.
This is indeed observed in Fig. \ref{peaks}, however the plateaus
seen at moderate $\lambda$ suggest some cross-over from one to
another type of wavefunction associated with these higher levels. As
$\lambda \rightarrow \infty$, an infinite sequence of such bound
states appear, as expected in the Lang-Firsov limit.

\begin{figure}[t]
\includegraphics[width=\columnwidth]{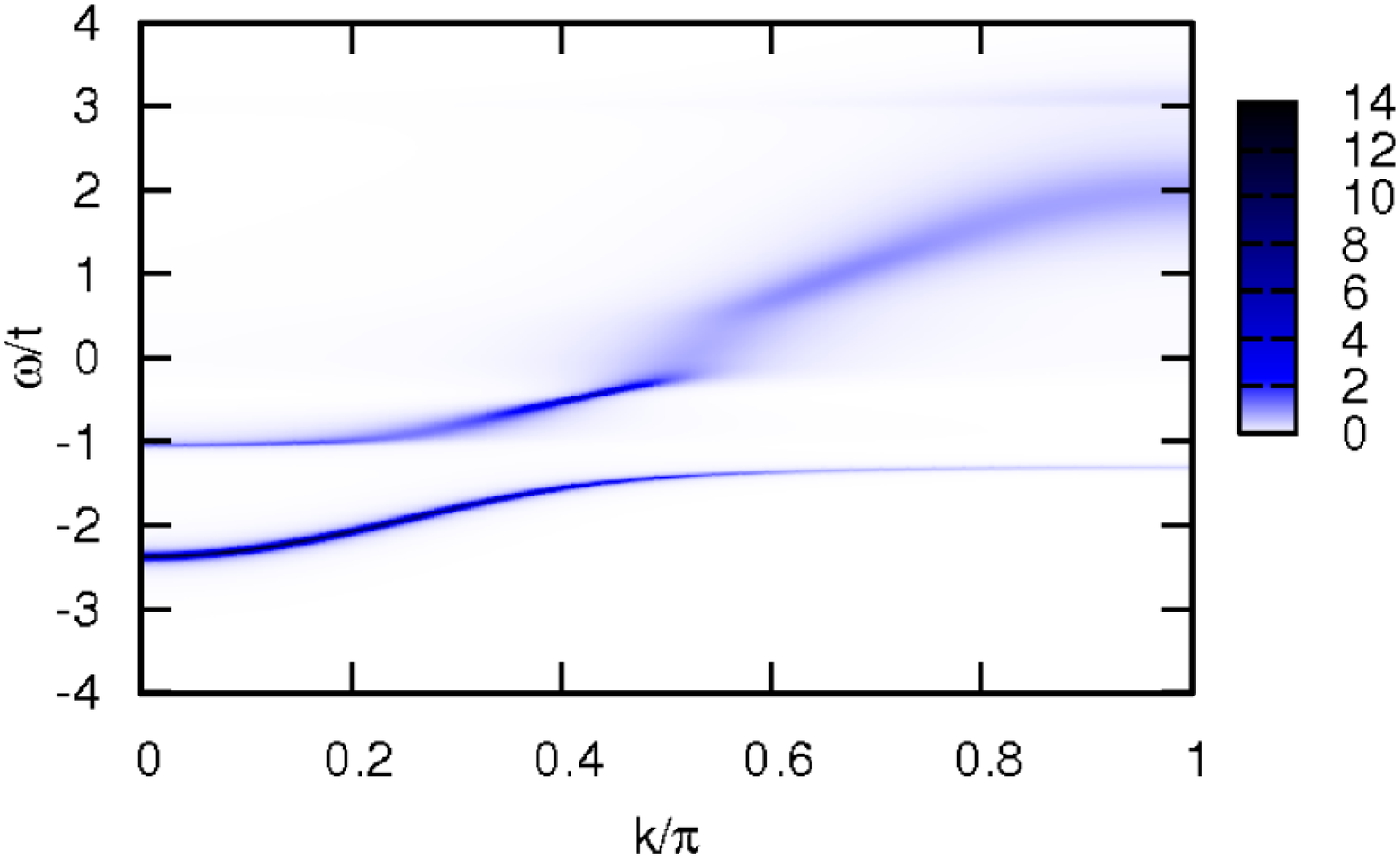}
\includegraphics[width=0.97\columnwidth]{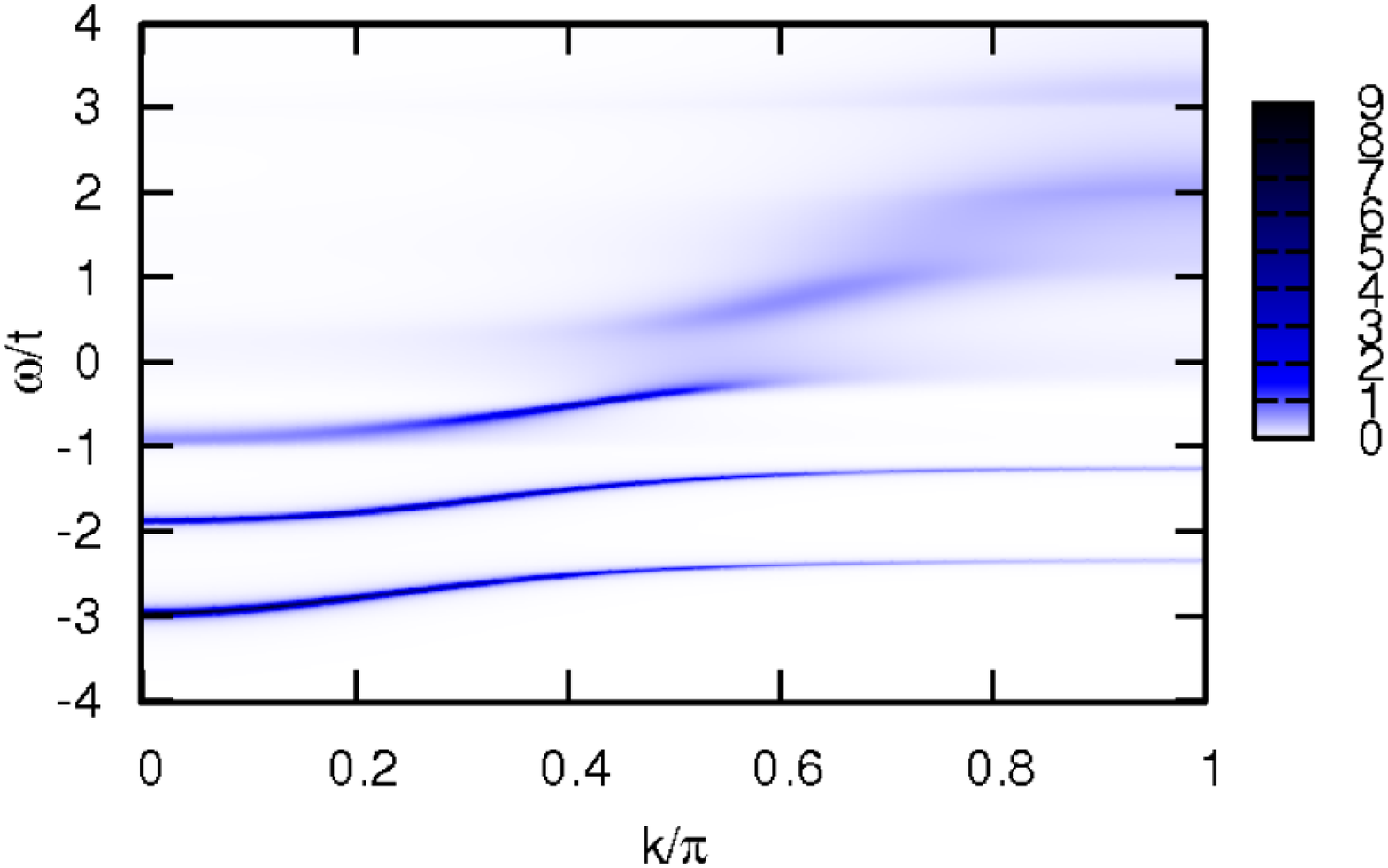}
\includegraphics[width=\columnwidth]{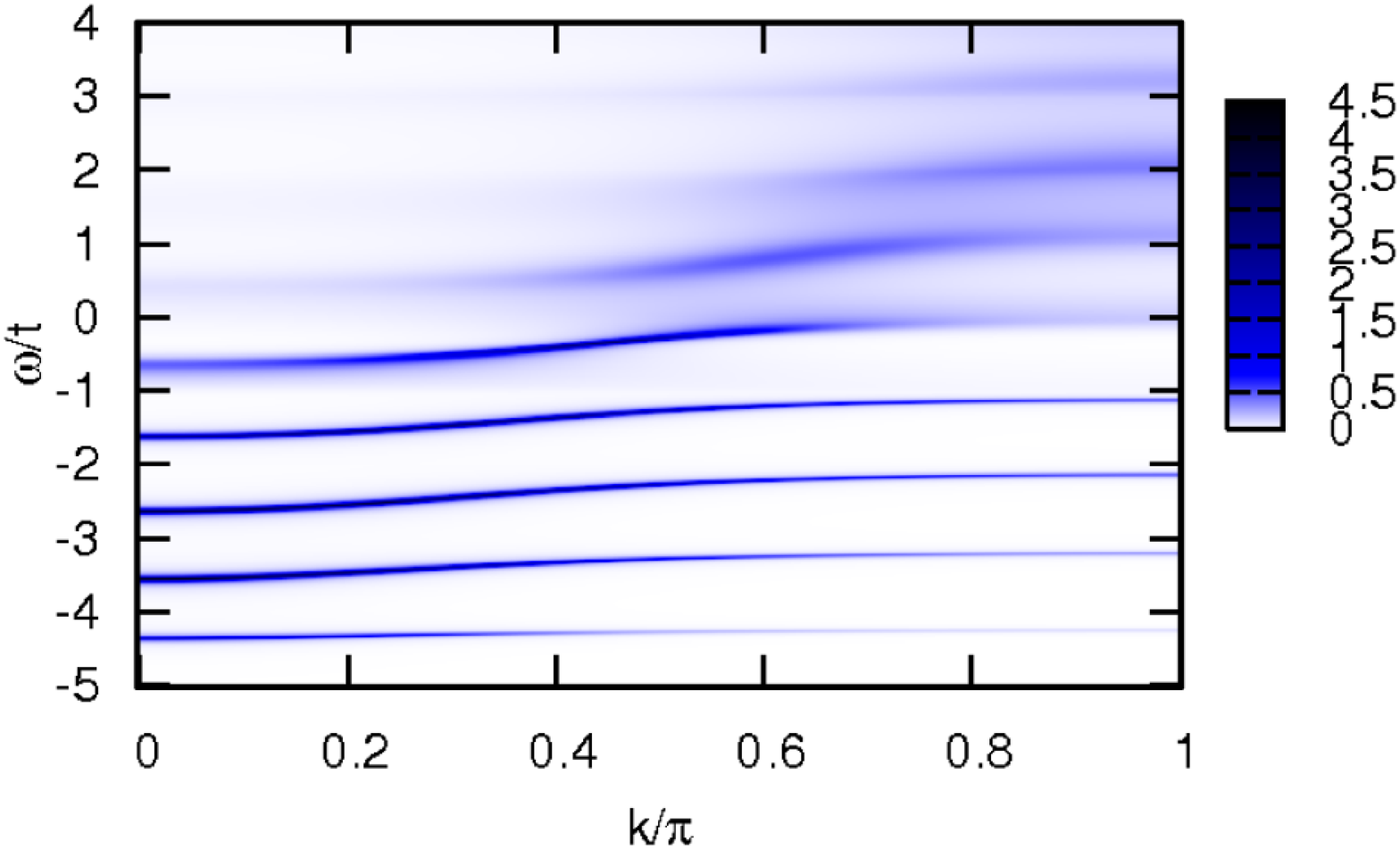}
\caption{(color online) Contour plots of the 1D spectral weight
  $A(k,\omega)$ as a function of $k$ and $\omega$. The intensity
  scales are shown to the right of each plot. Parameters are
  $\Omega=t$, $\lambda=0.4, 1$ and $2$, respectively, and $\eta =
  0.02$. These results are in excellent agreement with the results of
  Hohenadler \emph{et al.}, Ref. \onlinecite{hohenadler:2003}, and the
  GS results of Bonca \emph{et al.}, Ref. \onlinecite{bonca:1999}.  }
\label{1Dkcont}
\end{figure}

For a better illustration of the appearance of these bound states
and of their evolution, we show contour plots of the spectral weight
$A(\mb{k}, \omega)$. We begin by plotting $A(\mb{k}, \omega)$ as a
function of $\mb{k}$ and $\omega$, for fixed parameters $g,t$ and
$\Omega$.  In Fig. \ref{1Dkcont} we show 1D results corresponding to
$\Omega=t$ and $\lambda=0.4, 1$ and 2, respectively. Only half of
the BZ is shown, since time-invariance guarantees that
$G(\mb{k},\omega)=G(-\mb{k},\omega)$. Each of these MA contour plots
takes below 
ten seconds to generate. Note that similar plots for
the same parameters were provided by Hohenadler \emph{et al.} in
Ref. \onlinecite{hohenadler:2003}, based on a cluster perturbation
theory approach. The agreement between the main features of our and
their plots is excellent. As expected, their data does show a few
more low-weight features at lower energies, below $-2t+\Omega=-t$,
in this case, where our continuum starts. Such contour plots are
richer versions of plots like those shown in Figs. \ref{Cat05},
\ref{Cat1} and \ref{Cat2}. They illustrate basically the same
points, although one can now also see clearly the dispersion of
various features. For the lowest $\lambda$, the free electron
dispersion $\epsilon_k = -2t \cos(ka)$ is still almost visible,
except that electron-phonon interactions split it into the lower
polaron band, and the higher continuum. This continuum is not
featureless, instead one can already see weight accumulating near
its lower edge. As $\lambda$ increases, a new bound state will split
off from this. Indeed, this is seen for $\lambda=1$, where there are
2 bound state below the continuum starting at $-1$ (for these
parameters). The bandwidth of each of these states is now narrowed
below $\Omega$. The weight in the continuum at higher energies is
redistributed suggesting the impending formation of yet more bound
states. Indeed, the $\lambda=2$ data shows 4, even narrower bound
states below the continuum, which is showing signs of further
fragmentation at multiples of the phonon frequency.

\begin{figure}[t]
\includegraphics[width=\columnwidth]{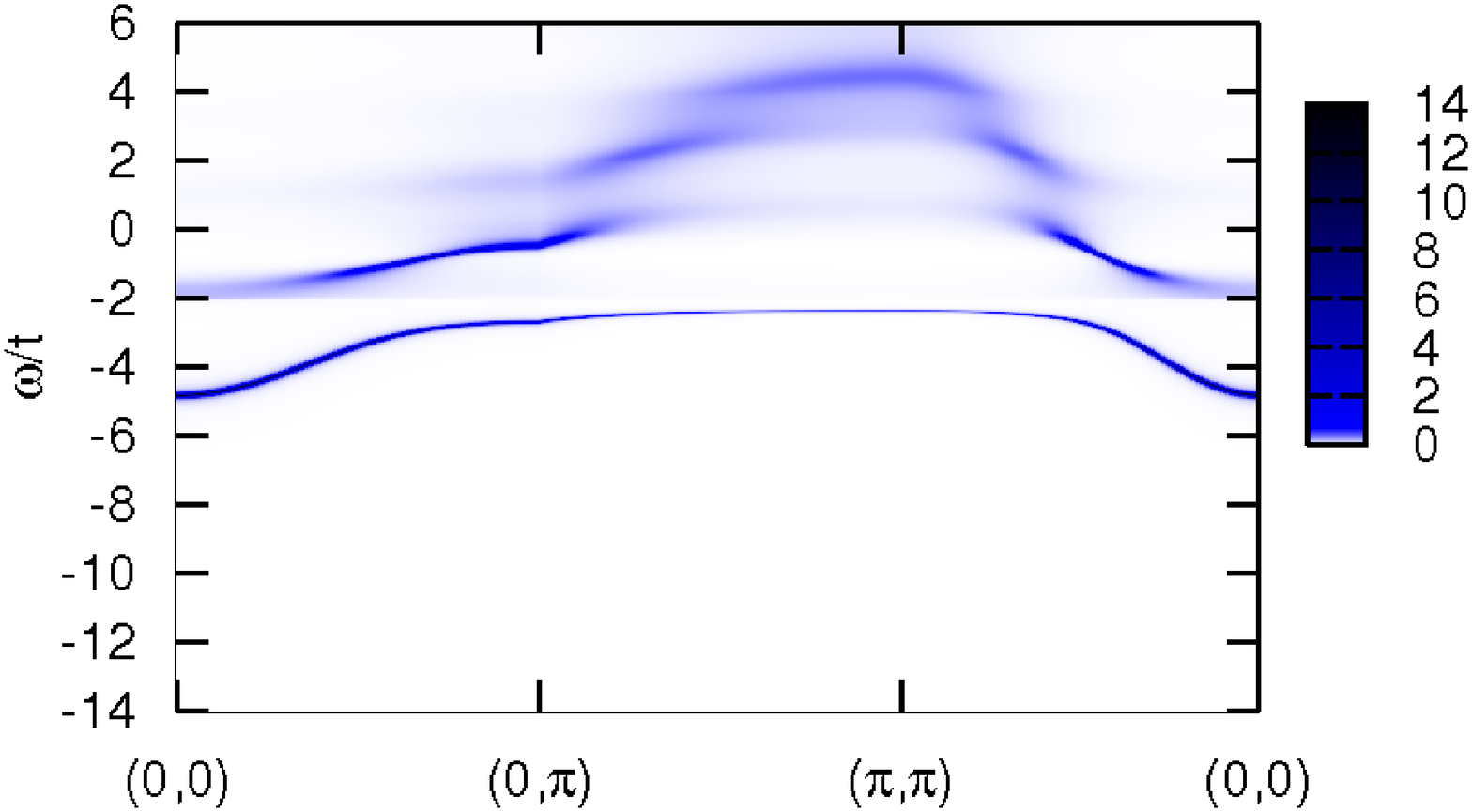}
\includegraphics[width=\columnwidth]{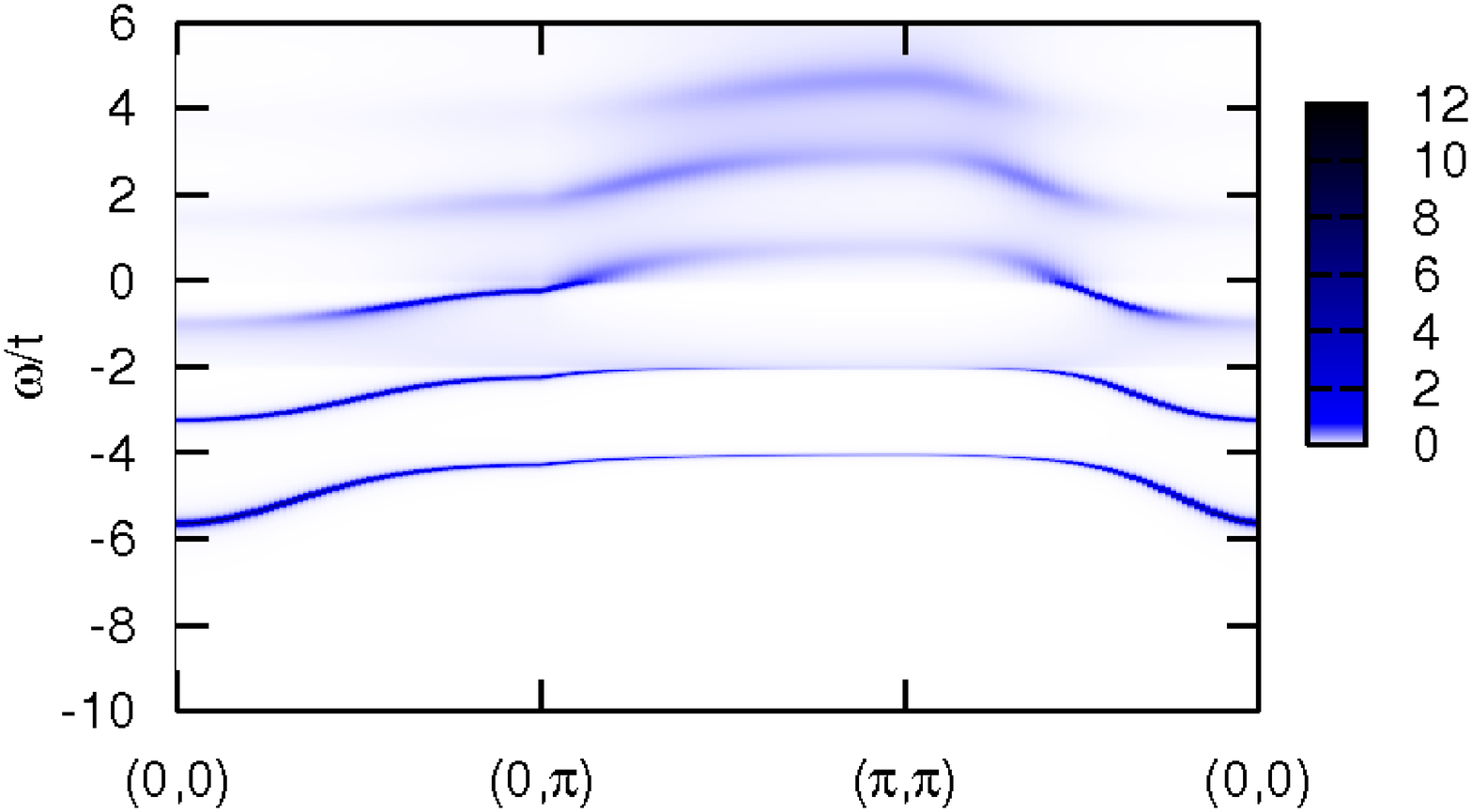}
\includegraphics[width=\columnwidth]{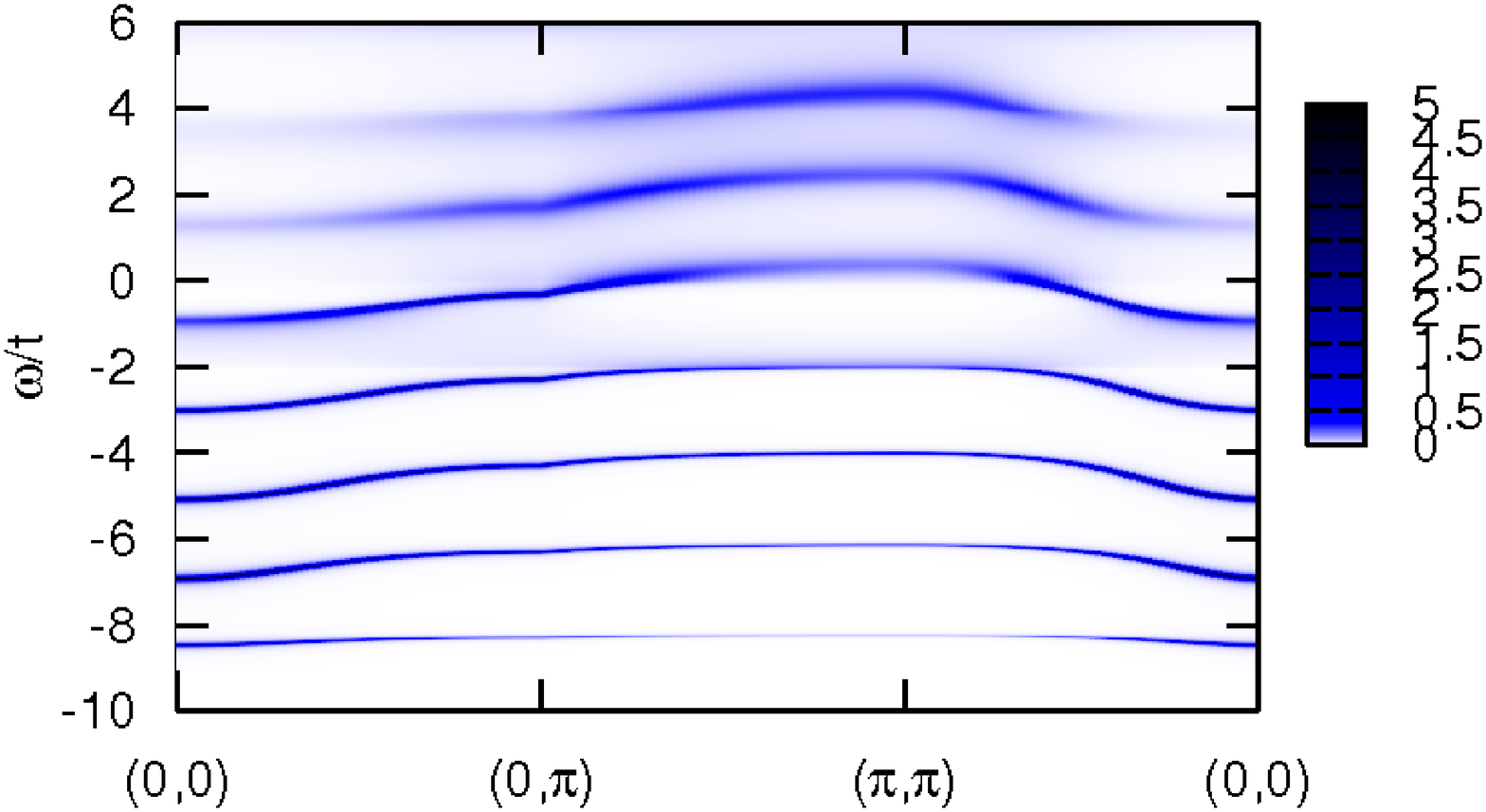}
\caption{(color online) Contour plots of the 2D spectral weight
  $A(\mb{k},\omega)$ as a function of $\mb{k}$ and $\omega$, along
  several cuts in the BZ. The intensity scales are shown to the right
  of each plot. Parameters are $\Omega=2t$, $\lambda=0.5, 0.945$ and
  $2$, respectively, and $\eta = 0.02$. The middle panel is in
  excellent agreement with the results of Hohenadler \emph{et al.}
  given in Ref. \onlinecite{hohenadler:2003}.  }
\label{2Dkcont}
\end{figure}

Similar behavior is expected, and indeed seen, in higher dimensions.
Here we only show similar 2D contour plots, for $\Omega=2t$ and
$\lambda=0.5, 0.945$ and $2$, in Fig. \ref{2Dkcont}. The middle
panel again agrees very well with data shown in Ref.
\onlinecite{hohenadler:2003}. In this case, the MA continuum starts
at $-4t+\Omega = -2$. As $\lambda$ increases, we see again first 1,
then 2 and then 4 bound states below the continuum. Their bandwidths
decrease with increasing $\lambda$, so that the lowest band for
$\lambda=2$ is already almost dispersionless, even though its weight
still varies with $\mb{k}$. As in the 1D case, as the interaction
becomes stronger, the weight in the continuum also redistributes
itself, with strong resonances seen around multiples of the phonon
frequency.

\begin{figure}[t]
\includegraphics[width=0.8\columnwidth]{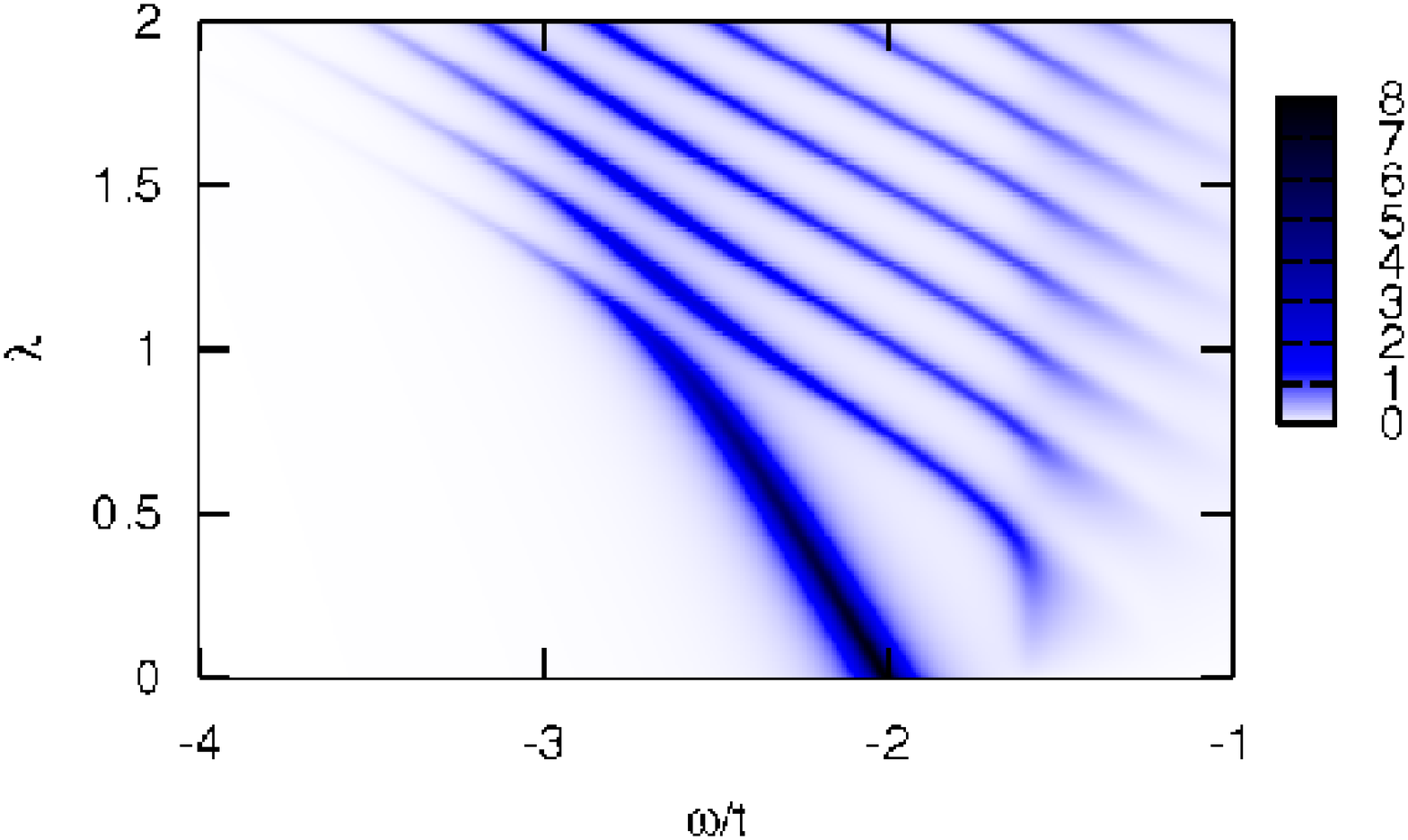}
\includegraphics[width=0.8\columnwidth]{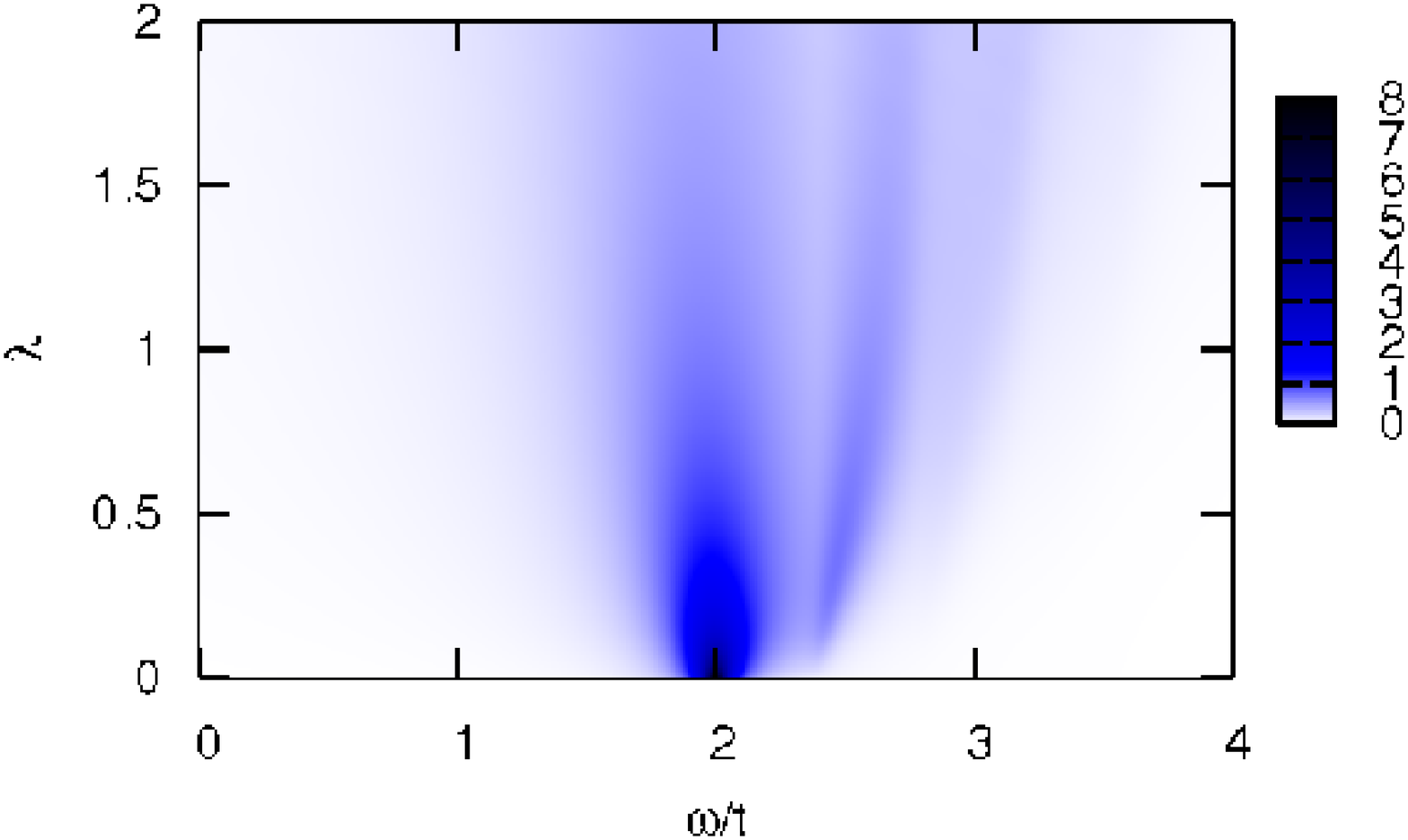}
\includegraphics[width=0.8\columnwidth]{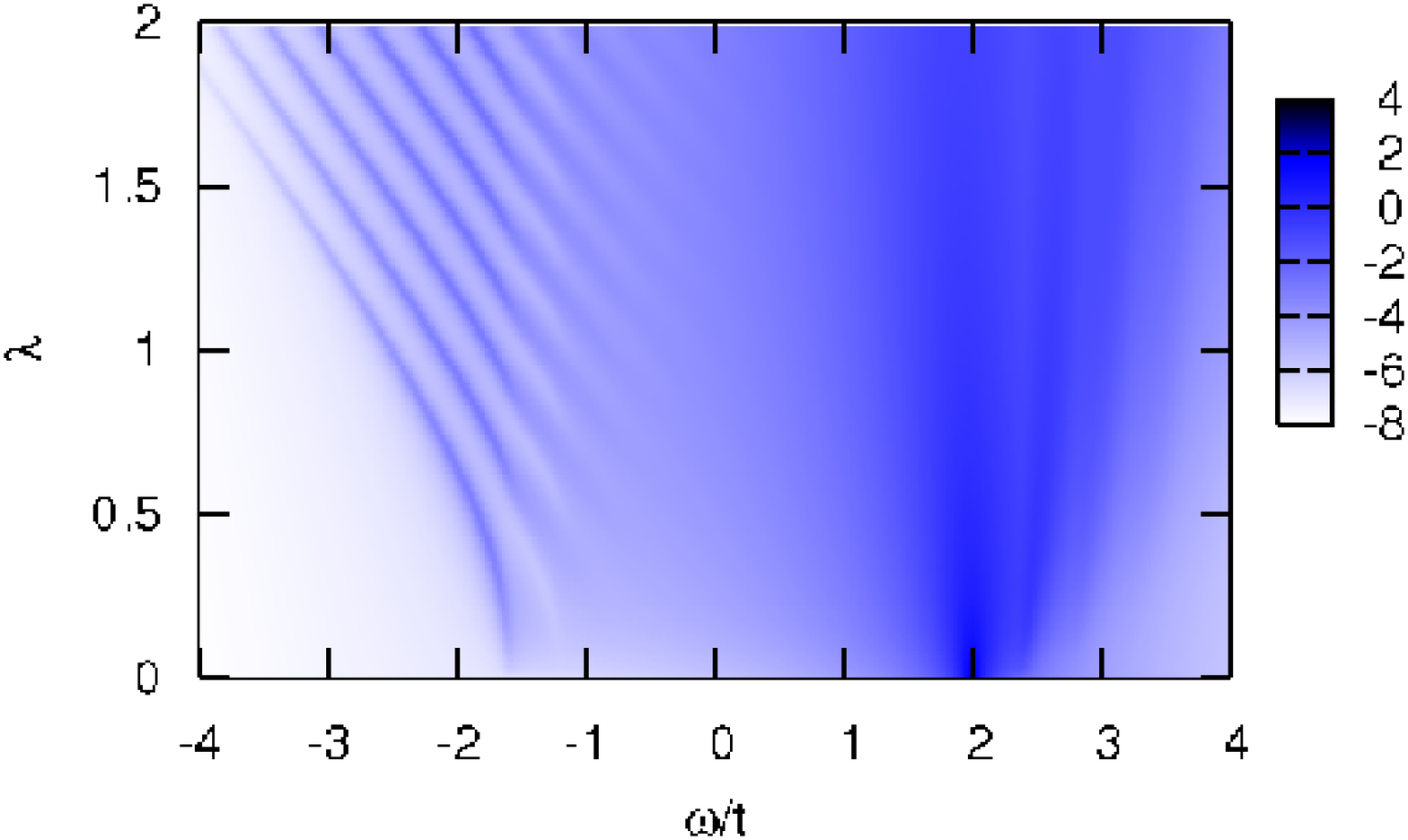}
\includegraphics[width=0.85\columnwidth]{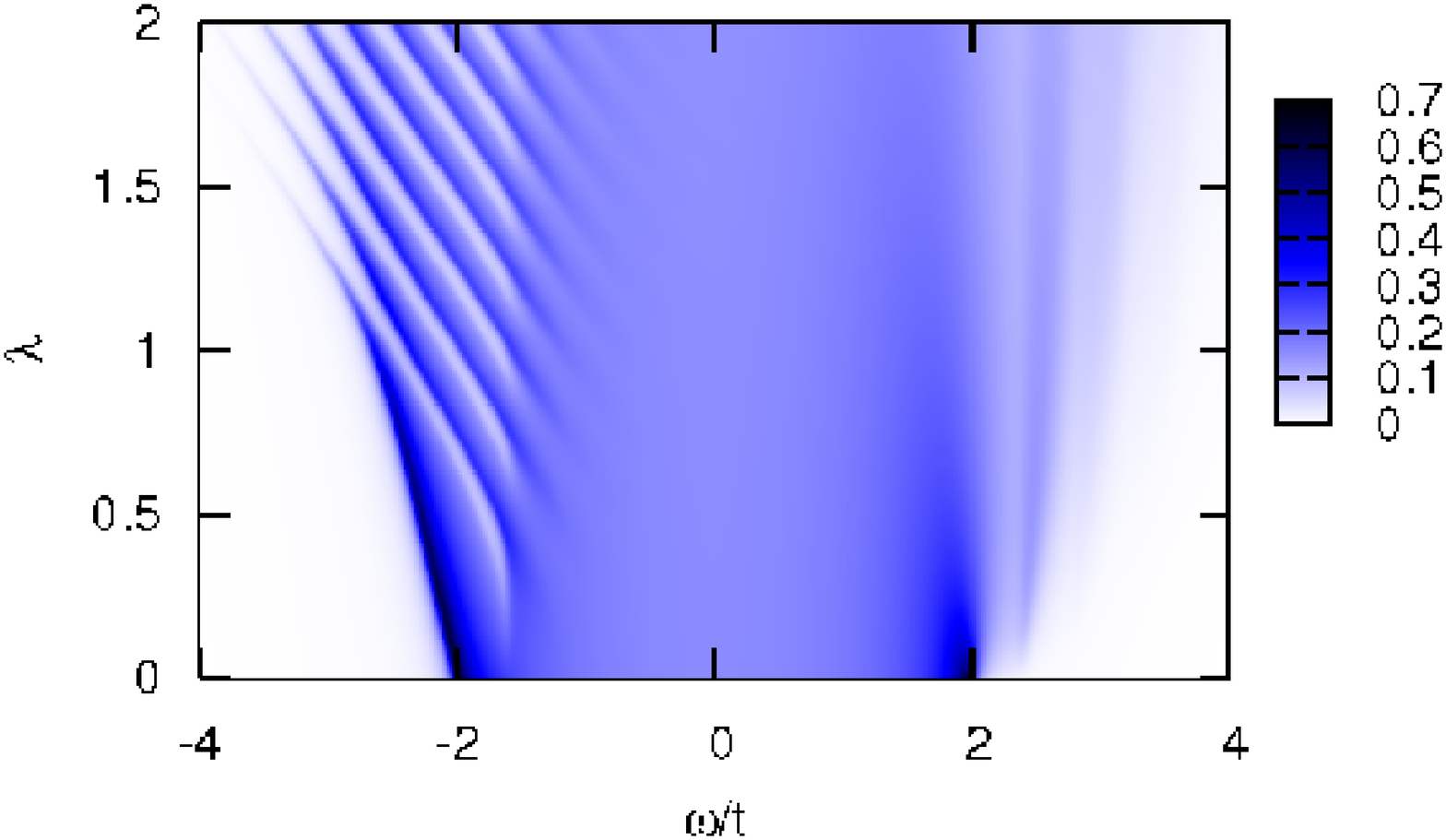}
\caption{(color online) 1D results for $\Omega=0.5t$ and $\eta =
  0.04t$. (a) $A(k=0, \omega)$ vs. $\omega$ and $\lambda$; (b)
  $A(k=\pi, \omega)$ vs. $\omega$ and $\lambda$, on a linear scale; (c)
  $A(k=\pi, \omega)$ vs. $\omega$ and $\lambda$, on a logarithmic
  scale; (d) total spectral weight $A(\omega)$ vs. $\omega$ and
  $\lambda$. The intensity scales are shown to the right of each plot.
} \label{1Dc}
\end{figure}

Another way to understand the dependence on the coupling $\lambda$
(or any other parameter) is to plot a contour of the spectral weight
$A(\mb{k}, \omega)$ vs. $\lambda$ and $\omega$ for a fixed value of
$\mb{k}$. Such a task is equally trivial at the MA level. In fact,
one can also just as easily calculate and plot the total density of
states, or spectral weight:
$$ A(\omega) = {1\over N}\sum_{\mb{k}}^{}A(\mb{k},\omega),
$$ since within the MA approximation this is given by:
\begin{eqnarray}
\nonumber A(\omega) &=& -\frac{1}{\pi}\textrm{Im} \, \left[
\frac{1}{N} \sum_{\mb{k}}
\frac{1}{\omega-\varepsilon_{\mb{k}}-\Sigma_{\textrm{MA}}
(\omega)+i\eta} \right] \\ \nonumber &=& -\frac{1}{\pi} \textrm{Im}
\, \left\{ \bar{g}_0 \left[ \omega - \Sigma_{\textrm{MA}}(\omega)
\right] \right\}.
\end{eqnarray}

In Fig. \ref{1Dc} we show 4 such contour plots for the 1D case. The
uppermost one shows $A(k=0, \omega)$ vs. $\omega$. For $\lambda=0$
(noninteracting case), only one state exists at $-2t$, as expected.
As the coupling turns on, the energy of this state (the
ground-state) decreases, but $k=0$ weight is also transferred to
higher energies, due to hybridization with the states in the
electron-plus-one-phonon continuum. The MA continuum here starts at
$-1.5t$. For moderate and larger $\lambda$ one can clearly see how
weight is re-arranged inside the continuum as $\lambda$ increases,
and new bound states split from it and move towards lower energies.
The apparent ``break'' in the slope of the GS energy, as $\lambda$
increases, is now seen to occur when the first bound state
approaches the GS, and is suggestive of an avoided crossing. From
this point on the GS lowers its energy much faster, but its weight
also decreases dramatically and it becomes difficult to see. The
second panel of Fig. \ref{1Dc} shows $A(k=\pi, \omega)$ vs.
$\omega$, on a linear scale. At $\lambda=0$, there is only one peak
at $+2t$, as expected. As the coupling is turned on, this weight
seems to spread around in a rather featureless, broad continuum. In
fact, on a logarithmic scale (panel (c) of Fig. \ref{1Dc}) one can
see that $k=\pi$ weight is pulled down into all the bound states, in
agreement with the previous data we showed. This weight, however, is
so small that it is not visible on the linear scale. Finally, the
lowest panel of Fig. \ref{1Dc} shows the total spectral weight, or
DOS. At $\lambda=0$ we see the usual 1D DOS, with the singularities
near the band-edges rounded off because of the finite $\eta$ used.
As $\lambda$ is turned on, one can recognize both the contribution
from the $k=0$ and $k=\pi$ states to the total DOS: each bound state
has a finite bandwidth due to its dispersion (this is to be
contrasted to the upper panels, where the bound states are true
delta-functions, with a width defined by the broadening $\eta$).  As
the coupling strength increases, the number of bound states
increases; they are spaced by roughly $\Omega$, their bandwidths
narrow down, and their weights also decrease. In other words, they
approach the expected Lang-Firsov behavior.

Thus, this figure actually answers the question posed in the
Introduction, regarding the evolution of the spectral weight from
that of a free electron towards that of the impurity limit. While
the MA results are certainly not exact, we can claim with a high
degree of certainty that the main features are accurately captured,
especially in the weak and in the strong coupling limits. This
suffices to understand the physics of this problem, and given the
simplicity of the approximation, to investigate in detail any number
of other quantities we have not shown here, such as the self-energy.
Of course, if one is interested in exact results for some particular
set of parameters, numerically-intensive methods have to be used.

\begin{figure}[t]
\includegraphics[width=0.88\columnwidth]{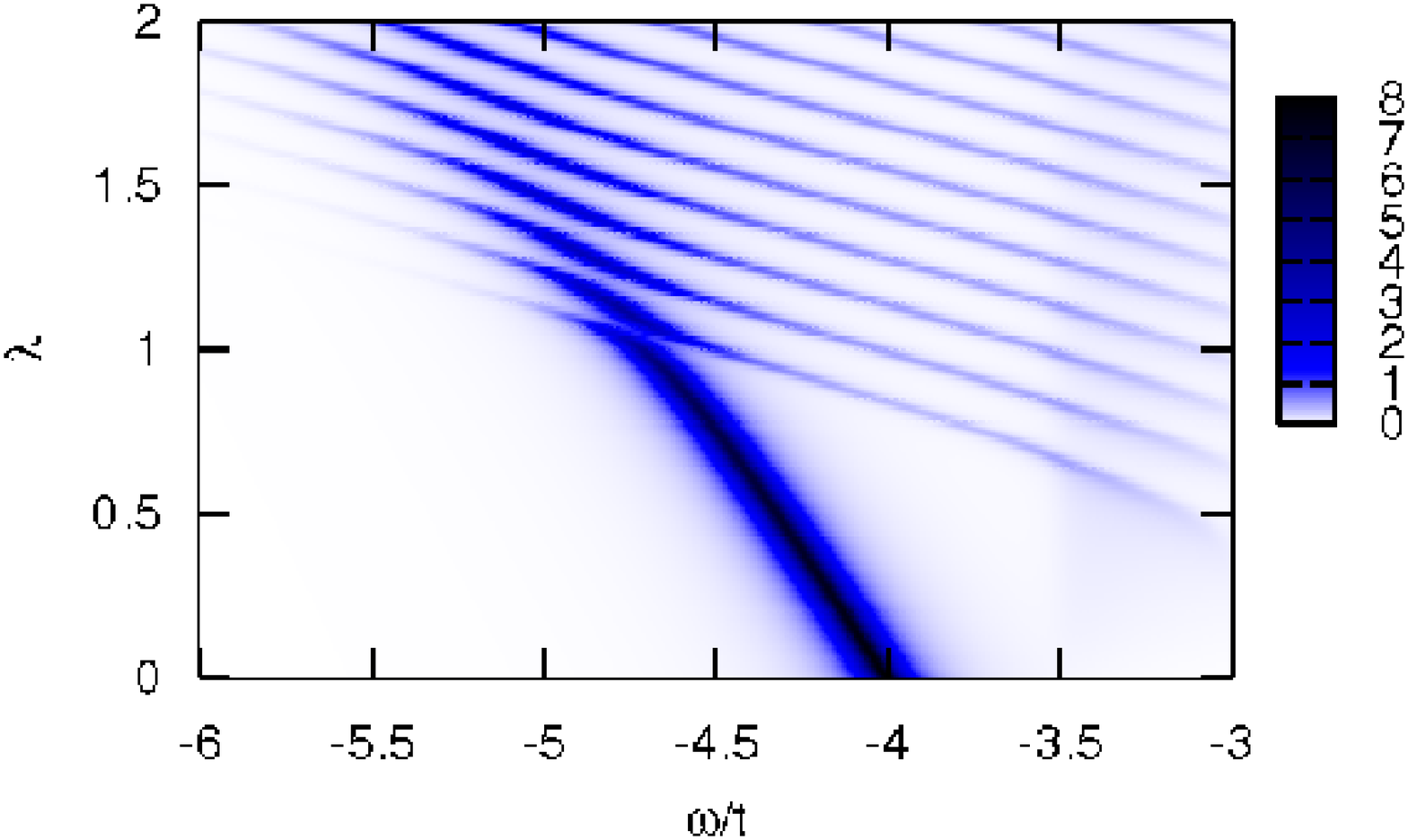}
\includegraphics[width=0.9\columnwidth]{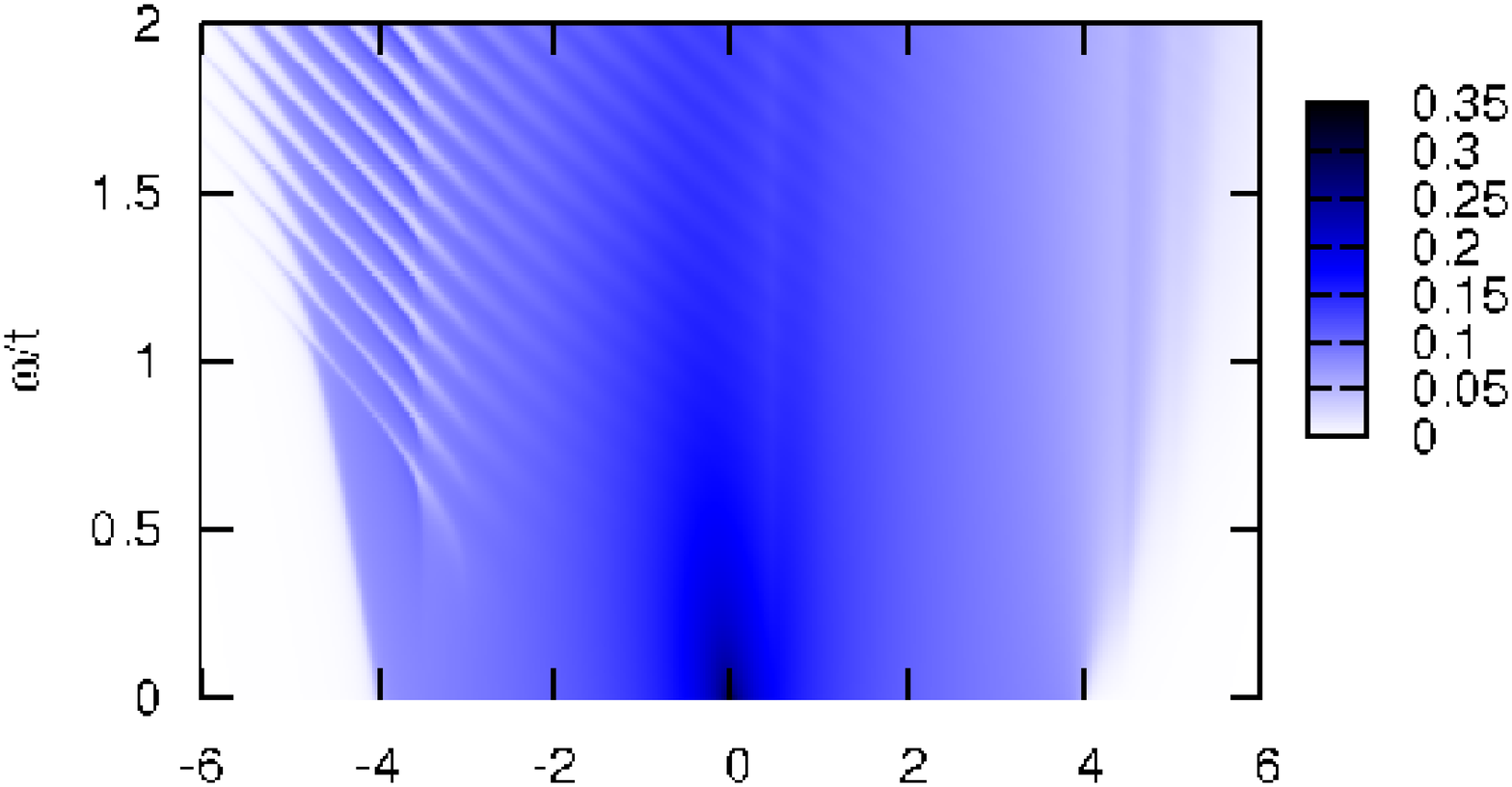}
\caption{(color online) 2D results for $\Omega=0.5t$ and $\eta =
  0.04t$. Top: $A({\mb k}=(0,0), \omega)$ vs. $\omega$ and $\lambda$.
  Bottom: total 2D spectral weight $A(\omega)$ vs. $\omega$ and
  $\lambda$. The intensity scales are shown to the right of each plot.
} \label{2Dc}
\end{figure}

\begin{figure}[t]
\includegraphics[width=0.88\columnwidth]{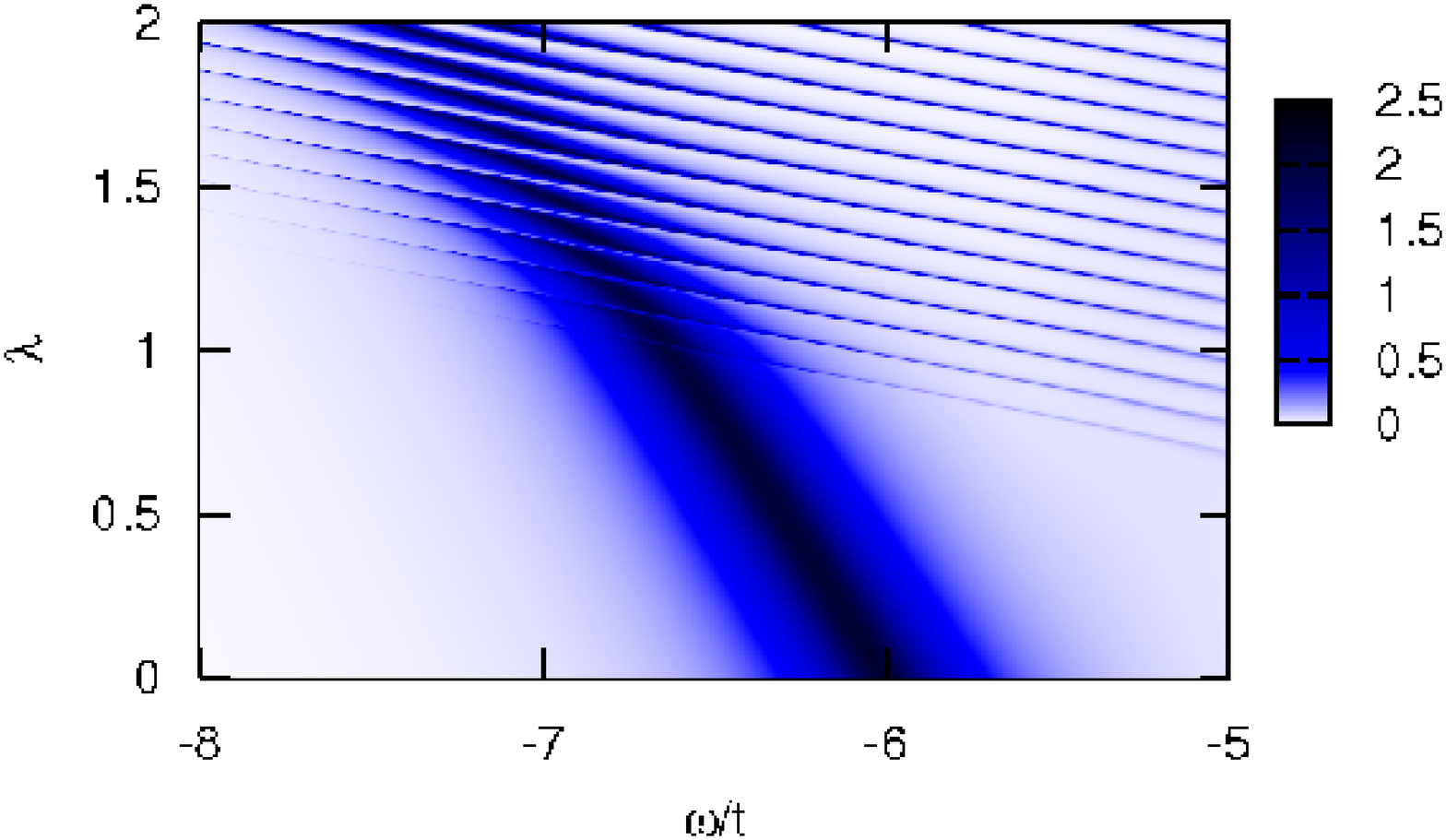}
\includegraphics[width=0.9\columnwidth]{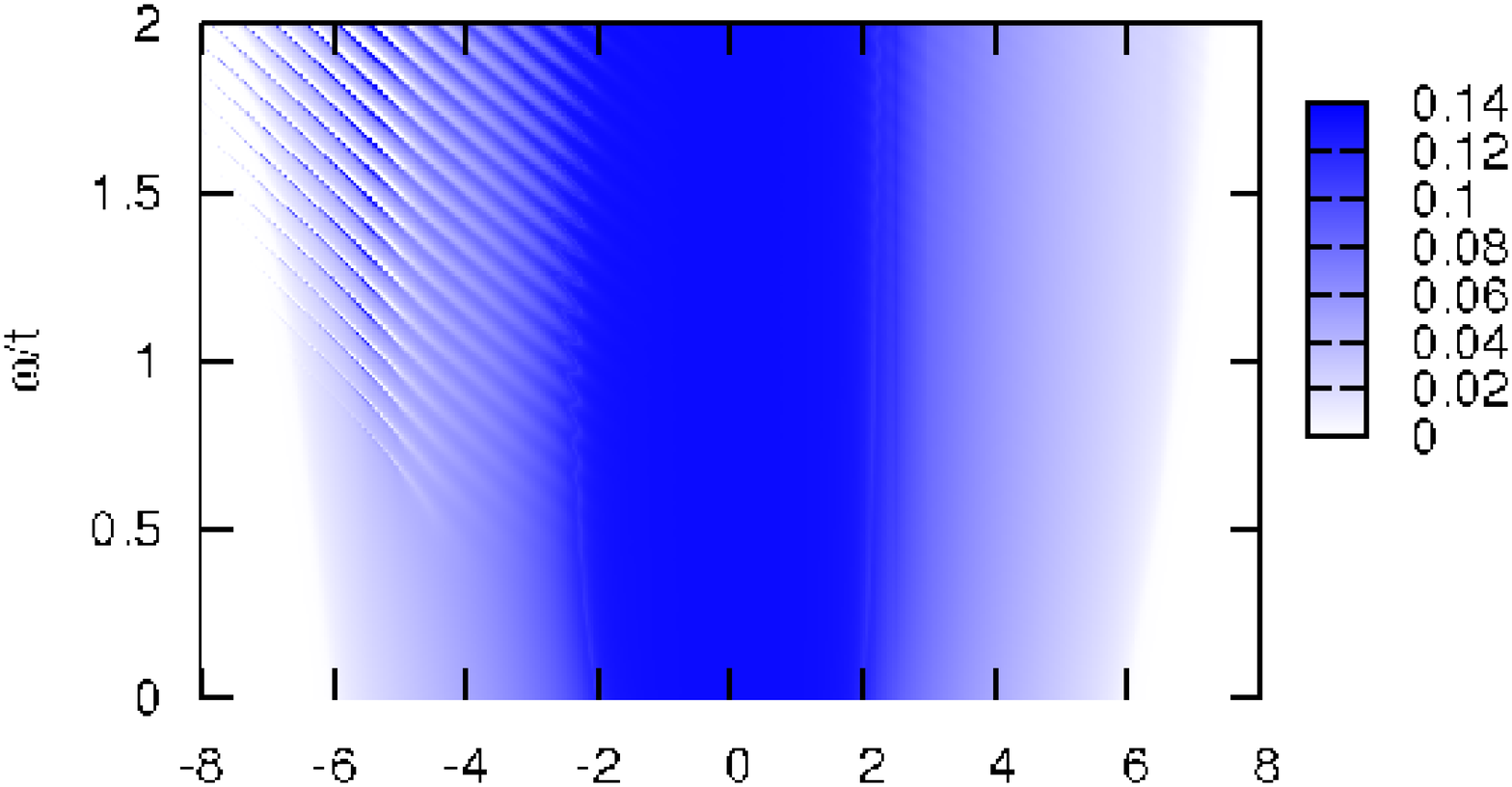}
\caption{(color online) 3D results for $\Omega=0.5t$. Top: $A({\mb
	k}=(0,0,0), \omega)$ vs. $\omega$ and $\lambda$, for $\eta =
  0.15t$.
  Bottom: total 3D spectral weight $A(\omega)$ vs. $\omega$ and
  $\lambda$, for $\eta =
  0.04t$. The intensity scales are shown to the right of each plot.
} \label{3Dc}
\end{figure}

Qualitatively similar plots are obtained in 2D and 3D, as shown in
Figs. \ref{2Dc} and \ref{3Dc}. Of course, the $\lambda=0$ DOS is
very different, with a van-Hove singularity at $\omega=0$ for the 2D
case, and the characteristic nearest-neighbor hopping DOS in the 3D
case. However, the appearance of multiple bands below the continuum
as $\lambda$ increases and all the remaining phenomenology is very
similar. Thus, we see no evidence of any qualitative differences in
the polaron physics due to different dimensionality.

\section{Summary and conclusions}

In this paper we analyzed the Green's function of the Holstein
polaron, using the {\em momentum average} approximation, which
consists in summing all the self-energy diagrams, but with each
individually averaged over all its free propagator momenta. The
resulting self-energy can be written as an infinite continuous
fraction that is numerically trivial to evaluate. This procedure
becomes exact in the limits $g=0$ and $t=0$.

We gauged the accuracy of this approximation by computing its
corresponding spectral weight sum rules and comparing them against
the exact sum rules, which are known for this type of Hamiltonian.
We showed that the MA spectral weight satisfies exactly the first 6
sum rules. Even though this is quite impressive at first sight, it
is actually no guarantee of overall accuracy, as the case of SCBA
demonstrates. The SCBA spectral weight always satisfies exactly the
first 4 sum rules, even at large couplings $\lambda$ where it
predicts very wrong results! The meaningful criterion of accuracy
for the sum rules is to show that the vast majority of terms in {\em
all sum rules}, and in particular the dominant terms in the various
asymptotic limits, are captured by the approximation. MA indeed
satisfies this very restrictive criterion.

The accuracy of the approximation was also tested by direct
comparison with data obtained through numerically intensive methods.
In all cases we obtain remarkable agreement, especially considering
the ease of evaluation of the MA results. The MA approximation is
not exact and some features are not correctly captured (for example,
the continuum starting at $E_0+\Omega$) however, in all cases, all
the higher-weight features in the spectral weight are quantitatively
and qualitatively well described by the MA approximation. Trading
some of the accuracy of numerically exact but time consuming methods
in exchange for very fast results which capture the main features
accurately is a useful approach when trying to understand the main
aspects of the physics of a problem, as well as when one is
concerned about comparison with experiments. It is very unlikely
that ARPES could capture very low-weight features, these would be
lost in noise statistics. Thus, an approximation like MA,  that
quickly but accurately estimates results is very useful, to be
followed, of course, by detailed numerics for the parameter sets of
interest.

It is important to note that  the existence and accuracy of
approximations like MA is not guaranteed, in fact it can be regarded
as a surprise for the case of the Holstein polaron, that has been
under investigation for almost five decades. However, this
demonstration of its existence and efficiency in the Holstein
polaron problem gives some hope for making progress for the general
class of strongly-correlated systems problems. One can always use
some flavor of perturbation theory to understand behavior in
asymptotic limits, but the really challenging problems are set in
regimes where perturbation does not apply. The MA approximation
suggests that one way to make non-trivial progress is to sum all
diagrams, with each simplified enough so as to make the calculation
feasible, but not so much as to really alter the physics. This is a
very different approach from the usually employed summation of a
subclass of diagrams. Note that there are many classes of problems
with diagrams similar to the ones arising in the single polaron
problem, although of course with different propagators and/or
vertices.

A first possible generalization of this work is to models with
several phonon branches, and/or momentum-dependent coupling $g_{\mb
q}$, and/or dispersive phonons, $\Omega_{\mb q}$. The way to achieve
this for a single polaron is briefly discussed in Ref.
\onlinecite{berciu:2006}. Given the length of this article, we
postpone this discussion for future publications where results can
also be shown. Other directions of generalization are to finite
particle densities and/or finite temperatures, and indeed to
Hamiltonians which also include electron-electron interactions.

Such work is currently in progress. It is still far from clear which
cases will admit useful generalizations, however the proof of
existence of this method for the
 Holstein polaron problem is in itself encouraging.

\vspace{5mm}

{\bf Acknowledgments:} We thank T. Devereaux, F.
Marsiglio, A. Mishchenko and N. Nagaosa for useful discussions, and
A. Macridin, V. 
Cataudella, G. De Filippis and B. Lau for allowing us to use their numerical
results. This work was supported by NSERC and CFI, and by CIAR
Nanoelectronics and the Alfred P. Sloan Foundation (M.B.) and CIAR
Quantum Materials (G.S.).  \appendix

\section{Comparison with DMFT}
\label{dmft}

The MA self-energy of Eq. (\ref{eq:sigma_MA}) looks similar to the
DMFT self-energy, discussed in Ref. \onlinecite{ciuchi:1997}. This
is not so surprising, since both become equal to the exact
Lang-Firsov limit if the bandwidth goes to zero, and this can be
re-written as an infinite continued fraction, as shown in Eq.
(\ref{t0}). However, this similarity may raise questions about the
relationship between the two approximations.  In this appendix, we
briefly show that the two approximations are very different at all
finite $t$ and $g$.

The meaning of the MA $\bar{g}_0(\omega)$ and of corresponding DMFT
$G_0(\omega)$ (notation of Ref. \onlinecite{ciuchi:1997}) is very
different. The DMFT $G_0(\omega)$ is obtained by solving exactly the
problem of an impurity coupled to an environment in the
$d\rightarrow \infty$ limit, and then imposing the self-consistency
condition that the impurity site behaves similarly to all other
sites in the environment. The DMFT $G_0(\omega)$ is calculated
self-consistently using the following steps:\cite{ciuchi:1997} (i)
with some initial guess for $G_0(\omega)$, one calculates the DMFT
self-energy, given by a formula similar to Eq. (\ref{eq:sigma_MA}),
with $\bar{g}_0(\omega)$ replaced by $G_0(\omega)$; this self-energy
is then used to calculate the total Green's function
$$ G(\omega) = \int_{-\infty}^{\infty}d\epsilon \rho_0(\epsilon)
{1\over \omega - \epsilon - \Sigma(\omega) + i \eta},
$$ where $\rho_0(\epsilon)$ is the density of states of the
non-interacting electrons. The usual procedure is to take
$d\rightarrow \infty$ and use as DOS the semielliptical formula
corresponding to an infinitely branched Bethe tree,
$$ \rho_0(\omega) = {2\over \pi \left({W\over 2}\right)^2}
\sqrt{\left({W\over 2}\right)^2-\epsilon^2},
$$ where $W$ is the bandwidth for the non-interacting system. Then,
(iii) the new $G_0(\omega)$ is extracted from the condition that
$G(\omega) = \left[G^{-1}_0(\omega)-\Sigma(\omega) \right]^{-1}$ and
the procedure is repeated until self-consistency is reached.
Reaching self-consistency is a non-trivial numerical task,
especially compared to obtaining the MA $\bar{g}_0(\omega)$ (see
below). More importantly, the DMFT $G_0(\omega)$ is an explicit
function of $g$ and $\Omega$.

\begin{figure}[t]
\includegraphics[width=0.8\columnwidth]{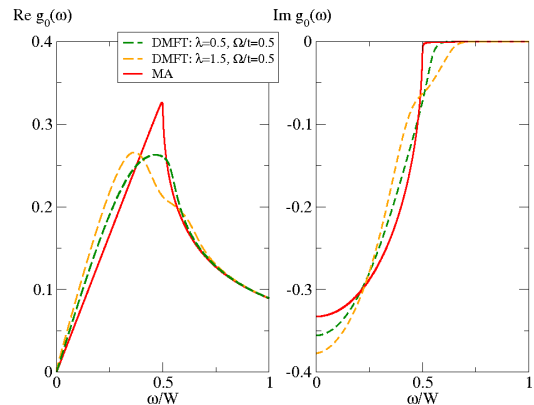}
\caption{(color online) Comparison between the function
  $G_0(\omega)$ entering the DMFT self-energy, for
  $\omega/t=0.5$ and $\lambda=0.5$ (green) and $\lambda=1.5$ (yellow),
  and $\bar{g}_0(\omega)$ entering the MA self-energy, for the
  $d\rightarrow \infty$ semi-elliptical DOS.  }
\label{fig:dmft}
\end{figure}

In contrast, the MA $\bar{g}_0(\omega)$ is the momentum average of
the free propagator: thus, it is a known function, independent of
the parameters $g$ and $\Omega$. In particular, for the
semi-elliptical DOS, we have simply:
\begin{eqnarray}
\nonumber & & \bar{g}_0(\omega)= \int_{-\infty}^{\infty}d\epsilon
  {\rho_0(\epsilon)\over \omega - \epsilon + i \eta} \\ \nonumber & &
  = \frac{8}{W^2}(\omega+ i\eta) \left[ 1 - \sqrt{ 1-
  \frac{W^2}{4(\omega+i\eta)^2} } \quad \right].
\end{eqnarray}

A comparison of these functions is provided in Fig. \ref{fig:dmft}.
They are clearly different.  Moreover, note that in the MA
approximation, $G(\mb{k},\omega)$ is an explicit function of the momentum. The
MA self-energy for the Holstein polaron happens to be independent of
the momentum, but this a consequence of the simplicity of the model,
not an ``in-built'' feature like in DMFT. Generalizations to models
which have a momentum-dependent coupling and/or dispersive phonons,
lead to momentum-dependent self-energies.\cite{berciu:2006}

\end{document}